\documentclass[a4paper,twoside,12pt]{book}

\pdfoutput=1

\usepackage[T1]{fontenc} 
\usepackage[cp1250]{inputenc} 

\usepackage[english]{babel} 
\usepackage{enumitem}

\usepackage[a4paper, hmarginratio=3:2]{geometry} 

\usepackage{pdfpages} 
\usepackage[hidelinks,final]{hyperref} 

\usepackage{graphicx} 
\usepackage{caption}
\usepackage{subcaption}
\usepackage{float} 

\usepackage{caption} 

\usepackage{tabularx} 
\usepackage{amsmath,amssymb,amsthm} 
\usepackage{mathrsfs} 

\usepackage{mathtools}

\usepackage{textcomp}

\usepackage{url}
\DeclareUrlCommand\url{\def\UrlLeft{<}\def\UrlRight{>} \urlstyle{tt}}

\newcommand{\tb}{\textbf} 

\usepackage{cleveref}

\usepackage{bm}

\makeatletter
\g@addto@macro\bfseries{\boldmath}
\makeatother

\renewcommand{\epsilon}{\varepsilon}

\renewcommand{\i}{\mathrm{i}}

\newcommand{\pd}{\partial}
\renewcommand{\d}{{\operatorname{d}}}

\newcommand{\pder}[1]{\ensuremath{\frac{\partial}{\partial #1}}}
\newcommand{\pderA}[2]{\ensuremath{\frac{\partial #1}{\partial #2}}}

\newcommand{\R}{\mathbb{R}}

\newcommand{\N}{\mathbb{N}}

\theoremstyle{plain}

\theoremstyle{remark}

\newcommand{\cvut}{Czech Technical University in Prague}
\newcommand{\fjfi}{Faculty of Nuclear Sciences and Physical Engineering}
\newcommand{\ksi}{Department of Physics}
\newcommand{\program}{Aplikace pøírodních vìd} 
\newcommand{\obor}{Mathematical Physics} 

\newcommand{\druh}{Master's thesis} 
\newcommand{\woman}{} 

\newcommand{\logoCVUT}{\includegraphics{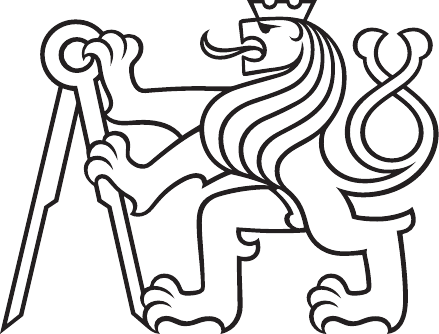}} 

\newcommand{\nazevcz}{Integrabilní a superintegrabilní systémy cylindrického typu v magnetic\-kých polích}  
\newcommand{\nazeven}{Integrable and superintegrable systems of cylindrical type in magnetic fields}     
\newcommand{\autor}{Bc. Ondøej Kubù}  
\newcommand{\vedouci}{doc. Ing. Libor Šnobl, Ph.D.} 
\newcommand{\konzultant}{Prof. Ing. Pavel Winternitz, Ph. D.} 
\newcommand{\pracovisteKonz}{Département de mathématiques et de statistique \& Centre de recherches mathématique, Université de Montréal} 

\newcommand{\rok}{2020} 
\newcommand{\kde}{Praze} 

\newcommand{\klicova}{integrabilita, superintegrabilita, kvantová korekce, magnetické pole, cylindrické souøadnice}  
\newcommand{\keyword}{integrability, superintegrability, quantum correction, magnetic field, cylindrical coordinates}    
\newcommand{\abstrCZ}{Cílem této práce je hledání integrabilních a superintegrabilních systémù cylindrického typu s~magnetickým polem. Po zformulování kvantovì mechanických urèujících rovnic pro integrály pohybu druhého øádu v cylindrických souøadnicích jsou nalezeny všechny kvadraticky integrabilní systémy cylindrického typu. Mezi nimi jsou hledány systémy, které pøipouštìjí dodateèné integrály pohybu. Nejprve jsou pøímým øešením urèujících rovnic nalezeny všechny systémy s~dodateèným integrálem prvního øádu v klasické i kvantové mechanice. Ukazuje se, že všechny tyto systémy již byly známé a žádné další neexistují. Nalezeny jsou také všechny klasické systémy s dodateèným integrálem typu $L^2+\ldots$, respektive $L_x p_y-L_y p_x+\ldots$, z~nichž vìtšina zatím nebyla publikována. Všechny nalezené superintegrabilní systémy pøipouštìjí integrál prvního øádu $L_z$ a jejich Hamiltonovy-Jacobiho, respektive Schr\"odingerovy, rovnice jsou vyøešeny separací promìnných v~cylindrických souøadnicích, u~systémù prvního øádu i v~kartézských.}

\newcommand{\abstrEN}{The goal of this thesis is the search for integrable and superintegrable systems with magnetic field. We formulate the quantum mechanical determining equations for second order integrals of motion in the cylindrical coordinates and we find all quadratically integrable systems of the cylindrical type. Among them we search for systems admitting additional integrals of motion. We find all systems with an additional first order integral both in classical and quantum mechanics. It turns out that all these systems have already been known and no other exist. We also find all systems with an additional integral of type $L^2+\ldots$, respectively $L_y p_y-L_x p_y+\ldots$, of which the majority is new to the literature. All found superintegrable systems admit the first order integral $L_z$ and we solve their Hamilton-Jacobi and Schr\"odinger equations by separation of variables in the cylindrical coordinates, for the first order systems in the Cartesian coordinates as well.} 

\newcommand{\prohlaseni}{Prohlašuji, že jsem svojí diplomovou práci vypracoval\woman{} samostatnì a použil\woman{} jsem pouze podklady (literaturu, projekty, SW atd.) uvedené v pøiloženém seznamu.

Nemám závažný dùvod proti použití tohoto školního díla ve smyslu $\S$ 60 Zákona è.~121/2000 Sb., o právu autorském, o právech souvisejících s právem autorským a~o~zmìnì nìkterých zákonù (autorský zákon).} 

\newcommand{\declaration}{I declare that I wrote my master's thesis independently and that I have not made use of any aid
	(literature, projects, SW etc.) other than those acknowledged.
	
	I agree with the usage of this thesis in the sense of the $\S$ 60 Act 121/2000 (Copyright Act of the Czech
	Republic).}

\newcommand{\podekovani}{First, I would like to thank my supervisor doc. Ing. Libor Šnobl, Ph.D. for perfect collaboration and support in writing the thesis. The weekly consultations were immensely helpful for overcoming smaller or bigger problems encountered during the work on the thesis. I am also very grateful for support and help during the not-so-smooth administration of my stay in Montréal.
	
Second, I would like to thank my consultant and supervisor during the stay at Université de Montréal, Prof. Ing. Pavel Winternitz, Ph.D. for warm hospitality during the stay. I am very grateful for the opportunity to study abroad, which is an invaluable experience, and it would not be possible if Prof. Ing. Pavel Winternitz, Ph. D. did not accept to supervise my thesis.

And third I would like to thank my parents whose support allows me to focus on my studies.
	
This work was supported by the Grant Agency of the Czech Technical University in Prague, grant No. SGS19/183/OHK4/3T/14, and by the Czech Science Foundation (Grant Agency of the Czech Republic), project 17-11805S.} 

\usepackage{microtype}

\begin{document}
	\thispagestyle{empty}
	
	\begin{center}
		{\LARGE
			\cvut\par
			\fjfi
		}
		\vspace{10mm}
		
		\begin{tabular}{c}
			\tb{\ksi} \\[3pt]  
			\tb{Field of study: \obor}\\
		\end{tabular}
		
		\vspace{10mm} \logoCVUT \vspace{15mm} 
		
		{\Large \tb{\nazevcz}\par}
		\vspace{5mm}  
		{\Large\tb{\nazeven}\par}
		
		\vspace{15mm}
		{\Large \MakeUppercase{\druh}}
		
		\vfill
		{\large
			\begin{tabular}{ll}
				Author: & \autor\\
				Supervisor: & \vedouci\\
				Consultant: & \konzultant\\
				Year: & \rok
			\end{tabular}
		}
	\end{center}
	
	\clearpage{\pagestyle{empty}\cleardoublepage} 
	
	\newpage 
	\thispagestyle{empty} 
	
	%
	\includepdf[pages={1,2}]{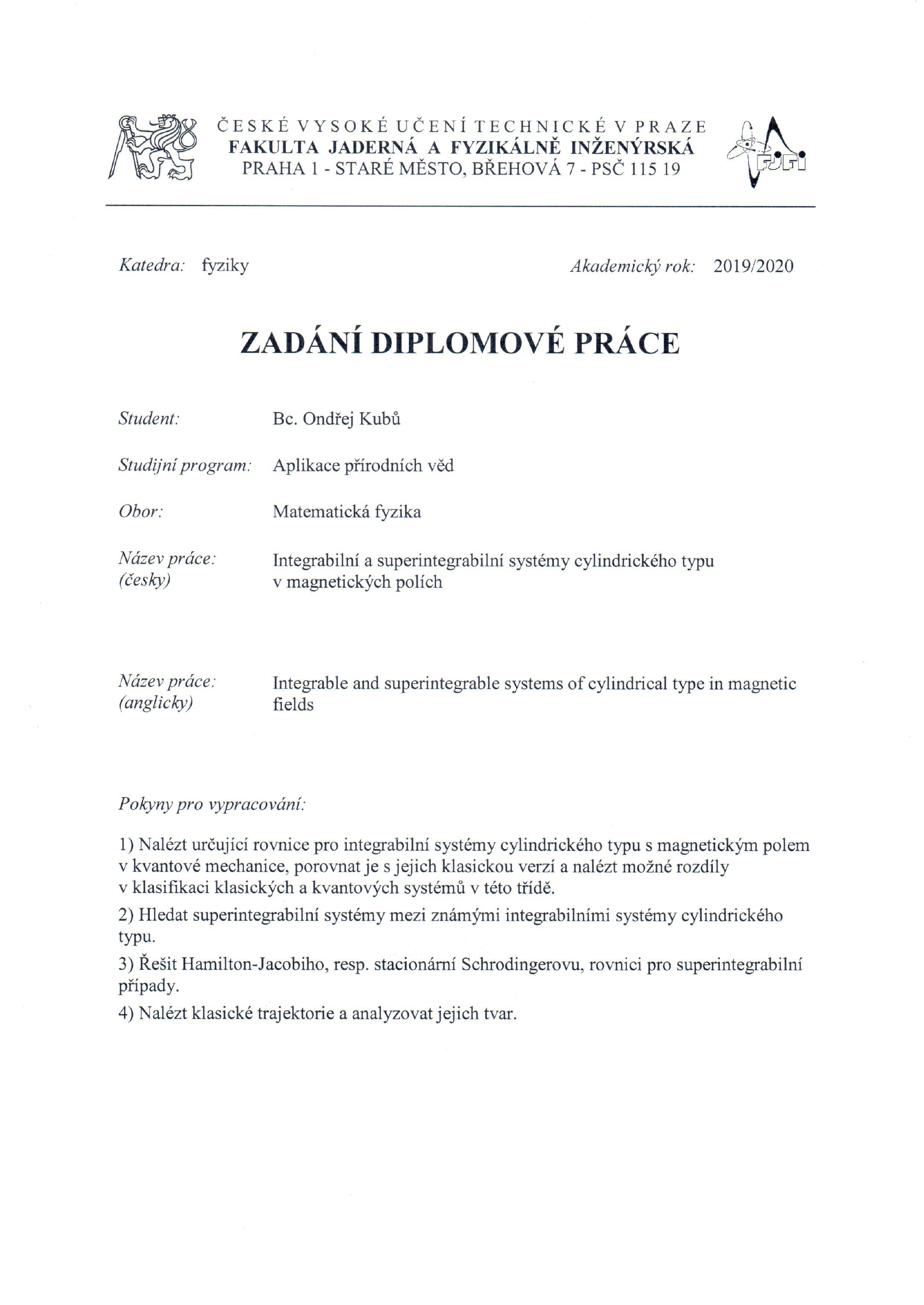} 
	%
	%

	\newpage 
	\thispagestyle{empty} 
	
	~ 
	\vfill 
	
	\tb{Prohlášení} 
	
	\vspace{1em} 
	\prohlaseni
	
	\vspace{3em} 
	\tb{Declaration} 
	
	\vspace{1em} 
	\declaration
	
	\vspace{2em} 
	\hspace{-0.5em}\begin{tabularx}{\textwidth}{X c} 
		V \kde\ dne .................... &........................................ \\	
		& \autor
	\end{tabularx}	

	\newpage
	\thispagestyle{empty}
	
	~
	\vfill 

	\tb{Acknowledgement}
	
	\vspace{1em} 
	\podekovani
	\begin{flushright}
		\autor
	\end{flushright} 

	\newpage  
	\thispagestyle{empty}  
	
	\newbox\odstavecbox
	\newlength\vyskaodstavce
	\newcommand\odstavec[2]{%
		\setbox\odstavecbox=\hbox{%
			\parbox[t]{#1}{#2\vrule width 0pt depth 4pt}}%
		\global\vyskaodstavce=\dp\odstavecbox
		\box\odstavecbox}
	\newcommand{\delka}{113mm} 
	
	\begin{tabular}{ll}
		{\em Název práce:} & ~ \\
		\multicolumn{2}{l}{\odstavec{\textwidth}{\bf \nazevcz}} \\[1em]
		{\em Autor:} & \autor \\[1em]
		{\em Studijní program:} & \program \\
		{\em Obor:} & Matematická fyzika \\
		{\em Druh práce:} & Diplomová práce \\[1em]
		{\em Vedoucí práce:} & \odstavec{\delka}{\vedouci\  Katedra fyziky, Fakulta jaderná a~fyzikálnì inženýrská, Èeské vysoké uèení technické v Praze}\\
		{\em Konzultant:} & \odstavec{\delka}{\konzultant\ \pracovisteKonz}\\ \\[0.1em]
		 
		\multicolumn{2}{l}{\odstavec{\textwidth}{{\em Abstrakt:} ~ \abstrCZ }} \\[0.125em] \\
		{\em Klíèová slova:} & \odstavec{\delka}{\klicova} \\[2em]
		
		{\em Title:} & ~\\
		\multicolumn{2}{l}{\odstavec{\textwidth}{\bf \nazeven}}\\[1em]
		{\em Author:} & \autor \\[1em]
		\multicolumn{2}{l}{\odstavec{\textwidth}{{\em Abstract:} ~ \abstrEN }} \\[0.125em] \\
		{\em Key words:} & \odstavec{\delka}{\keyword}
	\end{tabular}

	\newpage 
	\parskip=0pt
	\tableofcontents 
	\parskip=7pt
	\newpage 

	
	\chapter*{Introduction} 
	\addcontentsline{toc}{chapter}{Introduction} 
	This thesis is a contribution to the study of integrable and superintegrable systems with electric and magnetic fields. These are Hamiltonian systems distinguished by the existence of as many independent integrals of motion in involution as degrees of freedom (integrability) or even some additional not necessarily in involution with the rest (superintegrability) which give them extraordinary properties: in classical mechanics the equations of motion can be solved by quadratures (integrable) or even algebraically in closed form (maximally superintegrable), in quantum mechanics the energy levels are degenerate and it has been conjectured that all maximally superintegrable systems are exactly solvable \cite{TTW}. These properties make them invaluable as physical models which allow us to develop insight into the principles governing physical laws. They are also the basis for constructing more complicated models, often by the method of perturbations. The prime example in this regard is the periodic table \cite{Miller2013}, which is obtained as a perturbation to the Coulomb model. 
	
	The most well-known superintegrable systems are the Kepler-Coulomb system and the harmonic oscillator. As stated by Bertrand's theorem \cite{Bertrand}, see also \cite{Goldstein2001}, these are the only spherically symmetrical maximally superintegrable systems without magnetic field. For analysis of these systems in quantum mechanics see e.g.~\cite{Jauch}. 
	
	Most of the work in the field of superintegrability was done for the systems with so-called natural Hamiltonian
	\begin{equation}\label{natural H}
		H=\frac{1}{2}\left(\vec{p}\right)^2+W(\vec{x}),
	\end{equation}
	mainly with the assumption that the integrals of motion are polynomial in momenta. The best studied cases are those on the Euclidean spaces $\mathbb{E}_2$ and $\mathbb{E}_3$, for which all second order superintegrable systems were found \cite{WinternitzSmorodinski,FrisMandrosov,Evans}. The subsequent developments include studies in higher dimensional Euclidean spaces, more general spaces, e.g. Riemannian or pseudo-Riemannian, and of course higher order integrals, see the review article \cite{Miller2013} and references therein.
	
	Another natural generalization, which we consider in this thesis, is the Hamiltonian for the electromagnetic field, that is with the vector potential $\vec{A}(\vec{x})$ in addition to the scalar potential $W(\vec{x})$, namely
	\begin{equation}\label{HamMagn}
		H=\frac{1}{2}\left(\vec{p}+\vec{A}(\vec{x})\right)^2+W(\vec{x}).
	\end{equation} 
	(We consider an electron in electromagnetic field, so we choose the units of measurement so that $e=-1$ and $m=1$.)
	The earlier work, which focused mainly on the $\mathbb{E}_2$ case, see e.g.~\cite{Berube,McSween2000,Dorizzi}, was followed by the $\mathbb{E}_3$ case, see \cite{Zhalij_2015,Marchesiello2015,Marchesiello2017,Marchesiello2018,BertrandSnobl} and references therein.
	
	The case with the Hamiltonian \eqref{natural H} is related to the separation of Hamilton-Jacobi equation in classical mechanics and Schr\"odinger equation in quantum mechanics. More specifically, in \cite{WinternitzSmorodinski} it was shown that in $\mathbb{E}_3$ there are 11 pairs of commuting integrals corresponding to 11 coordinate systems in which Hamilton-Jacobi or Schr\"odinger equations separate determined in \cite{Eisenhart}. Even though the correspondence between second order integrals and separation no longer holds when magnetic field is present \cite{Benenti,Marchesiello2015}, we still talk about these 11 classes because the highest order terms have the same structure, namely lie in the enveloping algebra of the Euclidean Lie algebra.
	
	The focus of our thesis is one of the 11 coordinate systems, namely the cylindrical with the defining transformations $x=r \cos(\phi),\ y=r \sin(\phi),\ z=Z$. If we have vanishing magnetic field and scalar potential of the form
	\begin{equation}
		W(\vec{x})=W_1(r)+\frac{1}{r^2}W_2(\phi)+W_3(Z),
	\end{equation}
the Hamilton-Jacobi equation separates in these coordinates. In the article \cite{Fournier2019} the authors obtained all classical integrable cases with magnetic field in this class. In this thesis we will continue this research program by studying the quantum integrable cases and determine which of the integrable cases admit additional first order independent integrals of motion, hence being superintegrable. We use Maple\texttrademark~\cite{Maple} for the calculations and occasionally verify some results in Mathematica\textsuperscript{\textregistered}~\cite{Mathematica}.
	
	The structure of the thesis is as follows: After reviewing definitions of integrability and superintegrability in Section~\ref{sec:integr}, the first chapter is focused on integrable systems in quantum mechanics. In Section~\ref{sec: det eq CM} we review classical determining equations for integrals of second order and in Section~\ref{sec: det eq QM} we extend them to quantum mechanics by calculating the quantum correction in the cylindrical coordinates with the focus on the first order and cylindrical integrals. In Section~\ref{sec:quantum integr} we find all quadratic cylindrical integrable systems in quantum mechanics and analyse those differing from the classical case considered in \cite{Fournier2019}.
		
	In Chapter \ref{kap:sup} we search for additional integrals of motion to make the integrable systems superintegrable. In Section~\ref{sec:first ord} we find all cylindrical systems with at least one additional first order integral and in Section~\ref{sec:second ord} we consider second order integrals. Due to the computational complexity of the problem we were not able to solve the second order case in general. After analysing one system we found by chance in Subsection~\ref{sec:biquadr}, we restrict to a physically motivated \emph{ansatz} and find all cylindrical systems admitting integrals of the form $L^2+\ldots$ in Subsection~\ref{sec:L^2} and $L_x p_y- L_y p_x+\ldots$ in Subsection~\ref{sec:Runge-Lenz}. We conclude by a summary of results.


	%
	%
\chapter{Quantum integrable systems in cylindrical coordinates}\label{kap corr}
In this chapter, we consider quantum integrable systems in cylindrical coordinates. In particular, we derive the determining equations in cylindrical coordinates (Section~\ref{sec: det eq QM}), compare them with the classical ones from \cite{Fournier2019}, which we include in Section~\ref{sec: det eq CM} for reference, and solve them for the so-called cylindrical case, i.e. integrals with the highest order term $L_z^2$ or $P_z^2$ (Section~\ref{sec:quantum integr}). 

But first we review the key notions for our thesis: integrability and superintegrability.

\section{Integrability and superintegrability}\label{sec:integr}
We will use the standard definitions in the field, see e.g.~\cite{Marchesiello2015}, which are, however, slightly different from those in the review article \cite{Miller2013}.

We start in classical mechanics: A finite-dimensional classical Hamiltonian system in a $2n$-dimensional phase space is called integrable (or Liouville integrable) if it allows $n$ integrals of motion $\{X_0=H,..., X_{n-1} \}$ (including the Hamiltonian $H$), which are
\begin{enumerate}
	\item well-defined analytic functions on the phase space (possibly with exceptions of lower dimensional manifolds, such as the $z$-axis or $xy$-plane),
	\item in involution, that is pairwise Poisson commute, $ \{X_i, X_j \}_{\text{P.B.}} = 0$, where the Poisson bracket is defined in any canonical coordinates $(q_i, p_i)$ as
	\begin{equation}\label{PB}
		\{X_1, X_2 \}_{\text{P.B.}}=\sum_{i=1}^{n} \left(\pderA{X_1}{q_i}\pderA{X_2}{p_i}-\pderA{X_2}{q_i}\pderA{X_1}{p_i}\right),
	\end{equation}
	\item and are functionally independent, i.e. their Jacobian matrix $\left(\pderA{X_i}{q_j}, \pderA{X_i}{p_k}\right)$ has the maximal possible rank (here $n$) on the region where the integrals are well defined and locally analytic.
\end{enumerate}

A finite dimensional Hamiltonian system is superintegrable if it is integrable with integrals $\{X_0=H, X_1, ..., X_{n-1} \}$ and admits additional integrals of motion $Y_1, \ldots, Y_k, $ $1 \leq k \leq n-1$ so that the set $\{X_0=H, X_1, \ldots, X_{n-1}, Y_1, \ldots, Y_k \}$ is functionally independent. (Integrals $Y_i$ need not Poisson commute with any other integral except the Hamiltonian.) We call the system minimally superintegrable if $k=1$ and maximally superintegrable if $k=2n-1$. (In our case we will have $n=3$ and maximal superintegrability means $k=2$, so there is no other option.)

In quantum mechanics the notions in previous definitions must be slightly modified:
\begin{enumerate}
	\item The integrals of motion must be well-defined operators in the enveloping algebra of the Heisenberg Lie algebra, i.e. polynomials in phase space coordinate operators $\hat{Q}_i$ and $\hat{P}_i$, or convergent series in them. (Here we must work in the Cartesian coordinates because there are some problems with quantizing the momenta in other coordinate systems.)
	\item The Poisson bracket is replaced by the commutator of operators. (We proceed formally, ignoring complications arising from unbounded operators.)
	\item Functional independence is replaced by the so-called algebraic (or polynomial) independence which means that no non-trivial fully symmetrized (Jordan) polynomial of the integrals vanishes.
\end{enumerate}

\section{Classical determining equations}\label{sec: det eq CM}
We are working with magnetic fields so let us review a few basic notions: The magnetic field $B$, which is defined in the Cartesian coordinates as
\begin{equation}
\vec{B}=\nabla\times \vec A, \quad \text{i.e.} \quad B_j=\epsilon_{jkl}\frac{\pd A_l}{\pd x_k},
\end{equation}
where the Levi-Civita symbol $\epsilon_{jkl}$ is totally antisymmetric with $\epsilon_{123}=1$,
is gauge invariant, i.e. does not change under transformation of the potentials
\begin{equation}\label{gauge}
\vec A'(\vec x)=\vec A (\vec x)+\nabla \chi(\vec{x}), \quad V'(\vec x)=V(\vec x)
\end{equation}
with an arbitrary choice of the scalar function $\chi(\vec x)$. (We consider the time independent case only.)
For the effect of gauge transformation in quantum mechanics see \cref{gauge tramsf QM}.

Because the systems we consider are gauge invariant, it is useful to write their integrals of motion in terms of the so-called covariant momenta (using units $e=-1$, $m=1$)
\begin{equation}\label{cov mom CM}
		p_j^A=p_j+A_j.
\end{equation}
Using these momenta, the commutation relations in the Cartesian coordinates read
\begin{equation}\label{PB cov mom}
\{p_j^A, p_k^A\}_\text{P.B.}=\epsilon_{jkl}B_l,\quad \{p_j^A, q_k\}_\text{P.B.}=- \delta_{jk},
\end{equation}
where $\delta_{jk}$ is the Kronecker delta.

To work with momenta and magnetic fields in the cylindrical coordinates $(r, \phi, Z)$ defined by the transformation
\begin{equation}\label{cyl}
x=r \cos(\phi), \quad y=r\sin(\phi), \quad z=Z,
\end{equation}
it is convenient to use the formalism of differential forms introduced in \cite{Marchesiello2018Sph}:
Given the structure of the canonical 1-form
\begin{equation}
\lambda = p_x \mathrm{d}x + p_y \mathrm{d}y + p_z \mathrm{d}z = p_r \mathrm{d}r + p_\phi \mathrm{d}\phi + p_Z \mathrm{d}Z,
\end{equation}
we obtain the following transformation for the linear momenta
\begin{align}\label{transf p}
p_x = \cos(\phi)p_r-\frac{\sin(\phi)}{r}p_\phi, \quad p_y = \sin(\phi)p_r+\frac{\cos(\phi)}{r}p_\phi, \quad p_z = p_Z,
\end{align}
and the components of the vector potential $A$ transform the same way. (We directly associate the Cartesian vector components to the 1-form components.) This enables us to use the gauge invariant momenta from \cref{cov mom CM} in the cylindrical coordinates as well.

On the other hand, components of the magnetic field 2-form $B=\mathrm{d}A$ are
\begin{align}
\begin{aligned}
B &= B^x (\vec{x})\,\mathrm{d}y \wedge \mathrm{d}z + B^y (\vec{x})\, \mathrm{d}z \wedge \mathrm{d}x + B^z (\vec{x})\, \mathrm{d}x \wedge \mathrm{d}y \\
&= B^r (r, \phi, Z)\, \mathrm{d}\phi \wedge \mathrm{d}Z + B^\phi (r, \phi, Z)\, \mathrm{d}Z \wedge \mathrm{d}r + B^Z(r, \phi, Z)\, \mathrm{d}r \wedge \mathrm{d}\phi.
\end{aligned}
\end{align}
This leads to the following transformation
\begin{align} \label{transformB}
B^x (\vec{x}) &= \frac{\cos(\phi)}{r}B^r (r, \phi, Z) - \sin(\phi)B^\phi (r, \phi, Z), \nonumber\\
B^y (\vec{x}) &= \frac{\sin(\phi)}{r}B^r (r, \phi, Z) + \cos(\phi)B^\phi (r, \phi, Z), \\
B^z (\vec{x}) &= \frac{1}{r}B^Z (r, \phi, Z), \nonumber
\end{align}
so that the components of the magnetic field are computed in the same way as in the Cartesian coordinates, namely
\begin{equation}\label{B=dA}
B^r=\pderA{A_Z}{\phi}-\pderA{A_\phi}{Z}, \quad B^\phi=\pderA{A_r}{Z}-\pderA{A_Z}{r}, \quad B^Z=\pderA{A_\phi}{r}-\pderA{A_r}{\phi}.
\end{equation}

Now we search for integrals of motion. In classical mechanics, we use the determining equations from \cite{Fournier2019}, which were obtained by setting the Poisson bracket $\{X, H\}_\text{P.B.}=0$ in the cylindrical coordinates (for definition see \cref{PB}) with the Hamiltonian $H$ written in terms of the covariant momenta from \cref{cov mom CM},
\begin{equation}\label{HamMagnCyl}
H=\frac{1}{2}\left(\left(p_r^A\right)^2+\frac{\left(p_\phi^A\right)^2}{r^2}+\left(p_Z^A\right)^2\right)+W(r, \phi, Z),
\end{equation}
and with a general integral of motion of the second order
\begin{align}\label{IntCyl}
X ={}& h^r \left(r, \phi, Z\right)\left(p_r^A\right)^2 + h^\phi \left(r, \phi, Z\right)\left(p_\phi^A\right)^2 + h^Z \left(r, \phi, Z\right)\left(p_Z^A\right)^2 + \nonumber\\
&+ n^r \left(r, \phi, Z\right)p_\phi^A p_Z^A + n^\phi \left(r, \phi, Z\right)p_r^A p_Z^A + n^Z \left(r, \phi, Z\right)p_\phi^A p_r^A +\\
&+ s^r \left(r, \phi, Z\right)p_r^A + s^\phi \left(r, \phi, Z\right)p_\phi^A + s^Z \left(r, \phi, Z\right)p_Z^A + m\left(r, \phi, Z\right),
\nonumber\end{align}
where the functions $h, $ $n, $ $s, $ $m$ are to be determined. We separate the coefficients of obtained powers of momenta $p_r, $ $p_\phi, $ $p_Z$. The determining equations obtained from the third order terms are
\begin{align} \label{ord3}
\partial_r h^r &= 0, \quad \partial_\phi h^r = -r^2 \partial_r n^Z, \quad \partial_Z h^r = - \partial_r n^\phi, \nonumber\\
\partial_r h^\phi &= -\frac{1}{r^2}\partial_\phi n^Z - \frac{2}{r^3}h^r, \quad \partial_\phi h^\phi = -\frac{1}{r}n^Z, \quad \partial_Z h^\phi = -\frac{1}{r^2}\partial_\phi n^r - \frac{1}{r^3}n^\phi, \nonumber\\
\partial_r h^Z &= -\partial_Z n^\phi, \quad \partial_\phi h^Z = -r^2 \partial_Z n^r, \quad \partial_Z h^Z = 0, \\ \nonumber
\partial_\phi n^\phi &= -r^2(\partial_Z n^Z + \partial_r n^r).
\end{align}
We simplify the second order term equations using the third order \cref{ord3} and rewrite derivatives of $\vec A$ in terms of $\vec B$ using \cref{B=dA} to obtain
\begin{align} \label{ord2}
\partial_r s^r &= n^\phi B^\phi - n^Z B^Z, \nonumber\\
\partial_\phi s^r &= r^2(n^r B^\phi - 2h^\phi B^Z - \partial_r s^\phi) - n^\phi B^r + 2h^r B^Z, \nonumber\\
\partial_r s^Z &= n^Z B^r - \partial_Z s^r - n^r B^Z + 2h^Z B^\phi - 2h^r B^\phi, \nonumber\\
\partial_\phi s^\phi &= -n^r B^r + n^Z B^Z - \frac{1}{r}s^r, \\
\partial_\phi s^Z &= r^2(2h^\phi B^r - n^Z B^\phi - \partial_Z s^\phi) - 2h^Z B^r + n^\phi B^Z, \nonumber\\
\partial_Z s^Z &= n^r B^r - n^\phi B^\phi. \nonumber
\end{align}
After the same simplification the first and zeroth order terms imply
\begin{align} \label{ord1}
\partial_r m &= s^Z B^\phi - s^\phi B^Z + n^\phi \partial_Z W + n^Z \partial_\phi W + 2h^r \partial_r W, \nonumber\\
\partial_\phi m &= s^r B^Z - s^Z B^r + r^2(n^r \partial_Z W + 2h^\phi \partial_\phi W + n^Z \partial_r W), \\
\partial_Z m &= s^\phi B^r - s^r B^\phi + 2h^Z \partial_Z W + n^r \partial_\phi W + n^\phi \partial_r W, \nonumber
\end{align}
and
\begin{equation}\label{ord0}
s^r \partial_r W + s^\phi \partial_\phi W + s^Z \partial_Z W = 0,
\end{equation}
respectively.

The third order equations \eqref{ord3} can be readily integrated, because they do not depend on the magnetic field $\vec{B}$ nor the scalar potential $W$. The solution is therefore the same as in the vanishing magnetic field: The highest order terms lie in the enveloping algebra of the Euclidean Lie algebra \cite{Marchesiello2015} generated by $p_x,$ $p_y, $ $p_z, $ $L_x, $ $L_y, $ $L_z$. For convenience we write it in terms of gauge covariant expressions and in the Cartesian coordinates:
	\begin{equation}
	X=\sum_{1\leq a\leq b\leq 6}\alpha_{ab}Y^A_aY^A_b+\sum_{j=1}^3s_j(\vec{x})p_j^A+m(\vec{x}),
	\end{equation}
	where $Y^A=(p_x^A, p_y^A, p_z^A, L_x^A, L_y^A, L_z^A)$. The coordinates of the gauge covariant angular momentum vector $\vec L^A=(L_1^A, L_2^A, L_3^A)\equiv(L_x^A, L_y^A, L_z^A)$ are defined as
	\begin{equation}
	L_j^A=\sum_{k, l}\epsilon_{jkl}x_k p_l^A
	\end{equation}
with the covariant momenta from \cref{cov mom CM}.

Transforming the second order terms to the cylindrical coordinates using \cref{cyl} and \cref{transf p} followed by collecting the momenta $p^r$, $p^\phi$, $p^Z$, we obtain
\begin{align}\label{sol third 1}
\begin{split}
h^r ={}& \left[\left(\alpha_{55}-\alpha_{44} \right){Z}^2 + \left(\alpha_{24}+\alpha_{15}\right) Z+\alpha_{11}-\alpha_{22} \right] \left(\cos (\phi) \right)^2-\\
& - \left[\alpha_{45}{Z}^2+\left(\alpha_{14}-\alpha_{25} \right) Z-\alpha_{12} \right] \sin (\phi) \cos (\phi) +\alpha_{44}{Z}^2-\alpha_{24}Z+\alpha_{22},\raisetag{32 pt}
\end{split}\\
\begin{split} 
	{h^\phi} ={}&\frac{\left[\left(\alpha_{44}-\alpha_{55} \right){Z}^2- \left(\alpha_{24}+\alpha_{15} \right) Z+\alpha_{22}-\alpha_{11} \right] \left(\cos (\phi) \right)^2}{r^2}+\\
	&+\frac{\left[\left(\alpha_{45}{Z}^2+\left(\alpha_{14}-\alpha_{25} \right)Z -\alpha_{12} \right) \sin (\phi) -	\left(\alpha_{46}Z-\alpha_{26} \right) r\right] \cos (\phi)}{r^2}+\\
 	&+\frac{-\left(\alpha_{56}Z+\alpha_{16} \right) r \sin(\phi) +\alpha_{55}{Z}^2+\alpha_{66}r^2+ \alpha_{15}Z+\alpha_{11}}{r^2},
\end{split}\\
\begin{split}
{h^Z} ={}&r^2\left[\alpha_{44}-\alpha_{45}\cos (\phi) \sin (\phi) - \left(\alpha_{44}-\alpha_{55}\right)\left(\cos (\phi) \right)^2\right]+\\
&\alpha_{34}r\sin(\phi) -\alpha_{35}r\cos (\phi) +\alpha_{33},
\end{split}\\
\begin{split}
{n^r} ={}&\left(2 \alpha_{45}Z-\alpha_{25}+\alpha_{14} \right) \left(\cos (\phi) \right)^2+\\
&+ \left[2 \left(\alpha_{55}-\alpha_{44} \right)Zr + \alpha_{24}+\alpha_{15} \right] \sin (\phi) \cos(\phi)+\\
&+\frac{\left(-\alpha_{56}r^2-\alpha_{34}Z+\alpha_{23} \right) \cos (\phi) + \left(\alpha_{46}r^2-\alpha_{35}Z- \alpha_{13} \right) \sin (\phi)}{r}-\\ 
&-\left(\alpha_{45}Z+\alpha_{14}-\alpha_{36} \right)\label{sol third 6},
\end{split}\\
\begin{split} 
	{n^\phi} ={}& \left[2 \left(\alpha_{44}-\alpha_{55} \right) Z-\alpha_{24}-\alpha_{15} \right]r\left(\cos (\phi) \right)^2+\\ &+\left(2\alpha_{45}Z-\alpha_{25}+\alpha_{14} \right) r\sin (\phi)\cos (\phi)+\\ & +\left(\alpha_{35}Z+\alpha_{13} \right) \cos (\phi) - \left(\alpha_{34}Z-\alpha_{23}\right) r \sin (\phi) -\left(2\alpha_{44}Z-\alpha_{24} \right),
\end{split}\\
\begin{split} 
{n^Z} ={}&-\frac{2\left[\alpha_{45}{Z}^2+ \left(\alpha_{14} -\alpha_{25}\right) Z -\alpha_{12} \right] \left(\cos(\phi) \right)^2}{r}-\\
& -\frac{2\left[\left(\alpha_{55} -\alpha_{44}\right){Z}^2+ \left(\alpha_{24}+ \alpha_{15} \right) Z-\alpha_{22}+\alpha_{11}\right] \sin (\phi)\cos (\phi)}{r}-\\
&- \left(\alpha_{46}Z -\alpha_{26} \right) \sin (\phi)-\left(\alpha_{56}Z+\alpha_{16} \right) \cos (\phi)+\\
& +\frac{\alpha_{45}{Z}^2+ \left(\alpha_{14}-\alpha_{25} \right) Z-\alpha_{12}}{r}.
\end{split}
\end{align}

If we suppose that the integrals of motion are of the first order,
\begin{equation}
X= s^r \left(r, \phi, Z\right)p_r^A + s^\phi \left(r, \phi, Z\right)p_\phi^A + s^Z \left(r, \phi, Z\right)p_Z^A + m\left(r, \phi, Z\right),
\end{equation}
which corresponds to setting the functions $h, $ $n$ in \cref{IntCyl} to 0, the determining equations simplify as follows: There are no equations of the third order (they are satisfied identically).
The second order equations read
\begin{align}\label{ord2 lin}
\begin{aligned}
&\partial_r{s^r} =0, \quad \partial_\phi{s^r} =	-r^2 \partial_r{s^\phi}, \quad \partial_r{s^Z} =-\partial_Z{s^r}, \\
&\partial_\phi{	s^\phi} =-{\frac{{s^r}}{r}}, \quad \partial_\phi{s^Z}=-r^2 \partial_Z{s^\phi}, \quad \partial_Z{s^Z} =0.
\end{aligned}
\end{align}

The first order equations read
\begin{align}\label{ord1 lin}
\partial_r m &={s^Z}{B^\phi} -{s^\phi}{B^Z}, \nonumber\\
\partial_\phi m &={s^r}{B^Z} -{s^Z}{B^r}, \\
\partial_Z m &={s^\phi}{B^r} -{s^r}{B^\phi}, \nonumber
\end{align}
and the zeroth order equation is
\begin{equation}\label{ord0 lin}
{s^r} \partial_r W +{s^\phi} \partial_\phi W +{s^Z} \partial_Z W =0.
\end{equation}

The first thing we note is that the second order equations \eqref{ord2 lin} do not depend on either the magnetic field or the scalar potential in the same way as the third order equations in the previous case. We can therefore solve them for all the cases now. The solution reads
\begin{align}\label{res lin}
s^r &= (k_5 Z + k_1)\cos(\phi) - (k_4 Z- k_2) \sin(\phi), \\
s^\phi &= -\frac{(k_4 Z - k_2) \cos(\phi) + (k_5 Z + k_1)\sin(\phi) - k_6 r}{r}, \\
s^Z &=- k_5 r \cos(\phi)+ k_4 r \sin(\phi) + k_3,
\end{align}
which corresponds to the integral with the first order term in the Euclidean Lie algebra (in the Cartesian coordinates and gauge covariant form)
\begin{equation}
	Y=k_1 p_x^A+k_2 p_y^A+k_3 p_z^A+k_4 L_x^A+k_5L_y^A+k_6 L_z^A+m(x, y, z).
\end{equation}

\section{Determining equations in quantum mechanics}\label{sec: det eq QM}
Let us start with some basic notions regarding the considered systems in quantum mechanics.

We begin in the Cartesian coordinates. We work in the Schr\"odinger representation, where the coordinates and linear momenta are
\begin{equation}
	\hat{Q}_j=x_j, \quad \hat P_j=-\i \hbar \pd_{j}\equiv -\i \hbar \nabla_{j},
\end{equation}
respectively. (The operator $\hat{Q}$ is a multiplication by the corresponding coordinate.)

The convention for operator ordering in the field of integrability is to symmetrize properly \cite{Miller2013}, therefore the Hamiltonian for a system with magnetic field in the Cartesian coordinates reads
\begin{align}
	\hat H&=\frac{1}{2}\sum_{j}\left(\hat P_j \hat P_j+\hat{P}_j \hat A_j(\vec x)+ \hat A_j(\vec x)\hat{P}_j+ \hat A_j(\vec x) \hat A_j(\vec x)\right)+\hat V(\vec x)=\nonumber\\
	&=-\frac{\hbar^2}{2}\Delta -\i\hbar \hat{\vec A}\cdot \vec{\nabla}-\frac{\i\hbar}{2} \widehat{\textrm{div} \vec A}+\frac{1}{2}\hat{\vec A}\cdot\hat{\vec A}+\hat V(\vec x), \label{Ham quant}
\end{align}
where $\hat A_j(\vec x)$, $\widehat{\textrm{div} \vec A}$ and $\hat V(\vec x)$ are the operators of multiplication by the $j$-th coordinate of the vector potential $\vec A(\vec x)$, the divergence of $\vec A(\vec x)$ and the scalar potential $V(\vec x)$, respectively.

We follow the same convention for the integrals of motion as well, namely we write the general second order integral as
\begin{equation}
	\hat{X}=\sum_{j=1}^3\{h_j(\vec{x}), \hat P_j^A \hat P_j^A\}+\sum_{j, k, l=1}^3\frac{|\epsilon_{jkl}|}{2}\{n_j(\vec x), \hat P_k^A \hat P_l^A\}+\sum_{j=1}^3\{s_j(\vec x), \hat P_j^A\}+m(\vec x),
\end{equation}
where $\{, \}$ denotes the symmetrization
\begin{equation}
	\{\hat a, \hat b\}=\frac{1}{2}\left(\hat a\hat b+\hat b \hat a\right).
\end{equation}
All choices of symmetrization are equivalent up to redefinition of the lower order terms \cite{Marchesiello2015}, so the choice is without loss of generality.

In quantum mechanics the gauge transformation \eqref{gauge} demonstrates itself as a unitary transformation $\hat{U}$ of the underlying Hilbert space. Defining $\hat U$ as
\begin{equation}\label{gauge tramsf QM}
	\hat U \psi(\vec x)=\exp\left(\frac{\i}{\hbar}\chi(\vec x)\right)\cdot \psi(\vec x),
\end{equation}
the states and observables describing the same physical system are
\begin{equation}
	\psi\to\psi'=\hat U \psi, \quad \hat O\to\hat O'=\hat U \hat O\hat U^\dagger.
\end{equation}
The potentials and momenta transform the following way
\begin{equation}
	(\hat P_j+\hat A_j)\to\hat U(\hat P_j+\hat A_j)\hat U^\dagger =\hat P_j+\hat A'_j, \quad \hat V=\hat U\hat V\hat U^\dagger.
\end{equation}
We can therefore use the gauge covariant momenta
\begin{equation}\label{cov mom}
\hat P_j^A=\hat P_j+\hat A_j
\end{equation}
in quantum mechanics as well. Their commutation relations in the Cartesian coordinates read
\begin{equation}\label{com rel}
	[\hat P_j^A, \hat P_k^A]=-\i\hbar\epsilon_{jkl}B_l,\quad [\hat P_j^A, \hat{Q}_k]=-\i\hbar \delta_{jk}.
\end{equation}


Now let us turn to the cylindrical coordinates $(r, \phi, Z)$, see \cref{cyl}. Quantization in non-Cartesian coordinates poses a problem, because the Dirac quantization rule
\begin{equation}
	\{A, B\}_\text{P.B}\to \frac{1}{\i\hbar}[\hat A, \hat B]
\end{equation}
is not invariant under canonical transformations (not even coordinate transformations) in the sense that Schr\"odinger equations obtained from different canonical coordinates are inequivalent \cite[Exercise 7.4.10]{Shankar}. Experiments show that the correct Schr\"odinger equation is obtained by quantization of the classical Hamiltonian in the Cartesian coordinates and subsequent transformation into the other coordinate systems. Succinctly, the rule is to quantize and transform (in that order).

Let us therefore transform the quantized Hamiltonian \eqref{Ham quant}. The gradient operator in the cylindrical coordinates reads
\begin{equation}
	\nabla f=\left(\pderA{f}{r}, \frac{1}{r}\pderA{f}{\phi}, \pderA{f}{Z}\right)
\end{equation}
and the Laplace operator is
\begin{equation}
	\Delta=\frac{1}{r}\pder{r}\left(r\pder{r}\right)+\frac{1}{r^2}\frac{\pd^2}{\pd \phi^2}+\frac{\pd^2}{\pd Z^2}.
\end{equation}
The divergence is
\begin{equation}
	\vec\nabla \cdot \vec{A}=\frac{1}{r}\pderA{(r A_r)}{r}+\pderA{A_\phi}{\phi}+\pderA{A_Z}{Z},
\end{equation}
(the right hand side in term of 1-form components)
so the final form of the Hamiltonian in the cylindrical coordinates reads ($A_i$ and $V$ functions of $r, $ $\phi, $ $Z$)
\begin{align}
	\hat H={}&-\frac{\hbar^2}{2}\left(\frac{1}{r}\pder{r}\left(r\pder{r}\right)+\frac{1}{r^2}\frac{\pd^2}{\pd \phi^2}+\frac{\pd^2}{\pd Z^2}\right) -
	\i\hbar \left(\hat{A}_r\pder{r}+\frac{\hat{A}_\phi}{r^2}\pder{\phi}+\hat{A}_Z\pder{Z}\right)-\nonumber\\
	&-\frac{\i\hbar}{2}\left(\frac{1}{r}\pderA{(r A_r)}{r}+\pderA{A_\phi}{\phi}+\pderA{A_Z}{Z}\right)+
	\frac{1}{2}\left((\hat{A}_r)^2+\frac{(\hat{A}_\phi)^2}{r^2}+(\hat{A}_Z)^2\right)+\hat V.
\end{align}
The same procedure must be applied to the integrals of motion as well, we thus quantize in the Cartesian coordinates, symmetrize and transform into the cylindrical coordinates. For this we need the transformation rules for momenta form \cref{transf p}, which we readily invert to get
\begin{equation}
	p_r=\cos (\phi)p_x+\sin(\phi)p_y, \quad p_\phi=-r\sin(\phi)p_x+r\cos(\phi)p_y, \quad p_Z=p_z
\end{equation}
and the momentum operators $\hat P_i$ in the cylindrical coordinates read
\begin{equation}
	\hat P_x=-\i\hbar \left(\cos(\phi)\pd_{r}-\frac{\sin(\phi)}{r}\pd_{\phi}\right), \ \hat P_y=-\i\hbar \left(\sin(\phi)\pd_{r}+\frac{\cos(\phi)}{r}\pd_{\phi}\right), \ \hat P_z=-\i\hbar\pd_{Z}.
\end{equation}

The correct way to obtain the determining equations in the cylindrical coordinates is to transform the Cartesian equations from \cite{Marchesiello2015}. Let us list them for reference:

The third order equations read
\begin{align}\label{third QM cart}
	&\pd_x h_x =0, \quad \pd_y h_x =-\pd_x n_x, \quad \pd_z h_x =-\pd_x n_y, \nonumber\\
	&\pd_x h_y =-\pd_y n_z, \quad \pd_y h_y =0, \quad \pd_z h_y =-\pd_y n_x, \\
	&\pd_x h_z =-\pd_z n_z, \quad \pd_y h_z =-\pd_z n_x, \quad \pd_z h_z =0, \nonumber\\
	&\nabla\cdot \vec{n}=0.	\nonumber
\end{align}
The second order equations are
\begin{align}\label{second QM cart}
&\pd_x s_x=n_y B^y-n_z B^z, \nonumber\\
&\pd_y s_y=n_z B^z-n_x B^x, \nonumber\\
&\pd_z s_z=n_x B^x-n_y B^y, \nonumber\\
&\pd_y s_x+\pd_x s_y=n_x B^y-n_y B^y+2(h_x-h_y)B^z, \\
&\pd_z s_x+\pd_x s_z=n_z B^x-n_x B^z+2(h_z-h_x)B^y, \nonumber\\
&\pd_y s_z+\pd_z s_y=n_y B^z-n_z B^y+2(h_y-h_z)B^x\nonumber
\end{align}
with the following consequence
\begin{equation}
	\nabla \cdot \vec{s}=0.
\end{equation}
The first order
\begin{align}\label{first QM cart}
&\pd_x m=2h_x\pd_x W+n_z\pd_y W+n_y\pd_z W+ s_z B^y-s_y B^z, \nonumber\\
&\pd_y m=2h_y\pd_y W+n_z\pd_x W+n_x\pd_z W+ s_x B^z-s_z B^x, \\
&\pd_z m=2h_z\pd_z W+n_y\pd_x W+n_x\pd_y W+ s_y B^x-s_x B^y.\nonumber
\end{align}
The zeroth order equation containing the $\hbar^2$--proportional quantum correction reads
\begin{equation}\label{zeroth QM cart}
\begin{split}
	\vec{s}\cdot \nabla W+\frac{\hbar^2}{4}\left(\pd_z n_x\pd_zB^x-\pd_y n_x \pd_y B^x+\pd_x n_y\pd_x B^y-\pd_z n_y\pd_z B^y\right.+\\
	+\left.\pd_yn_z\pd_yB^z-\pd_x n_z\pd_x B^z+\pd_x n_x \pd_y B^y-\pd_yn_y\pd_x B^x\right)=0.
\end{split}
\end{equation}
Although the last two terms suggest that the quantum correction is not invariant under Euclidean transformations, which are symmetries the system possesses clearly, it is not the case due to the identity \cite{Marchesiello2015}
\begin{equation}\label{ident}
	\pd_x n_x \pd_y B^y-\pd_yn_y\pd_xB^x=\pd_y n_y \pd_z B^z-\pd_z n_z \pd_yB^y=\pd_z n_z \pd_x B^x-\pd_x n_x\pd_z B^z,
\end{equation}
which follows from
\begin{equation}
	\nabla \cdot \vec{B}=0, \quad \nabla \cdot \vec{n}=0.
\end{equation}

The last missing piece needed to perform the transformation is
\begin{equation}\label{hns transf}
\begin{split}
	h_x&=h^r(\cos(\phi))^2+h^\phi r^2 (\sin(\phi))^2-n^Z r\cos(\phi)\sin(\phi), \\
h_y&=h^r(\sin(\phi))^2+h^\phi r^2 (\cos(\phi))^2+n^Z r\cos(\phi)\sin(\phi), \\
h_z&=h^Z, \\
n_x&=n^r r\cos(\phi)+n^\phi \sin(\phi), \\
n_y&=-n^r r \sin(\phi)+n^\phi \cos(\phi), \\
n_z&=2 h^r \sin(\phi) \cos(\phi)-2 h^\phi r^2 \sin(\phi)\cos(\phi)+n^Z r[(\cos(\phi))^2-(\sin(\phi))^2], \\
s_x&=s^r\cos(\phi)-s^\phi r \sin(\phi), \\
s_x&=s^r\sin(\phi)+s^\phi r \cos(\phi), \\
s_z&=s^Z.
\end{split}
\end{equation}
It is obtained (in classical mechanics) by transforming the linear momenta using \cref{transf p} in the second order integral of motion $X$, see \cref{IntCyl}, and the definition of the functions $n, $ $h, $ $s$ as the coefficients of the corresponding momenta ($h_x$ is the coefficient of $p_x^2$, $n_x$ of $p_yp_z$, $s_x$ of $p_x$ etc. and analogously in the cylindrical coordinates).

Due to the transformation rules \eqref{hns transf}, the third, second and first order equations \eqref{third QM cart}, \eqref{second QM cart} and \eqref{first QM cart}, which do not have any quantum correction, get the same form as in classical mechanics, namely the form of equations \eqref{ord3}, \eqref{ord2} and \eqref{ord1}, respectively.
The transformed zeroth order equation with derivatives of the functions $h$ eliminated using the third order equations \eqref{ord3} reads
\begin{align}
\begin{split}\label{corr cyl}
	&s^r \partial_r W + s^\phi \partial_\phi W + s^Z \partial_Z W+\\
	&+\frac{\hbar^2}{4}\left[-{\frac{\pd_Z n^Z B^r}{r}} +{\frac{\left({r} \pd_\phi n^r + n^\phi \right) B^\phi}{r^2}}+{\frac{\left(r \pd_rn^Z +n^Z \right) B^Z}{r^2}}-\right.\\
	&-{\frac{\left(r\pd_\phi n^r +n^\phi \right) \pd_\phi B^r}{r^3}}+ \frac{\left(r \pd_Z n^Z + r \pd_r n^r +n^r \right)\pd_r B^r}{r}+ \left(\pd_Z n^r\right) \pd_Z B^r-\\
	&-\left(\pd_Z n^\phi \right)\pd_Z B^\phi + \frac{\left(r\pd_r n^r +n^r \right)\pd_\phi B^\phi}{r}+\left(\pd_r n^\phi\right) \pd_r B^\phi+\\
	&+\left.{\frac{\left(r \pd_\phi n^Z +2 h^r-2r^2 h^\phi\right) \pd_\phi B^Z}{r^3}}-{\frac{\left(r \pd_r n^Z + n^Z \right) \pd_r B^Z}{r}} \right]=0.
\end{split}
\end{align}

If we use the identity \cref{ident} to change the last 2 terms in the quantum correction in \cref{zeroth QM cart}, we find that only the terms with diagonal derivatives such as $\pd_r B^r$ change. By taking $\tfrac{1}{3}$ of each term in \cref{ident} instead of the last 2 terms in \cref{zeroth QM cart} and using the identity
\begin{equation}\label{id B}
\pd_r B^r+\pd_\phi B^\phi+\pd_Z B^Z=0,
\end{equation}
which follows from exactness of the 2-form $B=\d A$, we simplify the diagonal terms
\begin{equation}\label{diag}
 \frac{\left(r \pd_Z n^Z + r \pd_r n^r +n^r \right)\pd_r B^r}{r}+\frac{\left(r\pd_r n^r +n^r \right)\pd_\phi B^\phi}{r}=(\pd_Z n^Z) \pd_r B^r - \frac{(r \pd_r n^r+ n^r)\pd_Z B^Z}{r}.
\end{equation}
We note that brackets of the type $\left(r \pd_r n^Z + n^Z \right)$ vanish if $n^Z\sim\frac{1}{r}$. This means vanishing quantum correction for the integrals of motion with the highest order term $P_iP_j, \ i, j\in\{x, y, z\}$, taking into account \cref{hns transf}. (The result can be clearly seen from \cref{zeroth QM cart}, so this merely checks the transformation.)

For comparison, we include the quantum correction with collected functions $n$ and their derivatives. We simplified it using the identities \eqref{id B} and \eqref{diag}.
%
\begin{align}
\begin{split}
&\frac{\hbar^2}{4} \left[{\frac{\left(B^Z - r \pd_r B^Z\right) (n^Z+r\pd_r n_Z)}{r^2}}+
{\frac{\left(r B^\phi -\pd_\phi B^r \right) (n^\phi+r\pd_\phi n^r)}{r^3}}-\right.\\
&-\frac{\left(\pd_Z B^Z \right) (n^r+r\pd_r n^r)}{r}
-\frac{\left(B^r - r\pd_r B^r \right) \pd_Z n^Z}{r}+
{\frac{\left(\pd_\phi B^Z \right) \pd_\phi n^Z}{r^2}}-
\\
&- (\pd_Z B^\phi)\pd_Z n^\phi + \left(\pd_r B^\phi \right) \pd_r n^\phi	+ (\pd_Z B^r )\pd_Z n^r
\left.
+\frac{2 \pd_\phi B^Z \left(h^r -r^2h^\phi\right)}{r^3} \right].
\end{split}
\end{align}

Let us have a look at the effects of the correction \eqref{corr cyl} on the cylindrical integrable systems. From \cref{zeroth QM cart} follows that the determining equations for the first order integrals are the same as in the classical case ($n^i=0$, $i \in\{x, y, z\}$). As we have already mentioned, the correction vanishes for integrals of motion with the highest order term $\hat{P}_i\hat{P}_j,$ $i, j\in\{x, y, z\},$ whose functions $n^i$ are constant. This includes our integral $\hat{P}_Z^2+\ldots$
On the other hand, the integral $\hat{P}_\phi^2+\ldots$  with $h^\phi=1$ corresponds to
\begin{equation}\label{hn pphi}
	h^x=y^2, \quad h^y=x^2, \quad h^z=0, \quad n^z=-{2xy}, \quad n^x=0, \quad n^y=0
\end{equation}
and the quantum correction reads
\begin{equation}\label{corr pphi}
-\frac{\hbar^2}{2r}\pd_\phi B^Z.
\end{equation}
(It can be seen from \cref{corr cyl} directly.)
The correction is non-zero for the general form of $B^Z$ obtained from the highest order determining equations \cite{Fournier2019},
\begin{equation}
	B^Z=-\frac{1}{2r^2}(\psi(\phi)+\psi''(\phi))+r\mu(Z)-\frac{1}{2}\rho'(r),
\end{equation}
see \cref{Bcond} below.

There could be another quantum correction arising from $[\hat{X}_1, \hat{X}_2]=0$, which is the quantum analogue of involutivity condition. Let us show that is not the case:

We first assume that both integrals of motion are of the first order, that is
\begin{align}
	\hat{X}_1=\sum_{j=1}^3\frac{1}{2} \left(s_j(x, y, z) \hat{P}_j^A+\hat{P}_j^A s_j(x, y, z) \right)+m(x, y, z), \label{inv}\\
	\hat{X}_2=\sum_{k=1}^3\frac{1}{2} \left(\tilde{s}_k(x, y, z) \hat{P}_k^A+\hat{P}_k^A \tilde{s}_k(x, y, z)\right)+\tilde{m}(x, y, z).
\end{align}
After some calculations using the commutation relations from eq.~\eqref{com rel},
we obtain
\begin{align}
\begin{split}
	[X_1, X_2]={}&\sum_{j, k}\i\hbar\left(\tilde{s}_k\pd_k s_j \hat{P}_j^A-s_j\pd_j \tilde{s}_k \hat{P}_k^A
-\epsilon_{jkl}s_j\tilde{s}_k B^l-s_j\pd_j\tilde{m}+\tilde{s}_k\pd_k m\right)+\\
&+\sum_{j, k}\frac{\hbar^2}{2}(\tilde{s}_k\pd_{jk}s_j-s_j\pd_{jk}\tilde{s}_k).
\end{split}
\end{align}
The first line corresponds to the Poisson bracket, the second line is an apparent quantum correction.
Collecting the coefficients of $\hat{P}_j^A$ (renaming the dummy indices in the second term), we obtain the first order equations
\begin{equation}
	\sum_{k=1}^3(\tilde{s}_k\pd_k s_j-s_k\pd_k \tilde{s}_j)=0.
\end{equation}
The quantum correction can be rewritten as follows
\begin{equation}
	\sum_{j, k}\frac{\hbar^2}{2}\pd_j(\tilde{s}_k\pd_k s_j-s_k\pd_k\tilde{s}_j),
\end{equation}
because the terms with the first order derivatives cancel if we rename the indices. The bracket contains the first order equations; therefore, the quantum correction vanishes.

Now let us turn to the cylindrical case, i.e. the integrals of motion of the form
\begin{align}
	\hat{X}_1=\left(\hat{L}_z^A\right)^2+\sum_{j=1}^3\frac{1}{2} \left(s_j(x, y, z) \hat{P}_j^A+\hat{P}_j^A s_j(x, y, z) \right)+m(x, y, z),\label{X1 kart} \\
	\hat{X}_2=\left(\hat{P}_z^A\right)^2+\sum_{k=1}^3\frac{1}{2} \left(\tilde{s}_k(x, y, z) \hat{P}_k^A+\hat{P}_k^A \tilde{s}_k(x, y, z)\right)+\tilde{m}(x, y, z).\label{X2 kart}
\end{align}
We impose the relation $[\hat{X}_1, \hat{X}_2]=0$ and separate the equation according to the powers of $P$.
The second order terms
\begin{align}\label{second inv}
	\hat{P}_z \hat{P}_z &: \pd_zs_z&=0, \nonumber\\
	\hat{P}_z \hat{P}_x &:y(x\pd_y-y\pd_x)\tilde{s}_z+2y(yB_x+yB_y)+\pd_zs_x&=0, \nonumber\\
	\hat{P}_z \hat{P}_y &:x(x\pd_y-y\pd_x)\tilde{s}_z+2x(yB_x+yB_y)-\pd_zs_y&=0, \\
	\hat{P}_x \hat{P}_x &:y\tilde{s}_y+y(x\pd_y-y\pd_x)\tilde{s}_x&=0, \nonumber\\
	\hat{P}_y \hat{P}_y &:x\tilde{s}_x-y(x\pd_y-y\pd_x)\tilde{s}_y&=0, \nonumber
\end{align}
give the same equations as would be obtained in the classical case. (The equation for $\hat{P}_x\hat{P}_y$ vanishes as a consequence of those for $\hat{P}_x\hat{P}_x$ and $\hat{P}_y\hat{P}_y$.)

The first order equations obtain some rather complicated apparent corrections, which nevertheless vanish when we use the second order equations \eqref{second inv}. The equations (without the apparent corrections) read
\begin{align}\label{first inv}
&	2y[(x\pd_y-y\pd_x)\tilde{m}+(xB_x+yB_y)\tilde{s}_z-(x\tilde{s}_x+y\tilde{s}_y)B_z]
	+ \sum_{j=1}^3(\tilde{s}_j\pd_js_x-\tilde{s}_j\pd_js_x)&=0, \nonumber\\
&	2x[(x\pd_y-y\pd_x)\tilde{m}+(xB_x+yB_y)\tilde{s}_z-(x\tilde{s}_x+y\tilde{s}_y)B_z]
	+ \sum_{j=1}^3(\tilde{s}_j\pd_js_y-\tilde{s}_j\pd_js_y)&=0, \\	
&	2\pd_z m-2(s_yB_x-s_xB_y)+\sum_{j=1}^3(\tilde{s}_j\pd_js_z-s_j\pd_j\tilde{s}_z)&=0.\nonumber
%
\end{align}
The same happens in the zeroth order: All apparent corrections vanish when we use the higher order conditions \eqref{second inv} and \eqref{first inv}, so the equation has the classical form
\begin{equation}
\tilde{s}_x(s_yB_z-s_zB_y)+\tilde{s}_y(s_zB_x-s_xB_z)+\tilde{s}_z(s_yB_x-s_xB_y)+\sum_{j=1}^3(\tilde{s}_j\pd_j m-s_j\pd_j\tilde{m})=0.
\end{equation}
To sum up: The apparent quantum corrections to the involutivity condition $[\hat{X}_1, \hat{X}_2]=0$ vanish once we impose the higher order conditions from the same commutator for both the first order and the cylindrical integrals. We note that we do not need to use the determining equations for the integrals to prove the assertion.

In the cylindrical case there is, therefore, only one equation with quantum correction, namely the zeroth order equation for $X_1$, which now reads
\begin{equation}\label{ord 0 zmena}
	s_1^r \partial_r W + s_1^\phi \partial_\phi W + s_1^Z \partial_Z W -\frac{\hbar^2}{2r}\pd_\phi B^Z= 0.
\end{equation}
In order to consider the quantum correction for integrals containing angular momenta in the second order term, we suggest using the Cartesian form of the correction from \cref{zeroth QM cart} or transforming it into more suitable coordinates because using the correction from \cref{corr cyl} seems not tractable.

\section{Quantum integrable systems}\label{sec:quantum integr}
Here we solve the quantum determining equations for the case of cylindrical integrals in the cylindrical coordinates. As we have seen in the last section, the quantum correction has arisen in one equation only, namely the zeroth order equation for the integral $\hat{X}_1=(\hat{P}_\phi^A)^2+\ldots$, see eq.~\eqref{ord 0 zmena}. Because there are so little changes, we will closely follow the analysis from \cite{Fournier2019} and focus mainly on the cases which are affected by the correction.

We start with reducing the determining equations \eqref{ord3}--\eqref{ord0} to the case of cylindrical integrals, which in classical mechanics read
\begin{equation}
\begin{split} \label{cyl integrals}
& X_1=(p_\phi^A)^2+s_1^r(r, \phi, Z)p_r^A+s_1^\phi(r, \phi, Z)p_\phi^A+s_1^Z(r, \phi, Z)p_Z^A+m_1(r, \phi, Z), \\
& X_2=(p_Z^A)^2+s_2^r(r, \phi, Z)p_r^A+s_2^\phi(r, \phi, Z)p_\phi^A+s_2^Z(r, \phi, Z)p_Z^A+m_2(r, \phi, Z).
\end{split}
\end{equation}
To obtain their quantum mechanical form, we would have to transform the integrals into the Cartesian coordinates, quantize with proper symmetrization and transform back into the cylindrical coordinates. Because we have transformed the determining equations from the Cartesian form, we do not need the explicit form of the integrals here.

The reduction is done by substituting the appropriate values for the functions $h, $ $n$, namely the only non-zero functions
 \begin{equation}
	h^\phi_1=1, \quad h_2^Z=1,
\end{equation}
into the general determining equations \eqref{ord3}--\eqref{ord0} and taking into account the quantum correction \eqref{corr pphi} in the zeroth order equation. We obtain the following:

The third order equations \eqref{ord3} are trivially satisfied for both integrals. The second order equations for the integrals $\hat{X}_1$ and $\hat{X}_2$ read
\begin{equation}
\begin{split}\label{cyl1sec}
\partial_r s_1^r &= 0, \quad \partial_\phi s_1^\phi = -\frac{s_1^r}{r}, \\
\partial_\phi s_1^r &= -r^2(\partial_r s_1^\phi + 2 B^Z), \quad \partial_\phi s_1^Z = -r^2(\partial_Z s_1^\phi - 2 B^r), \\
\partial_r s_1^Z &= -\partial_Z s_1^r, \quad \partial_Z s_1^Z = 0,
\end{split}
\end{equation}
and
\begin{equation}
\begin{split}\label{cyl2sec}
\partial_r s_2^r &= 0, \quad \partial_\phi s_2^\phi = -\frac{s_2^r}{r}, \\
\partial_\phi s_2^r &= -r^2 \partial_r s_2^\phi, \quad \partial_\phi s_2^Z = -r^2 \partial_Z s_2^\phi - 2 B^r, \\
\partial_r s_2^Z &= -\partial_Z s_2^r + 2 B^\phi, \quad \partial_Z s_2^Z = 0,
\end{split}
\end{equation}
respectively.

The first order equations reduce to those for $\hat{X}_1$
\begin{align} \label{cyl 1 fir}
\partial_r m_1 &= s_1^Z B^\phi - s_1^\phi B^Z, \nonumber\\
\partial_\phi m_1 &= s_1^r B^Z - s_1^Z B^r + 2 r^2 \partial_\phi W, \\
\partial_Z m_1 &= s_1^\phi B^r - s_1^r B^\phi, \nonumber
\end{align}
and those for $\hat{X}_2$
\begin{align} \label{cyl 2 fir}
\partial_r m_2 &= s_2^Z B^\phi - s_2^\phi B^Z, \nonumber\\
\partial_\phi m_2 &= s_2^r B^Z - s_2^Z B^r, \\
\partial_Z m_2 &= s_2^\phi B^r - s_2^r B^\phi + 2 \partial_Z W \nonumber.
\end{align}
One of the zeroth order equations is the only one to change with respect to the classical case:
\begin{align} \label{cyl0}
&s_1^r \partial_r W + s_1^\phi \partial_\phi W + s_1^Z \partial_Z W -\frac{\hbar^2}{2r}\pd_\phi B^Z= 0, \\
&s_2^r \partial_r W + s_2^\phi \partial_\phi W + s_2^Z \partial_Z W = 0.
\end{align}

In addition to the commutation with the Hamiltonian, i.e. the condition on integrals of motion, we impose the commutation of the integrals $[\hat{X}_1, \hat{X}_2]=0$, which corresponds to the classical notion of involution. We obtain additional equations for each order in momenta, namely those of the second order
\begin{align} \label{extra2}
\partial_\phi s_2^\phi &= 0, \quad \partial_\phi s_2^r = 0, \quad \partial_Z s_1^r = 0, \quad \partial_\phi s_2^Z = \partial_Z s_1^\phi - 2 B^r,
\end{align}
of the first order
\begin{align} \label{extra1}
&s_2^Z \partial_Z s_1^r + s_2^\phi \partial_\phi s_1^r &= 0, \nonumber\\
&-s_1^\phi(2 B^r + \partial_\phi s_2^Z) + s_2^Z \partial_Z s_1^Z - s_1^Z \partial_Z s_2^Z +\nonumber\\
&+ s_2^\phi \partial_\phi 2_1^Z + s_1^r(2 B^\phi - \partial_r s_2^Z) + 2 \partial_Z m_1 &= 0, \\
&-s_2^Z(2 B^r - \partial_Z s_1^\phi) + s_2^\phi \partial_\phi s_1^\phi- \nonumber\\
&- s_1^Z \partial_Z s_2^\phi - s_1^r \partial_r s_2^\phi - 2 \partial_\phi m_2 &= 0, \nonumber
\end{align}
and of the zeroth order
\begin{align} \label{extra0}
&- s_1^r \partial_r m_2 + s_2^\phi \partial_\phi m_1 - s_1^\phi \partial_\phi m_2 + s_2^Z \partial_Z m_1 - s_1^Z \partial_Z m_2+&\nonumber\\
&+ B^r(s_2^\phi s_1^Z - s_1^\phi s_2^Z) + B^\phi s_1^r s_2^Z - B^Z s_1^r s_2^\phi &= 0.
\end{align}
None of them has any quantum correction, as was considered at the end of Section~\ref{sec: det eq QM} (in the Cartesian coordinates).

The second order equations \eqref{cyl1sec}, \eqref{cyl2sec} and \eqref{extra2} can be solved for the functions $s_j$ and the magnetic field $B$ in terms of 5 functions of one variable each, which we call the auxiliary functions:
\begin{align}
\begin{split}\label{scond}
s_1^r &=\frac{\mathrm{d}}{\mathrm{d}\phi}\psi(\phi), \quad s_1^\phi=-\frac{\psi(\phi)}{r}-r^2\mu(Z)+\rho(r), \quad s_1^Z=\tau(\phi), \\
s_2^r&=0, \quad s_2^\phi=\mu(Z), \quad s_2^Z=-\frac{\tau(\phi)}{r^2}+\sigma(r),
\end{split}\\
\begin{split} \label{Bcond}
B^r&=-\frac{r^2}{2}\frac{\mathrm{d}}{\mathrm{d}Z}\mu(Z)+\frac{1}{2r^2}\frac{\mathrm{d}}{\mathrm{d}\phi}\tau(\phi), \quad B^\phi=\frac{\tau(\phi)}{r^3}+\frac{1}{2}\frac{\mathrm{d}}{\mathrm{d}r}\sigma(r), \\
B^Z&=\frac{-\psi(\phi)}{2r^2}+r\mu(Z)-\frac{1}{2}\frac{\mathrm{d}}{\mathrm{d}r}\rho(r)-\frac{1}{2r^2}\frac{\mathrm{d}^2}{\mathrm{d}\phi^2}\psi(\phi).
\end{split}
\end{align}
From now on we use primes for derivatives of functions of one variable, but we use dot for derivatives with respect to time.

We substitute the result into the remaining determining equations to replace the functions $s$ and the magnetic field $B$ by the auxiliary functions $\rho(r), $ $\sigma(r), $ $\tau(\phi), $ $\psi(\phi), $ $\mu(Z)$. The first order equations \eqref{cyl 1 fir}, \eqref{cyl 2 fir} and \eqref{extra1} give us one direct condition on the auxiliary functions $\psi(\phi)$ and $\mu(Z)$ and several conditions on the derivatives of $m_1$ and $m_2$:
\begin{align} \label{1_0cond}
& \left(r \rho(r) - \psi(\phi)-r^3 \mu(Z) \right) \left(\psi''(\phi) + r^2 \rho'(r) \right) +\nonumber&\\
&+ \left(r^3 \mu(Z) + r \rho(r) \right) \psi(\phi) - \psi(\phi)^2 + r^3 \tau(\phi) \sigma'(r) +\nonumber&\\
&+ 2 r^6 \mu(Z)^2 - 2r^4 \rho(r) \mu(Z) + 2 \tau(\phi)^2 - 2r^3 \pd_r m_1 &{}={}&0, \nonumber&\\
& \psi'(\phi) \left(2r^3 \mu(Z) - r^2 \rho'(r) - \psi(\phi) - \psi''(\phi) \right) +\nonumber&\\
& + \tau(\phi) \left(r^4 \mu'(Z) - \tau'(\phi) \right) + 4r^4 W_\phi - 2r^2 \pd_\phi m_1 &{}={}& 0, \nonumber&\\
& \left(\tau'(\phi) - r^4 \mu'(Z) \right) \left(r \rho(r) - \psi(\phi) -r^3 \mu(Z) \right) -\nonumber&\\
& - \psi'(\phi) \left(r^3 \sigma'(r) + 2\tau(\phi) \right) -2r^3 \pd_Z m_{1} &{}={}& 0, \nonumber&\\
& r^3 \mu(Z) (\psi''(\phi)+ \psi(\phi)) + r^3 \sigma'(r) \left(r^2 \sigma(r) - \tau(\phi) \right) - 2r^6 \mu(Z)^2+&\\
& + r^5 \mu(Z) \rho'(r) + 2r^2 \sigma(r) \tau(\phi) - 2\tau(\phi)^2 - 2r^5 \pd_r m_2 &{}={}& 0, \nonumber&\\
& \left(r^4 \mu'(Z) - \tau'(\phi) \right) \left(r^2 \sigma(r) - \tau(\phi) \right) -2r^4 \pd_\phi m_{2} &{}={}&0, &\nonumber\\
& -r^4 \mu(Z) \mu'(Z) + \mu(Z) \tau'(\phi) + 4r^2 W_Z - 2r^2 \pd_Z m_{2} &{}={}&0, \nonumber&\\
& \mu(Z)\psi''(\phi)&{}={}&0, \nonumber&\\
& \left(r^2 \rho(r) - r \psi(\phi)-r^4 \mu(Z) \right) \mu'(Z) + \mu(Z) \tau'(\phi) + 2 \pd_Z m_1 &{}={}& 0, &\nonumber\\
&\tau'(\phi) \left(r^2 \sigma(r) - \tau(\phi) \right) + r^4 \tau(\phi) \mu'(Z) + r^3 \mu(Z) \psi'(\phi) + 2r^4 \pd_\phi m_{2} &{}={}& 0. \nonumber &
\end{align}

We see that we have two equations for each of the derivatives $\pd_Z m_1 $ and $\pd_\phi m_2$, so we obtain constraints for the auxiliary functions (the two values of $\pd_Z m_1 $ and $\pd_\phi m_2$ must coincide). Assuming $m_1$ and $m_2$ to be sufficiently smooth, we impose the Clairaut compatibility conditions $\pd_{ba}m_i=\pd_{ab}m_i$ on the second derivatives and obtain equations for the mixed second derivatives of the scalar potential $W$, which we list shortly, after we analyse the zeroth order equations.

Only one of the zeroth order equations \eqref{cyl0} and \eqref{extra0} obtains the quantum correction:
\begin{align} \label{0_0cond}
\begin{aligned}
& \left(r^2 \sigma(r) - \tau(\phi)\right)W_Z + r^2 \mu(Z) W_\phi &{}={}& 0, &\\
& \left( r \rho(r) - \psi(\phi)-r^3 \mu(Z)\right)W_\phi + r\left(\psi'(\phi)W_r + \tau(\phi)W_Z\right)+ &\\ &+\hbar^2\frac{\psi'''(\phi)+\psi'(\phi)}{4r^2} &{}={}& 0,  &\\
& 2r^4 \left(r^3 \mu(Z) - r \rho(r) + \psi(\phi) \right) \pd_\phi m_{2} + 2 r^3 \left(r^2 \sigma(r) - \tau(\phi) \right) \pd_Z m_1 + &\\
& + r^3 \mu(Z) \psi'(\phi)\psi''(\phi) + 2r^5 \mu(Z) \pd_\phi m_1 - 2r^5 \tau(\phi)\pd_Z m_{2} + &\\
& + \left[r^3 \left(r^2 \sigma(r) - \tau(\phi) \right)\sigma'(r) -2r^6 \mu(Z)^2 +r^5 \mu(Z) \rho'(r) +\right.&\\
& \left. + r^3 \mu(Z) \psi(\phi) + 2r^2 \sigma(r) \tau(\phi) -2\tau(\phi)^2 - 2r^5 \pd_r m_{2} \right]\psi'(\phi)- &\\
& - \left[r^2 \left(r^3 \mu(Z) - r \rho(r) + \psi(\phi) \right)\sigma(r) \right.+ &\\
& + \left. \tau(\phi) \left(r \rho(r) - \psi(\phi) \right) \right]\left(r^4 \mu'(Z) - \tau'(\phi) \right) &{}={}& 0.  &
\end{aligned}
\end{align}
Substituting for the derivatives of $m$ from \cref{1_0cond} into eq.~\eqref{0_0cond}, we obtain a system of linear inhomogeneous algebraic equations for the first derivatives of the scalar potential $W$.

The final form of the reduced equations, which we analyse hereafter, is as follows. (Indexes of $W$ mean partial derivatives.)
\begin{align}
\psi'(\phi) \left(r^3 \sigma'(r) + 2 \tau(\phi) \right) - \tau'(\phi) \left(r \rho(r) - \psi(\phi) \right) ={}& 0, \label{reducedAa}\\
\mu(Z) \psi'(\phi) + r^3 \sigma(r)\mu'(Z) ={}& 0, \label{reducedAb}
\end{align}
\begin{align} \label{reducedB}
 W_{r \phi} ={}&-\frac{2}{r}W_\phi + \frac{1}{4r^5}\left[\psi'(\phi) \left( r^3 (\rho''(r)- \mu(Z)) - r^2 \rho'(r) + r \rho(r) -3\psi''(\phi) - 4\psi(\phi) \right)+ \right. \nonumber\\
 &+ \left. \tau'(\phi) \left(r^3 \sigma'(r) + 2\tau(\phi) \right) - 2r^4 \tau(\phi) \mu'(Z) - \psi'''(\phi) \left(\psi(\phi) - r \rho(r) \right) \right], \nonumber\\
 W_{\phi Z} ={}&-\frac{1}{4r^2} \left[r^2 \mu''(Z) \left(\tau(\phi) - r^2 \sigma(r) \right) + \tau''(\phi) \mu(Z) \right], \\
 W_{r Z} ={}&\frac{1}{4r^3} \left[r \mu'(Z) \left(r^2 \rho'(r) + \psi(\phi) -2r^3 \mu(Z)\right) + 2 \mu(Z) \tau'(\phi) \right], \nonumber
\end{align}

\begin{align} \label{matrixform2}
\begin{pmatrix}
0 & r^2 \mu(Z) & r^2 \sigma(r) - \tau(\phi)\\
\psi'(\phi) & \rho(r) -r^2 \mu(Z) - \frac{\psi(\phi)}{r} & \tau(\phi)\\
0 & 4 r^7 \mu(Z) & -4 r^5 \tau(\phi)
\end{pmatrix}
\cdot \begin{pmatrix}
W_r\\
W_\phi\\
W_Z
\end{pmatrix}=
\begin{pmatrix}
	0\\
	-\frac{\hbar^2(\psi'''(\phi)+\psi'(\phi))}{4r^3}\\
	\alpha(r, \phi, Z)
\end{pmatrix}
\end{align}
with
\begin{align} \label{alpha}
 \alpha(r, \phi, Z) ={}& \psi'(\phi)
\left[\left(r^5 \sigma(r) - r^3 \tau(\phi)\right)\sigma'(r) + r^5 \mu(Z) \rho'(r) - 2 \tau(\phi)^2 \nonumber \right.+\\
 & \left. + 2 r^2 \sigma(r) \tau(\phi) - r^3 \mu(Z)\left(r^3 \mu(Z) + r \rho(r) - 2 \psi(\phi)\right)\right]\nonumber +\\
 & +\tau'(\phi)\left[\left(r \rho(r) - \psi(\phi)\right)\tau(\phi) + r^2 \sigma(r)\left(r^3 \mu(Z) - r \rho(r) + \psi(\phi)\right)\right] \nonumber -\\
 &- r^4 \mu'(Z) \tau(\phi)\left(r \rho(r) - \psi(\phi)\right).
\end{align}
We denote the matrix in \cref{matrixform2} by $M$.

The only change with respect to the classical case from \cite{Fournier2019} is the non-zero RHS in the system of linear inhomogeneous algebraic equations \eqref{matrixform2}, corresponding to eq.~(39) in the original article.

We separate the analysis of the reduced system into several cases with respect to the rank of the matrix $M$, following \cite{Fournier2019}. It can be either $3$, $2$ or $1$. Rank $0$ would mean that all the auxiliary functions vanish and with them the magnetic field as well, see \cref{Bcond}, so we rule this case out.

If the rank is $3$, then the determinant of $M$,
\begin{equation} \label{detM}
\det(M) = 4 r^9 \psi'(\phi)\mu(Z)\sigma(r),
\end{equation}
is not zero and it implies a unique solution for each first derivative of $W$. The analysis of this case from \cite{Fournier2019} shows that this assumption leads to a contradiction and the reduced system is inconsistent. The analysis remains valid because it uses the matrix $M$ and equations \eqref{reducedAa} and \eqref{reducedAb} only, which are unaffected by the quantum correction. \newpage

If the rank is either $2$ or $1$ instead, then $\det(M)=0$, and from \cref{detM} we see that there are {\it a priori} three possible cases:
\begin{enumerate}[label=\alph*)]
	\item $\psi'(\phi)=0$,
	\item $\psi'(\phi) \neq 0$ and $\mu(Z)=0$,
	\item $\psi'(\phi) \neq 0$, $\mu(Z) \neq 0$ and $\sigma(r)=0$. We rule this case out due to inconsistency with~\cref{reducedAb}.
\end{enumerate}

The assumption of case a) implies that the quantum correction vanishes; thus, we obtain the classical systems from \cite{Fournier2019}. Therefore, we continue with the case b) only, i.e. $\psi'(\phi) \neq 0$ and $\mu(Z)=0$. (The key results for case a) are summarised in the corresponding subsections of Section \ref{sec:first ord}, where we search for additional integrals of motion.)

Before we go to the specific subcases, we show that $\alpha=0$ for $\alpha$ from \cref{alpha}. In the considered case it follows from \cref{reducedAa} and \cref{reducedAb} only, so the considerations are the same as in \cite{Fournier2019}:

Assuming $\psi'(\phi) \neq 0$ and $\mu(Z)=0$, \cref{reducedAb} is satisfied trivially. By differentiating \cref{reducedAa} with respect to $r$, we obtain
\begin{equation}\label{reducedAadr}
\left(r^3 \sigma'(r)\right)' = \frac{\tau'(\phi)}{\psi'(\phi)} \left(r \rho(r)\right)'.
\end{equation}
This leads to 3 possibilities:
\begin{itemize}
	\item $\left(r^3 \sigma'(r)\right)' =\left(r \rho(r)\right)'=0$, i.e.
	\begin{equation}\label{reducedAadr1}
	\sigma(r)=\frac{C_\sigma}{r^2}+\tilde{C}_\sigma, \qquad \rho(r)= \frac{C_\rho}{r}.
	\end{equation}
	Substituting~\cref{reducedAadr1} into \cref{reducedAa} we find
	\begin{equation}
	2 \left(\tau(\phi) -C_{\sigma}\right) \psi'(\phi)+ \left(\psi(\phi)-C_{\rho}\right ) \tau'(\phi)=0
	\end{equation}
	which directly implies that $\alpha$ defined in~\cref{alpha} vanishes.
	\item $\left(r^3 \sigma'(r)\right)' =\tau'(\phi)=0$, i.e.
	\begin{equation}\label{reducedAadr2}
	\sigma(r)=\frac{C_{\sigma}}{r^2}+\tilde{C}_{\sigma}, \qquad \tau(\phi)=C_{\tau}.
	\end{equation}
	Substituting~\cref{reducedAadr2} into \cref{reducedAa} we find $C_{\sigma}=C_{\tau}$ and that together with \cref{reducedAadr2} implies again that $\alpha=0$ in~\cref{alpha}.
	\item $\frac{\left(r^3 \sigma'(r)\right)'}{\left(r \rho(r)\right)'} = \frac{\tau'(\phi)}{\psi'(\phi)} = \lambda \neq 0, $ implying that
	\begin{equation}\label{reducedAadr3}
	\rho(r)=\frac{1}{\lambda} r^2 \sigma'(r)+\frac{C_\rho}{r}, \qquad \tau(\phi)= \lambda \psi(\phi)+C_\tau.
	\end{equation}
	Substituting~\cref{reducedAadr3} into~\cref{reducedAa} and differentiating it with respect to $\phi$ we arrive at a contradiction with our assumptions, namely $\lambda \psi'(\phi) =0$.
\end{itemize}

\subsection{Case b) $\mathrm{rank}\, (M)=2$: $\psi'(\phi) \neq 0$ and $\mu(Z)=0$}
Here the second and third equations in~\eqref{reducedB} reduce to
\begin{equation}
	\pd_{rZ}W=0, \quad \pd_{\phi Z}W=0,
\end{equation}
implying separation of the $Z$ coordinate from $r$ and $\phi$ in the scalar potential $W$, i.e.
\begin{equation}\label{W123}
W(r, \phi, Z) = W_{12}(r, \phi) + W_3(Z).
\end{equation}
The reduced row echelon form of the extended matrix of the system of equations~\eqref{matrixform2} reads
\begin{equation}\label{redMcase2b}
\begin{pmatrix}
r \psi'(\phi) & r \rho(r) - \psi(\phi) & 0 &-\frac{\hbar^2(\psi'''(\phi)+\psi'(\phi))}{4r^2}\\
0 & 0 & \sigma(r) &0\\
0 & 0 & \tau(\phi)&0
\end{pmatrix}
\end{equation}
Our assumption $\psi'(\phi) \neq 0$ implies that we have two possibilities to have $\mathrm{rank}\, (M)=2$, namely either $\sigma(r) \neq 0$ or $\tau(\phi) \neq 0$. Both of them imply $W_3(Z)'=0$ and without loss of generality we can absorb the constant into redefinition of $W_{12}$, so we have $W(r, \phi, Z)=W_{12}(r, \phi)\equiv W(r, \phi)$.
\begin{enumerate}[label=\arabic*)]
	\item $\sigma(r)\neq0$:
	We follow the further splitting into the subcases from \cite{Fournier2019}.
	We use our assumption $\psi'(\phi)\neq0$ to rewrite~\cref{reducedAa} in the following way:
	\begin{equation}
	r^3 \sigma'(r) + 2 \tau(\phi) - \frac{\tau'(\phi)}{\psi'(\phi)}(r \rho(r) - \psi(\phi)) = 0.
	\end{equation}
	Differentiation with respect to $\phi$ leads to the equation:
	\begin{equation} \label{separablemess}
	3 \tau'(\phi) + \psi(\phi) \frac{\tau''(\phi) \psi'(\phi) - \tau'(\phi) \psi''(\phi)}{\psi'(\phi)^2} = r \rho(r) \frac{\tau''(\phi) \psi'(\phi) - \tau'(\phi) \psi''(\phi)}{\psi'(\phi)^2}.
	\end{equation}
	If $\tau''(\phi) \psi'(\phi) - \tau'(\phi) \psi''(\phi) \neq 0$, we can separate the variables $r$ and $\phi$, else the expression vanishes and we conclude from \cref{separablemess} that $\tau'(\phi)=0$, thus $\tau(\phi)=\tau_0$ is a constant. We treat the subcases separately.
	\begin{enumerate}[label=1.\arabic*)]
		\item\label{1.1} $\tau''(\phi) \psi'(\phi) - \tau'(\phi) \psi''(\phi) = 0$, so $\tau(\phi)=\tau_0$: From \cref{reducedAa}, which now reads $r^3\sigma'(r)=-2\tau_0$, follows
		\begin{equation}
			\sigma(r) = \frac{\tau_0}{r^2} + \sigma_0.
		\end{equation}
		The yet unsolved equations from eq.~\eqref{reducedAa}--\eqref{matrixform2} constraining the scalar potential $W$ read
		\begin{align}
		r\psi'(\phi)W_r + \left(r \rho(r) - \psi(\phi) \right)W_\phi +\frac{\hbar^2(\psi'''(\phi)+\psi'(\phi))}{4r^2}&= 0, \label{mcsween1}\\
		\psi'(\phi) \left( r^3 \rho''(r) - r^2 \rho'(r) + r \rho(r) -3 \psi''(\phi) - 4 \psi(\phi) \right)+\nonumber\\
		+ \psi'''(\phi) \left(r \rho(r) - \psi(\phi) \right) - 4 r^5 W_{r \phi} - 8 r^4 W_\phi &= 0.\label{mcsween2}
		\end{align}
		The magnetic field takes the form
		\begin{equation}\label{r2mu0B}
		B^\phi = 0, \quad B^r = 0, \quad B^Z = -\frac{\rho'(r)}{2} -\frac{\psi''(\phi)+\psi(\phi)}{2r^2}.
		\end{equation}
		Thus, we have a free motion in the $z$-direction (in the direction of the $z$-axis) and motion in the $xy$-plane under the influence of the scalar potential $W(r, \phi)$ constrained by \cref{mcsween1} and \cref{mcsween2} and the perpendicular magnetic field $B^Z(r, \phi)$. Such 2D problem was discussed by McSween and Winternitz \cite{McSween2000} in classical mechanics and by Bérubé and Winternitz \cite{Berube} in quantum mechanics.
		
		For both motions we have one integral of motion in addition to the separable Hamiltonian, namely the integral $X_1$ for the motion in the $xy$-plane, which we list shortly together with corresponding magnetic field $B$ and potential $W$, and $X_2$ for the $z$-direction. The integral $X_2$ reduces to the first order one
		\begin{equation}
		\tilde{X}_2=p_Z^A+\frac{\sigma_0}{2},
		\end{equation}
		which in a suitably chosen gauge reads $p_Z$.
		
		In \cite{Berube} it was shown that there are 2 solutions to equations \eqref{mcsween1} and \eqref{mcsween2} for $\psi'(\phi)\neq 0$, in both cases the magnetic field $B^Z$ does not depend on $Z$. Omitting the lengthy details, we present the results only.
		\begin{enumerate}[label=\roman*.]
			\item\label{i} The magnetic field independent of $\phi$ and $Z$ corresponds to
			\begin{equation}
			\begin{split}
				\psi(\phi)&=\psi_0+\psi_1\cos(\phi)+\psi_2\sin(\phi),\\
				\rho(r)&=3 \rho_2 r^4- \rho_1 r^2+ \rho_0
			\end{split}
			\end{equation}
			which means that this quantum system is the same as the classical (the correction vanishes).
			The resulting magnetic field $B^Z$ and potential $W$ are
			\begin{equation}\label{11a}
			\begin{split}
				B^Z&=-6\rho_2 r^3+\rho_1 r,\\
				W&=-2\rho_2 r(\psi_1\cos(\phi)+\psi_2\sin(\phi))-\rho_2^2 r^6+
				\frac{\rho_2 \rho_1}{2}r^4-\rho_2 W_0r^2.\raisetag{2\baselineskip}
			\end{split}
			\end{equation}
			The corresponding integral of motion in the Cartesian coordinates is
			\begin{flalign}
			\begin{aligned}
			X_1={}&(L_z^A)^2+(3\rho_2 r^4-\rho_1 r^2+W_0)L_z^A-\psi_2 p_x^A+\psi_1 p_y^A+\\
			&+ \psi_1(2\rho_2 r^2-\rho_1 )x+\psi_2(2\rho_2 r^2-\rho_1 )y+\\
			&+\frac{(3 \rho_2 r^4 - \rho_1 r^2 + 2 W_0) (3 \rho_2 r^2- \rho_1) r^2}{4},
			\end{aligned}
			\end{flalign}
			where we write $r=\sqrt{x^2+y^2}$ for brevity.
			
			\item \label{ii} The magnetic field depends on $\phi$ as well as $r$. We have
			\begin{equation}
				\rho(r)=\frac{\rho_0}{r}.
			\end{equation}
			This leads to the special case of the following subcase~\ref{1.2} with $\tau_1=0$. We therefore refer the reader there for details and list the results only.
			
			The magnetic field $B$ together with the potential $W$ read
			\begin{align}\label{11b}
			B^r ={}&0, \quad	B^\phi ={}0, \quad B^Z = \frac{2 \beta_1 \beta(\phi)^2+\beta_2}{4 r^2 \beta(\phi)^5}, \\
			W ={}& \frac{W_0}{r^2 \beta(\phi)^2}-\frac{\beta_2}{32 \beta(\phi)^4 r^4}
			+\hbar^2\frac{f_1 \beta(\phi)^4 -12 \beta_1 \beta(\phi)^2 - 5 \beta_2}{32 r^2 \beta(\phi)^6},\label{1.1W}
			\end{align}
			where  $\beta_1$, $\beta_2$ are constants and $\beta(\phi)=\psi(\phi)-\rho_0$ must satisfy
			\begin{equation}
				4 \beta(\phi)^4 \beta'(\phi)^2 + 4 \beta(\phi)^6 - 4 \beta_1 \beta(\phi)^2+ f_1 \beta(\phi)^4 = \beta_2.
			\end{equation}
			The corresponding integral $X_1$ is defined by
			\begin{align}
				s_1^r ={}&\frac{\sqrt{4 \beta_1 \beta(\phi)^2+\beta_2-4 \beta(\phi)^6-\beta(\phi)^4 f_1}}{2 \beta(\phi)^2}, \nonumber\\
				s_1^\phi ={}&-\frac{\beta(\phi)}{r}, \qquad s_1^Z =0, \label{1.1 sm}\\
				m_1 ={}&\frac{2 W_0}{\beta(\phi)^2}-\frac{2 \beta_1 \beta(\phi)^2+\beta_2}{8 \beta(\phi)^4 r^2}
				+\hbar^2\frac{f_1 \beta(\phi)^4 -12 \beta_1 \beta(\phi)^2- 5 \beta_2}{16 \beta(\phi)^6}.\nonumber
			\end{align}			
		\end{enumerate}
		
		\item\label{1.2} $\tau''(\phi) \psi'(\phi) - \tau'(\phi) \psi''(\phi) \neq 0$. Separating the variables $r$ and $\phi$ in~\cref{separablemess},
		\begin{equation}
		\frac{3 \tau'(\phi) \psi'(\phi)^2}{\tau''(\phi) \psi'(\phi) - \tau'(\phi) \psi''(\phi)} + \psi(\phi) = \rho_0 = r \rho(r)
		\end{equation}
		with separation constant $\rho_0$, we obtain 
		\begin{equation}
		\rho(r)=\frac{\rho_0}{r}, \qquad\tau(\phi) = \tau_0 + \frac{\tau_1}{(\psi(\phi)-\rho_0)^2}.
		\end{equation}
		Using these results in \cref{reducedAa}, we get $\sigma(r)=\frac{\tau_0}{r^2}+\sigma_0$.

There are two unsolved equations among eq.~\eqref{reducedB}--\eqref{matrixform2}. Simplifying them using the previous results, we get
\begin{align}\label{12remeqs}
 r \psi'(\phi) W_r + \left( \rho_0 - \psi(\phi) \right)W_\phi+\hbar^2\frac{\psi'''(\phi)+\psi'(\phi)}{4r^2} &= 0, \nonumber\\
 3 \psi'(\phi) \psi''(\phi) + 4 \psi'(\phi) \left( \psi(\phi) - \rho_0 \right) + \psi'''(\phi)(\psi(\phi) - \rho_0)+\\
 + \frac{4 \tau_1^2}{\left(\psi(\phi) - \rho_0 \right)^5} \psi'(\phi) + 4r^5 W_{r \phi} + 8r^4 W_\phi &= 0.\nonumber
\end{align}
We substitute $\beta(\phi) = \psi(\phi) - \rho_0$ and integrate the second equation with respect to $\phi$ to obtain
\begin{align}
r \beta'(\phi) W_r - \beta(\phi)W_\phi+\hbar^2\frac{\beta'''(\phi)+\beta'(\phi)}{4r^2} ={}& 0, \label{betabeta1}\\
\beta(\phi) \beta''(\phi) + \beta'(\phi)^2 + 2 \beta(\phi)^2 - \frac{\tau_1^2}{\beta(\phi)^4} +&\nonumber\\
+ 4r^5 W_r + 8r^4 W(r, \phi) + f(r) ={}& 0, \label{betabeta2}
\end{align}
where $f(r)$ is an arbitrary function arising in the integration.

Substituting for $W_r$ from \cref{betabeta2} into \cref{betabeta1} we find expressions for both $W_r$ and $W_\phi$, where only the expression for $W_\phi$ has the correction,
\begin{equation}\label{W phi}
	\pd_\phi W =-\frac{2 \beta' W}{\beta}	-\frac {\beta' (2 \beta^6+\beta^5\beta''+ \beta^4 \beta'^2+
	\beta^4 f \left( r \right) -\tau_1^2 )}{4r^4 \beta^5} +\hbar^2\frac{\beta''' +\beta'}{4r^2\beta}.
\end{equation}
Substituting them into~\cref{12remeqs} we obtain the following equation
\begin{equation}\label{betarcesf}
\beta'(\phi)\left(12 \beta(\phi)^2 + 7 \beta \beta''(\phi) + 4 \beta'(\phi)^2 - f'(r) r + 4 f(r)\right)+ \beta(\phi)^2 \beta'''(\phi)=0,
\end{equation}
which we differentiate by $r$ and get
\begin{equation}\label{f}
f(r) = \frac{f_1}{4}+f_2 r^4,
\end{equation}
where $f_1$, $f_2$ are integration constants. With the form of $f(r)$ from \cref{f}, we can determine the dependence of the scalar potential $W$ on $r$ from \cref{betabeta2}, treating the yet unknown function $\beta(\phi)$ as a parameter:
\begin{equation}\label{12Wsolved}
 W = -\frac{f_2}{8}+\frac{\tilde{W}(\phi)}{r^2}+\frac{\beta(\phi) \beta''(\phi)+\beta'(\phi)^2+\frac{f_1}{4}-\frac{\tau_1^2}{\beta(\phi)^4}+2 \beta(\phi)^2}{8 r^4}.
\end{equation}
We determine the function $\tilde{W}(\phi)$ from an equation obtained by inserting $W$ from eq.~\eqref{12Wsolved} into \cref{W phi} and subtracting \cref{betarcesf} with $f(r)$ from \cref{f}, i.e. we subtract
\begin{align} \label{betasimple}
\beta'(\phi) \left( 7 \beta(\phi)\beta''(\phi)+4\beta'(\phi)^2 + 12 \beta(\phi)^2 + f_1 \right) + \beta(\phi)^2\beta'''(\phi)=0.
\end{align}
The resulting equation
\begin{equation}
	\frac{\hbar^2 \beta'''(\phi) + (\hbar^2 - 8\tilde{W}(\phi)) \beta'(\phi) -
	4(\tilde{W}'(\phi))\beta(\phi)}{4r^2\beta(\phi)}=0
\end{equation}
 yields
\begin{equation}
\tilde{W}(\phi) = \frac{W_0}{\beta(\phi)^2}+\hbar^2\frac{2\beta(\phi) \beta''(\phi) + \beta(\phi)^2- \beta'(\phi)^2}{8\beta(\phi)^2},
\end{equation}
where $W_0$ is an arbitrary constant. The final form of the scalar potential after a constant shift eliminating $-\frac{f_1}{8}$ is 
\begin{equation}\label{12Wsolved2}
\begin{split}
 W ={}& \frac{W_0}{r^2 \beta(\phi)^2}+\frac{\beta(\phi) \beta''(\phi)+\beta'(\phi)^2+\frac{f_1}{4}-\frac{\tau_1^2}{\beta(\phi)^4}+2 \beta(\phi)^2}{8 r^4}+\\&+\hbar^2\frac{2\beta(\phi) \beta''(\phi) + \beta(\phi)^2- \beta'(\phi)^2}{8r^2 \beta(\phi)^2}.
\end{split}
\end{equation}
We have therefore almost explicit form of the potential, the only undetermined function is $\beta(\phi)$, which must satisfy the only remaining equation~\eqref{betasimple}. We reduce its order as follows \cite{Fournier2019}: we multiply by $\beta(\phi)$ and integrate, then multiply by $\beta'(\phi)\beta(\phi)$ and integrate again. The result is
\begin{equation} \label{betafirstorder}
4 \beta(\phi)^4 \beta'(\phi)^2 + 4 \beta(\phi)^6 - 4 \beta_1 \beta(\phi)^2+ f_1 \beta(\phi)^4 = \beta_2
\end{equation}
where $\beta_1, $ $\beta_2$ are the constants of integration. According to \cite{Berube}, \cref{betafirstorder} can be written as a quadrature expressing the independent variable $\phi$ as a function of $\beta$ in terms of elliptic integrals but the authors analyse further some special cases only. For more details see the cited article, we will later consider only the special case $\beta_2=0$.

We can use \cref{betafirstorder} to further simplify the scalar potential~\eqref{12Wsolved2}
\begin{equation}\label{12W}
W = \frac{W_0}{r^2 \beta(\phi)^2}-\frac{(4 \tau_1^2 + \beta_2)}{32 r^4 \beta(\phi)^4}
+\hbar^2\frac{f_1 \beta(\phi)^4 -12 \beta_1 \beta(\phi)^2 - 5 \beta_2}{32 r^2 \beta(\phi)^6}.
\end{equation}

The magnetic field is also expressed in terms of the function $\beta(\phi)$ and its derivatives
\begin{align}\label{B beta der}
B^r = -\frac{\tau_1 \beta'(\phi)}{r^2\beta(\phi)^3}, \quad B^\phi = \frac{\tau_1}{r^3\beta(\phi)^2},\quad B^Z = -\frac{\beta(\phi) + \beta''(\phi)}{2 r^2},
\end{align}
or substituting for the derivatives from eq.~\eqref{betasimple} and its integrated form eq.~\eqref{betafirstorder}
\begin{align}\label{12B}
B^r ={}&-\tau_1 \frac{\sqrt{4 \beta_1 \beta(\phi)^2+\beta_2-4\beta(\phi)^6- f_1 \beta(\phi)^4}}{2 r^2 \beta(\phi)^5}, \nonumber\\
B^\phi ={}&\frac{\tau_1}{r^3 \beta(\phi)^2}, \quad B^Z = \frac{2 \beta_1 \beta(\phi)^2+\beta_2}{4 r^2 \beta(\phi)^5}.
\end{align}
(The square root appears due to substitution for $\beta'(\phi)$ from eq.~\eqref{betafirstorder}, so the sign depends on the choice of the branch in that equation.)

We note that eq.~\eqref{betafirstorder} is the same in classical and quantum cases. This implies that the magnetic field is the same in both cases, only the scalar potential $W$ obtains an $\hbar^2$-proportional correction depending on $\beta$.

The integrals from \cref{cyl integrals} are determined by
\begin{align}
s_1^r ={}&\frac{\sqrt{4 \beta_1 \beta(\phi)^2+\beta_2-4 \beta(\phi)^6-\beta(\phi)^4 f_1}}{2 \beta(\phi)^2}, \nonumber\\
s_1^\phi ={}&-\frac{\beta(\phi)}{r}, \qquad s_1^Z = \tau_0+\frac{\tau_1}{\beta(\phi)^2}, \nonumber\\
m_1 ={}&\frac{2 W_0}{\beta(\phi)^2}-\frac{4 \beta(\phi)^2 \tau_0 \tau_1+2 \beta_1 \beta(\phi)^2+4 \tau_1^2+\beta_2}{8 \beta(\phi)^4 r^2}+\nonumber\\
&+\hbar^2\frac{f_1 \beta(\phi)^4 -12 \beta_1 \beta(\phi)^2 - 5 \beta_2}{16 r^2 \beta(\phi)^6},
\\ \nonumber
s_2^r ={}&0, \qquad s_2^\phi = 0, \qquad s_2^Z = \sigma_0-\frac{\tau_1}{r^2 \beta(\phi)^2}, \\ \nonumber
m_2 ={}&\frac{\tau_1}{\beta(\phi)^2 r^2} \left( \frac{\tau_1}{4 \beta(\phi)^2 r^2} - \frac{\sigma_0}{2} \right).
\end{align}
Writing the integral $X_2$ in full, we get
\begin{equation}
X_2=\left(p_Z^A-\frac{\tau_1}{2 r^2 \beta(\phi)^2}\right)^2+\sigma_0\left(p_Z^A-\frac{\tau_1}{2 r^2 \beta(\phi)^2}\right).
\end{equation}
so this integral of motion reduces to the first order one
\begin{equation}
\tilde{X}_2= p_Z^A-\frac{\tau_1}{2 r^2 \beta(\phi)^2}
\end{equation}
and is in suitable gauge $X_2=p_Z$, as can be checked in eq.~\eqref{B beta der}.

We can rewrite \cref{betafirstorder} using a substitution $\gamma(\phi)=\beta(\phi)^2$ to obtain
\begin{equation} \label{gammaeq}
\gamma(\phi) \gamma'(\phi)^2+4 \gamma(\phi)^3-4 \beta_1 \gamma(\phi)+f_1 \gamma(\phi)^2 = \beta_2.
\end{equation}

Let us analyse the special case $\beta_2=0$. We obtain the following solution:
\begin{equation}\label{gammaeqsolspec}
\gamma(\phi) = \frac{\sqrt{64 \beta_1+f_1^2}\sin\left( 2 (\phi-\phi_0) \right)-f_1}{8}.
\end{equation}
Under the assumptions that $f_1<0$, $-\frac{f_1^2}{64}<\beta_1<0$, the corresponding $\beta(\phi)$ is well defined, bounded and positive,
\begin{equation}\label{beta beta2=0}
\beta(\phi)=\sqrt{\frac{\sqrt{64 \beta_1+f_1^2}\sin\left( 2 (\phi-\phi_0) \right)-f_1}{8}}.
\end{equation}
The magnetic field now reads:
\begin{align}
B^r ={}&-\frac{8 \tau_1 \sqrt{64 \beta_1+f_1^2} \cos\left(2 (\phi- \phi_0)\right)}{r^2 \left(\sqrt{64 \beta_1+f_1^2} \sin\left(2 (\phi- \phi_0)\right)-f_1\right)^2}, \nonumber\\
B^\phi ={}&\frac{8 \tau_1}{r^3 \left( \sqrt{64 \beta_1+f_1^2}\sin\left( 2 (\phi-\phi_0) \right)-f_1 \right)}, \\ \nonumber
B^Z ={}&\frac{\beta_1}{2 r^2} \left(\frac{\sqrt{64 \beta_1+f_1^2}\sin\left( 2 (\phi-\phi_0) \right)-f_1}{8}\right)^{-\frac{3}{2}}
\end{align}
For solution from~\cref{gammaeqsolspec} we find 
\begin{align}
W &= \frac{8 W_0}{r^2 \left(a \sin\left(2(\phi-\phi_0)\right)-f_1\right)}- \nonumber\frac{8 \tau_1^2}{r^4 \left(a\sin\left(2(\phi-\phi_0)\right)-f_1\right)^2}+\\
&+\hbar^2\frac{a^2 f_1 (\sin(2( \phi - \phi_0))^2 +1)-2a^3 \sin(2( \phi - \phi_0)) + 32 \beta_1 f_1 - 320 \beta_2}{4 r^2 \left(a \sin(2( \phi - \phi_0))- f_1\right)^3},\raisetag{3\baselineskip}
\end{align}
where $a=\sqrt{f_1^2+64 \beta_1}$.

Choosing the coordinates so that $\phi_0=0$ and transforming the magnetic field into the Cartesian coordinates using \cref{transformB}, we get
\begin{align}
	B^x(\vec{x})& = -\frac{8 \tau_1 \left(\sqrt{f_1^2 + 64 \beta_1} x - f_1 y\right)}
	{\left(f_1 (x^2 + y^2) - 2 \sqrt{f_1^2 + 64 \beta_1} x y\right)^2},\\
	B^y(\vec{x})& = \frac{8 \tau_1 \left(\sqrt{f_1^2 + 64 \beta_1} y- f_1 x\right)}
	{\left(f_1 (x^2 + y^2) - 2 \sqrt{f_1^2 + 64 \beta_1} x y\right)^2},\\
	B^z(\vec{x})&= - \frac{8 \sqrt{2}\beta_1}{\left(f_1 (x^2 + y^2)-2 \sqrt{f_1^2 + 64 \beta_1} x y\right)^\frac{3}{2}},\\
	W(\vec{x})&=-\frac{8 W_0\left(f_1 (x^2 + y^2) - 2 \sqrt{f_1^2 + 64 \beta_1} x y\right)+8 \tau_1^2}
	{\left(f_1 (x^2 + y^2)-2 \sqrt{f_1^2 + 64 \beta_1} x y\right)^2}+\\
	&+\hbar^2 \frac{(r^4 + 4 x^2 y^2) f_1^3-4 a^3 r^2 x y + \beta_1  f_1 (96 r^4 + 256 y^2 x^2) - 320 r^4 \beta_2}{4 (f_1 r^2-2 a y x )^3}.\nonumber
\end{align}
	\end{enumerate}

	\item $\tau(\phi)\neq0$: We can use considerations from the previous case as in the classical case, because we have never divided by $\sigma(r)$, which may now vanish, and never assumed $\tau(\phi)=0$. We only need some constraints on the constants to assure $\tau(\phi)\neq 0$. We therefore obtain the system which splits into independent motions in $xy$-plane and $z$-direction (the motion in 2D was considered in \cite{Berube}) with magnetic field from \cref{11a} and the system with the magnetic field from \cref{12B} and scalar potential from \cref{12W}. (If $\tau_1=0$ in the second system, we get the separating system from \cref{11b} in subcase~\ref{i})
\end{enumerate}

\subsection{Case b) $\mathrm{rank}\, (M)=1$: $\psi'(\phi) \neq 0$ and $\mu(Z)=0$}\label{sec:rank 2}
From \cref{reducedB} follows the separation of the scalar potential
\begin{equation}
	W(r, \phi, Z)=W_{12}(r, \phi)+W_3(Z).
\end{equation}
Let us have a look at extended matrix \cref{redMcase2b} for the system of equations \eqref{matrixform2}. For the (non-extended) matrix of the system $M$ to have rank 1, the auxiliary functions $\tau(\phi)$ and $\sigma(r)$ must be zero, therefore $W_3(Z)$ remains unconstrained, which is in contrast with respect to rank $(M)=2$.
The remaining equations of the system \eqref{reducedAa}--\eqref{matrixform2} are the same as \cref{mcsween1} and \cref{mcsween2}, which read
\begin{align} \label{rhopsisys}
r\psi'(\phi)W_r + \left( r \rho(r) - \psi(\phi) \right)W_\phi +\hbar^2\frac{\psi'''(\phi)+\psi'(\phi)}{4r^2}= 0, \nonumber\\
\psi'(\phi) \left( r^3 \rho''(r) - r^2 \rho'(r) + r \rho(r) -3 \psi''(\phi) - 4 \psi(\phi) \right) +\\
+ \psi'''(\phi) \left( r \rho(r) - \psi(\phi) \right) - 4 r^5 W_{r \phi} - 8 r^4 W_\phi = 0. \nonumber
\end{align}
The magnetic field reads
\begin{equation}\label{r1mun0B}
B^r = 0, \quad B^\phi = 0, \quad B^Z = -\frac{1}{2 r^2} \left( \rho'(r) r^2+\psi''(\phi) +\psi(\phi) \right).
\end{equation}
We have therefore obtained a separable motion -- motion in the $xy$-plane under the influence of the scalar potential $W(r, \phi)$, which has a quantum correction, plus the perpendicular magnetic field $B^Z(r, \phi)$ and the motion in the direction of the $z$-axis under the influence of the scalar potential $W_3(Z)$, which is not constrained, in contrast with the case of rank $(M)=2$. The $xy$-plane motion was analysed in \cite{Berube} and \cite{McSween2000}. For both motions we have one integral of motion in addition to the separable Hamiltonian, namely integral $X_1$ for the motion in the $xy$-plane and $X_2$ for the $z$-direction.

There are 2 systems in 2D that are interesting for us in this case \cite{Berube}. They are those in case \ref{1.1}, so namely with the potential $W_{12}(r,\phi)$ and magnetic field $B$ from \cref{11a} and \cref{11b}, the form of the integral $X_1$ can be seen under the cited equations. The $Z$ component of the potential, $W_3(Z)$, remains unconstrained and the corresponding integral of motion $X_2$ will not reduce to a linear one, 
\begin{equation}
	X_2=(p_Z^A)^2+2W_{3}(Z),
\end{equation}
as can be seen from \cref{cyl 2 fir}, where all $s^i$ vanish.

\bigskip

To sum up: In this section we obtained all quantum quadratically integrable systems of cylindrical type by solving the corresponding determining equations. We followed the analysis from the classical case \cite{Fournier2019} because only one of the equations differs form the classical case, namely eq.~\eqref{cyl0}. The determining equations were reduced to equations \eqref{reducedAa}--\eqref{matrixform2} containing 5 auxiliary functions of one variable $\rho(r)$, $\sigma(r)$, $\psi(\phi)$, $\tau(\phi)$ and $\mu(Z)$ which determine the functions $s^i$ and $B^i$, see eq.~\eqref{scond} and \eqref{Bcond}. 

Equation \eqref{matrixform2} contains a matrix which depends on the auxiliary functions only and allows us to split the considerations according to its rank. In \cite{Fournier2019} it was shown that rank 0 and 3 are either impossible (assuming a non-vanishing magnetic field) or inconsistent and the arguments remain valid in quantum mechanics. Ranks 1 and 2 split further into subcases a) $\psi'(\phi)=0$ and b) $\psi'(\phi)\neq0$, $\mu(Z)=0$. Case a) implies vanishing of the quantum correction and the obtained systems are therefore the same as in classical mechanics. Those were not analysed further here, and we refer the reader to \cite{Fournier2019}. (The key results are cited in the corresponding subsections of  Section~\ref{sec:first ord}, where we search for the first order superintegrable system among them.) In case b) the quantum correction is \emph{a priori} non-trivial, but vanishes due to the consistency conditions on the scalar potential $W$ in case \ref{1.1} subcase~\ref{i} and the corresponding subcase in this subsection. The analysis shows that the remaining subcases have the same magnetic field as their classical counterparts and only the scalar potential $W$ is modified by a $\hbar^2$-proportional correction.

In all cases there is at least one free parameter, a constant or even a function. In the next chapter we search for values of the constants and functions so that the system admits additional integrals of motion, which makes the system superintegrable.

\chapter{Found superintegrable systems}\label{kap:sup}
After the analysis of the cylindrical integrable systems both in classical mechanics \cite{Fournier2019} and in quantum mechanics (Section~\ref{sec:quantum integr}) has been completed, we go one step further, namely look for superintegrable systems among the integrable ones.

The goal is to find additional integrals of motion. We use the usual \emph{ansatz} that the integrals of motion are first or second order polynomials in momenta. We tried to solve the second order case in general, but we encountered considerable computational difficulties. Although the third order equations \eqref{ord3} do depend on neither magnetic nor electric field and can be solved, see eq.~\eqref{sol third 1}--\eqref{sol third 6} for the solution, we still have to solve 10 differential equations \eqref{ord2}--\eqref{ord0} with 20 arbitrary constants $\alpha_{ij}$ from eq.~\eqref{sol third 1}--\eqref{sol third 6} and 4 unknown functions $s^r,$ $s^\phi,$ $s^Z,$ $m$ of 3 variables $r,$ $\phi,$ $Z$ each in addition to the magnetic field $B$ and scalar potential $W$ (simplified by assuming integrability).

Despite a lot of effort, we obtained a rather limited results by analysing the second order case in general, which we present in Subsection~\ref{sec:biquadr}. (We obtained some first order systems here as well). Therefore, we have to be less ambitious and consider the first order case (Section~\ref{sec:first ord}) and a physically motivated second order \emph{ansatz}, namely $L^2+\ldots$ in Subsection~\ref{sec:L^2} and $L_x p_y-L_y p_x+\ldots$ in Subsection~\ref{sec:Runge-Lenz}. We solve these in a systematic way, albeit the second order systems in classical mechanics only because the quantum corrections for these integrals are non-trivial.

\section{First order integrals}\label{sec:first ord}
\subsection{General considerations}\label{sec:general}
Before we start analysing the cases, let us outline the general procedure and prove that the first order integrals can contain neither $L_x$ nor $L_y$. 

The task is to solve equations \eqref{ord2 lin}--\eqref{ord0 lin}, where we substitute the magnetic field from eq.~\eqref{Bcond} with the form of the auxiliary functions corresponding to the considered case.

We start from the highest order equations \eqref{ord2 lin}, which do not depend on the magnetic field nor the scalar potential and can be solved in general. As we have already mentioned in Section~\ref{sec: det eq CM}, the solution \eqref{res lin} implies that the first order terms are in the Euclidean Lie algebra generated by 
\begin{equation}
	p_x,\ p_y,\ p_z,\ L_x,\ L_y,\ L_z,
\end{equation}
the solution therefore depends on $6$ constants $k_1,\ldots,$ $k_6.$ We associate the constants with the gauge invariant form of generators of the Euclidean algebra, namely
\begin{equation}
	Y=k_1 p_x^A+k_2 p_y^A+k_3 p_z^A+k_4 L_x^A+k_5L_y^A+k_6 L_z^A+m(x,y,z).
\end{equation}

We next solve the first order equations~\eqref{ord1 lin} with the solution \eqref{res lin} substituted. The best way to proceed is to assume that the function $m$ is smooth enough and impose the Clairaut compatibility conditions $\partial_{ab} m=\partial_{ba} m$ on the second derivatives of $m(r,\phi,Z)$. Taking into account the solution \eqref{res lin} to the second order equations \eqref{ord2 lin} and the form of magnetic field in terms of the auxiliary functions \eqref{Bcond}, the compatibility equations read
\begin{align}
	&[(k_4 Z - k_2) \cos(\phi) + (k_5 Z + k_1) \sin(\phi) - k_6 r] (\psi'''(\phi)+\psi'(\phi))+\nonumber \\
	&+ 3[(k_5 Z +k_1) \cos(\phi) - (k_4 Z - k_2)\sin(\phi) ]\psi''(\phi) +\nonumber \\
	&+[(k_5 Z +k_1) \cos(\phi) - (k_4 Z - k_2)\sin(\phi)] r^2 (\rho'(r)-r\rho''(r)) + \label{komp 1}\\
	& + [-3 k_5 r \cos(\phi) + 3k_4 r \sin(\phi) + 2 k_3]r^4 \mu'(Z) +\nonumber \\
	&+ r (k_5\cos(\phi) - k_4 \sin(\phi)) \tau'(\phi) - r^4 (k_4 \cos(\phi) +k_5\sin(\phi))\sigma'(r)+ \nonumber \\
	&+ 3[(k_5 Z + k_1) \cos(\phi)- (k_4 Z - k_2) \sin(\phi)]\psi(\phi) -\nonumber\\
	&- 2[ k_4 \cos(\phi)+ k_5\sin(\phi)]r\tau(\phi)&=0,\nonumber\\ 
	& [-(k_5 Z + k_1) \cos(\phi) + (k_4 Z - k_2) \sin(\phi)] r^4 \sigma''(r) +\nonumber \\
	&+ (k_4\cos(\phi) +k_5 \sin(\phi)) r \psi''(\phi) + \nonumber \\
	&+[3(k_4 Z - k_2) \cos(\phi) +3 (k_5 Z + k_1) \sin(\phi) - 2 k_6 r] \tau'(\phi) -\nonumber \\
	&-((k_4 Z - k_2) \cos(\phi) + \sin(\phi) (k_5 Z + k_1))r^4\mu'(Z)+\label{komp 2}\\
	&+ [k_4\cos(\phi) +k_5 \sin(\phi)] r^3 \rho'(r)+\nonumber\\
	&+6 [(k_5 Z+ k_1) \cos(\phi)-(k_4 Z - k_2)\sin(\phi)] \tau(\phi)-\nonumber \\
	&- [k_4 \cos(\phi) + k_5 \sin(\phi) ]r (2 r^3\mu(Z) - \psi(\phi)) &=0,\nonumber \\
	&[(k_4 Z - k_2) \cos(\phi) + (k_5 Z + k_1) \sin(\phi) - k_6 r] \tau''(\phi) + \nonumber \\
	&+ (- k_5 r \cos(\phi)+ k_4 r \sin(\phi)+ k_3) r^5\mu''(Z)-\nonumber\\
	&- (k_5\cos(\phi) - k_4 \sin(\phi))r \psi''(\phi)+\nonumber \\
	& + 3[(k_5 Z + k_1) \cos(\phi) - \sin(\phi) (k_4 Z - k_2)] \tau'(\phi) -\nonumber \\
	&- [(k_4 Z - k_2) \cos(\phi) + (k_5 Z + k_1)\sin(\phi) ] r^3 \sigma'(r)+\label{komp 3}\\
	& + [(k_5 Z + k_1) \cos(\phi) - (k_4 Z - k_2) \sin(\phi)] r^4 \mu'(Z)-\nonumber \\
	& -4 (k_5\cos(\phi) -k_2\sin(\phi))(r^3 \rho'(r)) - \nonumber\\
	&-2[(k_4 Z + k_2)\cos(\phi)+(k_5 Z + k_1)\sin(\phi)] \tau(\phi) -\nonumber \\
	&- [ k_4\sin(\phi)- k_5\cos(\phi) ]r (2r^3\mu(Z) -\psi(\phi))&=0\nonumber.
\end{align}
(We have multiplied the equations by the power of $r$ which was originally in the denominator. Because the equations must be satisfied for all $r$, $\phi$, $Z$, we implicitly assume $r\neq 0$ and we do these simplifications automatically, often without mentioning them.)

Although the equations are long, they are manageable because they can be separated by taking derivatives with respect to $r$ and/or $Z$ (and dividing/multiplying by $r$).

There is one thing that can be proved using the equations above only, i.e. for all integrable cases: If we assume that the magnetic field is real and does not vanish, these systems do not admit the first order integrals corresponding to the constants $k_4$, $k_5$, namely $L_x$ and $L_y$. We prove this fact by contradiction and at the same time illustrate the procedure used in the following subsections, which consists in repeated solving of equations with the highest powers of $Z$ and $r$ ($\mu(Z)$ reduces to a polynomial) and substituting the results.

To prove the previous assertion, we assume that at least one of the constants $k_4$ and $k_5$ does not vanish and we show it implies zero or complex magnetic field. Taking the second derivative with respect to $Z$ of equation \eqref{komp 1}, we obtain (dividing by $r^4$)
\begin{equation}
	[-3 k_5 r \cos(\phi) + 3 k_4 r \sin(\phi) + 2 k_3] \mu'''(Z)=0.
\end{equation}
Differentiating it with respect to $r$, our assumption together with the fact that the equation must be satisfied for all $r,$ $\phi,$ $Z$ implies 
\begin{equation}
	\mu(Z)=\mu_2 Z^2+\mu_1 Z+\mu_0.
\end{equation}
Inserting it into eq.~\eqref{komp 2}--\eqref{komp 3} and differentiating twice with respect to $Z$, we obtain
\begin{equation}
	k_5 \mu_2 \sin(\phi) + k_4 \mu_2 \cos(\phi) =0, \quad k_5 \mu_2 \sin(\phi) -k_4 \mu_2 \cos(\phi) =0,
\end{equation}
which imply linearity of $\mu(Z)$, $\mu_2=0$. 

Using the result in eq.~\eqref{komp 1} differentiated with respect to $Z$ and $r$, we obtain
\begin{equation}
	k_5(r^3 \rho'''(r) + 2 r^2 \rho''(r) - 2 r \rho'(r)) \cos(\phi) - k_4(r^3 \rho'''(r) + 2 r^2 \rho''(r) - 2 r \rho'(r)) \sin(\phi)=0,
\end{equation}
which has the following solution
\begin{equation}
	\rho(r)=\rho_0+\frac{\rho_1}{r}+\rho_2 r^2.
\end{equation}

Differentiating eq.~\eqref{komp 2} with respect to $Z$ and $r$, we get
\begin{equation}
	(k_5 r \sigma'''(r)+ 4 k_5 \sigma''(r) + 12 k_4 \mu_1) \cos(\phi) - (k_4 r \sigma'''(r) + 4 k_4 \sigma''(r) - 12 k_5 \mu_1) \sin(\phi)=0.
\end{equation}
Solving the brackets separately, the solutions are consistent only if $\mu_1=0$ and read
\begin{equation}
	\sigma(r) = \frac{\sigma_2}{r^2} + \sigma_1 r + \sigma_0.
\end{equation}
Inserting $\sigma(r)$ in into eq.~\eqref{komp 3}, we get
\begin{equation}
\sigma_1\left(k_4\cos(\phi)+k_5\sin(\phi)\right)=0,
\end{equation}
so we see that $\sigma_1=0$.

We now solve eq.~\eqref{komp 1}--\eqref{komp 3} differentiated with respect to $Z$ only to obtain $\psi(\phi)$ and $\tau(\phi)$ with the result
\begin{align}
		\psi(\phi) &= \rho_1 +\frac{c_1 \cos(2\phi) + c_2 \sin(2 \phi) + c_3}{2 (k_4 \cos(\phi) +k_5 \sin(\phi))},\\
	\tau(\phi) &= \sigma_2+\frac{2 c_4-( k_4^2 + k_5^2)}{2(k_4 \cos(\phi) +k_5 \sin(\phi))^2}.
\end{align}

With this solution, the coefficient of $Z$ in equations \eqref{komp 1}--\eqref{komp 3} vanishes, but the equations are not satisfied yet.
Differentiating eq.~\eqref{komp 2} with respect to $r$ four times, we obtain
\begin{equation}
	(\mu_0 - \rho_2)(k_4 \cos(\phi) +k_5 \sin(\phi))=0,
\end{equation}
therefore $\rho_2=\mu_0$.

Inserting this result, we differentiate eq.~\eqref{komp 1}--\eqref{komp 3} once with respect to $r$, take the numerators only and collect the coefficients of the polynomial in sines and cosines. Thus, we obtain a system of 7 algebraic equations:
\begin{flalign}
\scalebox{0.8}{$\sigma_2 k_4^4 k_5 + 3 (c_1 - c_3) k_4^3 k_6 +2 (\sigma_2 k_5^2 + 3 c_2 k_6 - c_4) k_4^2 k_5- 3 (c_1 + c_3) k_4 k_5^2 k_6+ \sigma_2 k_5^5- 2 c_4 k_5^3=0,$}\\
\scalebox{0.8}{$\sigma_2 k_4^5 + 2 (\sigma_2 k_5^2 - c_4) k_4^3 - 3(c_1 - c_3) k_4^2 k_6 k_5 + [\sigma_2 k_5^4 -2 (3 c_2 k_6 - c_4) k_5^2] k_4 + 3 (c_1 + c_3) k_5^3 k_6=0,$}\\
\scalebox{0.8}{$2 \sigma_2 k_4^3 k_6 + (c_3-c_1) k_4^2 k_5 + 2[(\sigma_2 k_6- c_2) k_5^2 - 2 c_4 k_6] k_4 + (c_1 + c_3) k_5^3=0,$}\\
\scalebox{0.8}{$(c_3-c_1 ) k_4^3 -2 (\sigma_2 k_6+ c_2) k_4^2 k_5+ (c_1 + c_3) k_4 k_5^2 - 2 \sigma_2 k_5^3 k_6 + 4 c_4 k_5 k_6=0,$}\\
\scalebox{0.8}{$3\sigma_2 k_4^4 k_6 + (c_3-c_1) k_5 k_4^3 +2 [(2 \sigma_2 k_6- c_2) k_5^2 - 3 c_4 k_6] k_4^2 + (c_1 + c_3)k_4 k_5^3 + k_5^4 k_6 \sigma_2 - 2 c_4 k_5^2 k_6=0,$}\\
\scalebox{0.8}{$(c_3-c_1) k_4^4 - 2(2 k_6 \sigma_2 + c_2) k_5 k_4^3 + 2 c_1 k_4^2 k_5^2 + 2[(c_2-2\sigma_2 k_6) k_5^3 + 4 c_4 k_5 k_6] k_4 - (c_1 + c_3) k_5^4=0,$}\\
\scalebox{0.8}{$(c_1 - c_3) k_5 k_4^3 -\sigma_2 k_4^4 k_6 +2[2(c_2 k_5^2 + c_4 k_6) k_4^2 - (c_1 + c_3) k_5^3 k_4 +\sigma_2 k_5^4 k_6 - 2 c_4 k_5^2 k_6]	=0.$}
\end{flalign}
It has 3 real solutions excluding $k_4=k_5=0$, namely
\begin{enumerate}
	\item $c_1 = c_3, \quad c_4 =\frac{\sigma_2}{2} k_4^2 ,\quad k_5 = 0,$
	\item $c_1 = -c_3, \quad c_4 =\frac{\sigma_2}{2} k_5^2 ,\quad k_4 = 0,$
	\item $c_4 = \frac{\sigma_2}{2} (k_4^2 + k_5^2) ,\quad c_2= \frac{(-c_1 + c_3) k_4^2 + (c_1 + c_3) k_5^2}{2 k_5 k_4}.$
\end{enumerate}
(The remaining constants are unconstrained in all three cases.) All of them, however, lead to vanishing of the magnetic field.
Therefore, we have proved that in order to have real non-vanishing magnetic field we need
\begin{equation}\label{k4k5}
k_4=k_5=0.
\end{equation}

The remaining part of equations \eqref{komp 1}--\eqref{komp 3} can be solved for each integrable system of cylindrical type in the same manner. In each step we have to consider which of the constants $k_i$ from solution \eqref{res lin} should be set to zero and which non-zero, because any vanishing constant means that the corresponding integral is not allowed by the system, but non-vanishing constants constrain auxiliary functions in such a way that the corresponding systems have zero magnetic field.

Having solved the compatibility conditions \eqref{komp 1}--\eqref{komp 3}, we solve eq.~\eqref{ord1 lin} and eq.~\eqref{ord0 lin} to obtain $m(r,\phi,Z)$ and $W(r,\phi,Z)$, respectively. We have to check compatibility of the obtained scalar potential $W(r,\phi,z)$ with the restrictions imposed by the considered cylindrical system.

Let us now treat the specific cases from \cite{Fournier2019} and Section~\ref{sec:quantum integr}.

\subsection{Case $\mathrm{rank}(M)=2$ subcase 2a)}\label{sec:case 2a lin}
The assumptions of this subcase read
\begin{equation}
	\psi(\phi)=0,\quad \mu(Z)=\mu_0,\quad \tau(\phi)=\tau_0,
\end{equation}
with the corresponding magnetic field and scalar potential
\begin{equation}\label{rank_2psi0cyl}
	W = W(r), \quad B^r =0, \quad B^\phi =\frac{\tau_0}{r^3} + \frac{1}{2} \sigma'(r),\quad B^Z = \mu_0 r - \frac{1}{2}\rho'(r).
\end{equation}
In this case we assume $\psi(\phi)=0$, so the quantum correction vanishes and therefore the system is the same in classical and quantum mechanics.

From \cite{Fournier2019} we know, that this system admits the integrals
\begin{equation}
	\tilde{X}_1=p_\phi^A+\frac{\rho(r)}{2}-\frac{\mu_0 r^2}{2},
	\quad \tilde{X}_2 = p_Z^A+\frac{\sigma(r)}{2}-\frac{\tau_0}{2 r^2},
\end{equation}
so we set $k_3=k_6=0$.

The additional conditions which split the case into further subcases in \cite{Fournier2019} all imply the scalar potential $W$ and magnetic field $B$ from eq.~\eqref{rank_2psi0cyl} with different constraints on the functions and constants. However, we lose no generality by redefining $\sigma(r)$ and $\rho(r)$ to set $\mu_0=\tau_0=0$ and proceeding with the assumption $\sigma(r)\rho(r)\neq 0$. (We can always choose the constant term which does not appear in the magnetic field and can be eliminated by the choice of gauge in the integrals.) Therefore, we do not need to split into subcases.


Recalling eq.~\eqref{k4k5}, the compatibility conditions \eqref{komp 1}--\eqref{komp 3} can be solved at once with the result
\begin{equation}
	\rho(r)=\rho_1 r^2+\rho_0,\quad \sigma(r)=\sigma_0,
\end{equation}
where $\rho_1\neq 0$ because otherwise the corresponding magnetic field
\begin{equation}
	B^r(r,\phi, Z) = 0,\quad B^\phi(r, \phi, Z) = 0, \quad B^Z(r,\phi, Z) =-r \rho_1
\end{equation}
vanishes. We write the magnetic field also in the Cartesian coordinates (transformed using eq.~\eqref{transformB})
\begin{equation}
	B^x(x,y,z) = 0,\quad B^y(x,y,z) = 0, \quad B^z(x,y,z) =-\rho_1.
\end{equation}
We can now solve eq.~\eqref{ord1 lin} to get
\begin{equation}
	m=\rho_1 r(-k_1 \sin(\phi) +k_2 \cos(\phi))-k_3\frac{\sigma_0}{2}-k_6\frac{\rho_0}{2}.
\end{equation}
(The last term is the integration constant, which we choose so that $\tilde{X}_1$ and $\tilde{X}_2$ do not contain unnecessary constants.)
The zeroth order equation \eqref{ord0 lin} becomes
\begin{equation}
	\left(k_1 \cos(\phi) + k_2 \sin(\phi) \right) W'(r)=0,
\end{equation}
we thus see that in order to have additional integrals (one of the constants $k_1,$ $k_2$ non-zero) we need $W$ constant, without loss of generality
\begin{equation}
	W=0.
\end{equation}
This is a well-known system, considered e.g. in \cite{Landau}. It was considered in \cite[eq. (42)]{Marchesiello2015}, with different choice of the frame of reference.
The integrals read (in the Cartesian coordinates)
\begin{equation}\label{IP konst pole}
Y_1=p_x^A-\rho_1 y,\quad Y_2=p_y^A+\rho_1 x,\quad \tilde{X}_1=L_z^A+\frac{\rho_1 (x^2+y^2)}{2},\quad \tilde{X}_2=p_z^A,
\end{equation}
where $L_z^A=xp_y^A- y p_x^A$. The integrals are clearly mutually independent but the Hamiltonian can be rewritten as
	\begin{equation}\label{H 2a}
		H=\frac{1}{2}\left(Y_1^2+Y_2^2+\tilde{X}_2^2\right)-\rho_1 X_4=
		\frac{1}{2}\left(\left(p_x^A\right)^2+\left(p_y^A\right)^2+\left(p_z^A\right)^2\right).
	\end{equation}
The fifth independent integral, which makes the system maximally superintegrable, is not polynomial in momenta \cite{Marchesiello2015}, namely 
\begin{equation}\label{paty konst pole}
	X_5=(p_y^A+\rho_1 x)\sin\left(\frac{\rho_1 z}{p_z^A}\right)-(p_x^A-\rho_1 y)\cos\left(\frac{\rho_1 z}{p_z^A}\right),
\end{equation}
where we assume that $p_z^A$. (Otherwise the system collapses to 2D.)

Let us follow \cite{Marchesiello2015} and solve the Hamiltonian equations of motion and the Schr\"odinger equation. We choose the gauge so that two of the integrals $Y_1,$ $Y_2,$ $\tilde{X}_2$ are simply momenta
\begin{equation}\label{gauge 2a}
\vec{A}(\vec{x})=\left(0,-\rho_1 x,0\right).
\end{equation} 
The Hamiltonian equations of motion read
\begin{gather}
	\dot{x}=p_x,\quad \dot{y}=p_y+\rho_1 x,\quad \dot{z}=p_z,\\
	\dot{p}_x=-\rho_1(p_y+\rho_1 x),\quad \dot{p}_y=0,\quad \dot{p}_z=0.
\end{gather}
The solution to these equations with the usual initial conditions $x(0)=x_0,$ $p_x(0)=p_x^0$ etc. is
\begin{align}
\begin{aligned}
	x(t)&=\frac{p_x^0}{\rho_1}\sin(\rho_1 t)+\left(x_0+\frac{p_y^0}{\rho_1}\right)\cos(\rho_1 t)-\frac{p_y^0}{\rho_1},\\
y(t)&=\left(x_0+\frac{p_y^0}{\rho_1}\right)\sin(\rho_1 t)-\frac{p_x^0}{\rho_1}\cos(\rho_1 t)+y_0+\frac{p_x^0}{\rho_1},\\
z(t)&=p_z^0 t+z_0,\\
p_x(t)&=-\left(\rho_1 x_0+p_y^0\right)\sin(\rho_1 t)+p_x^0\cos(\rho_1 t),\quad p_y(t)=p_y^0,\quad p_z(t)=p_z^0.
\end{aligned}
\end{align}
If $p_z^0\neq 0$, the trajectory is a helix with its axis parallel to the $z$-axis, otherwise it collapses to a circle in the plane $z=z_0$.

Let us now turn to the Schr\"odinger equation. Writing the Hamiltonian \eqref{H 2a} in full
\begin{equation}
	\hat{H}=-\frac{\hbar^2}{2}\left(\pd_{xx}+\pd_{yy}+\pd_{zz}\right)-\i \hbar \rho_1 x\pd_y+\frac{\rho_1^2}{2} x^2,
\end{equation}
we see that the stationary Schr\"odinger equation $\hat{H}\psi=E\psi$ separates in the Cartesian coordinates 
\begin{gather}\label{red schr}
	\psi(\vec{x})=f(x)\exp\left(\frac{\i}{\hbar}\lambda_2 y\right)\exp\left(\frac{\i}{\hbar}\lambda_3 z\right),\\
	\hbar^2 f''(x)=\left((\rho_1 x-\lambda_2)^2+\lambda_3^2-2E\right)f(x),\\
	Y_2\psi(\vec{x})=\lambda_2\psi(\vec{x}),\quad \tilde{X}_2\psi(\vec{x})=\lambda_3\psi(\vec{x}).
\end{gather}
The reduced Schr\"odinger equation corresponds to the 1D harmonic oscillator with energy $E-\frac{\lambda^2_3}{2}$, angular frequency $\omega=\rho_1$ with the centre of force located at $x=\frac{\lambda_2}{\rho_1}$. Therefore, the energy spectrum of the Hamiltonian \eqref{H 2a} 
\begin{equation}
	E=\frac{\lambda_3^2}{2}+\hbar\rho_1\left(n+\frac{1}{2}\right),
\end{equation}
is continuous due to the arbitrary momentum $\lambda_3$ and the generalized eigenvectors are
\begin{equation}
	\psi_{n,\lambda_2,\lambda_3}(\vec{x})=K_{n}\exp\left(\frac{\i}{\hbar}(\lambda_2 y+\lambda_3 z)\right) H_n\left(\sqrt{\frac{\rho_1}{\hbar}}\left(x-\frac{\lambda_2}{\rho_1}\right)\right)\exp\left(-\frac{\rho_1}{2\hbar}\left(x-\frac{\lambda_2}{\rho_1}\right)^2\right),
\end{equation}
where $H_n$ are the Hermite polynomials and $K_n$ is their normalization constant.

Let us see if the Schr\"odinger equation separates in the cylindrical coordinates as well. Using the gauge \eqref{gauge 2a}, the Hamiltonian reads
\begin{equation}
\begin{split}
\hat{H}={}&-\frac{\hbar^2}{2}\left(\frac{1}{r}\pd_r(r\pd_r)+\frac{1}{r^2}\pd_{\phi\phi}+\pd_{ZZ}\right)-\\
&-\i\hbar\rho_1 r \cos(\phi)\left(\sin(\phi)\pd_r+\frac{\cos(\phi)}{r}\pd_\phi\right)+\frac{\rho_1}{2} r^2(\cos(\phi))^2.
\end{split}
\end{equation}
It is clear that we can separate the $Z$ coordinate from the $r$ and $\phi$. However, it is not possible to separate $r$ and $\phi$: The equation to separate is (divided by $f(r)g(\phi)$)
\begin{equation}\label{sep r phi}
\begin{split}
	-\frac{\hbar^2}{2}\left(\frac{g''(\phi)}{g(\phi)} - \frac{f''(r)}{f(r)}\right) -\i \hbar \rho_1 r \sin(\phi) \cos(\phi) \frac{f'(r)}{f(r)} = \\
	=\i \hbar \rho_1 \cos(\phi)^2 \frac{g'(\phi)}{g(\phi)} - \frac{\rho_1^2}{2} r^2 \cos(\phi)^2 +E,
\end{split}
\end{equation}
where the last term on the left-hand side and the second term on the right-hand side will always depend both on $r$ an $\phi$. 

However, if we choose the gauge which makes $\tilde{X}_1$ simply $L_z$ \cite{Charest}, namely
\begin{equation}\label{konst pole schr cyl}
	\vec{A}(\vec{x})=\left(\frac{\rho_1 y}{2},-\frac{\rho_1 x}{2},0\right),\qquad A_r=0,\quad A_\phi=-\frac{\rho_1 r^2}{2},\quad A_Z=0,
\end{equation}
the Hamiltonian reads
\begin{equation}
	\begin{split}
	\hat{H}={}&-\frac{\hbar^2}{2}\left(\frac{1}{r}\pd_r(r\pd_r)+\frac{1}{r^2}\pd_{\phi\phi}+\pd_{ZZ}\right)+\i\hbar\frac{\rho_1}{2} \pd_\phi+\frac{\rho_1^2 r^2}{8}.
	\end{split}
\end{equation}
This time $L_z=-\i\hbar\pd_\phi$ commutes with the Hamiltonian and therefore the wave-function
\begin{equation}
	\psi(r,\phi,Z)=f(r)\exp\left(\frac{\i}{\hbar}\eta_2\phi\right)\exp\left(\frac{\i}{\hbar}\eta_3 Z\right)
\end{equation}
leads to a separation of variables, with the $r$ coordinate equation reading
\begin{equation}
	-\hbar^2\left(f''(r)+\frac{1}{r}f'(r)\right)+\left(\frac{\eta_2^2}{r^2}+\eta_3^2-\rho_1 \eta_2 \right)f(r)+\frac{\rho_1^2 r^2 f(r)}{4}=2 E f(r).
\end{equation}
Its solution in terms of Whittaker functions $M_{\mu,\nu}(z),$ $W_{\mu,\nu,}(z)$ \cite[Chapter 13]{AbramowitzStegun}, \cite[Section 13.14]{DLMF} reads
\begin{equation}
	f(r)=\frac{1}{r}\left(c_1 M_{\mu,\nu}\left(\frac{\rho_1}{\sqrt{2}\hbar}r^2\right)+c_2 M_{\mu,\nu}\left(\frac{\rho_1}{\sqrt{2}\hbar}r^2\right)\right),
\end{equation}
where $\mu$, $\nu$ are constants
\begin{equation}
	\mu=\frac{(\rho_1 \eta_2-\eta_3^2+2E)}{\sqrt{2}\hbar \rho_1},\quad \nu=\frac{\eta_2}{\sqrt{2}\hbar}.
\end{equation}

We solve the stationary Hamilton-Jacobi equation as well. Since we work with time-independent Hamiltonians, we can separate the time coordinate $t$ from the Hamilton's principal function $S$ with separation constant $-E$ and we use Hamilton's characteristic function $U$, namely $S=U - Et$. Using the Hamiltonian \eqref{H 2a} and the gauge \eqref{gauge 2a} the equation reads
	\begin{equation}
	\frac{1}{2}\left[\left(\pderA{U}{x}\right)^2+\left(\pderA{U}{y}-\rho_1 x\right)^2+\left(\pderA{U}{z}\right)^2\right]=E.
	\end{equation}
	We have 2 cyclic coordinates $y$ and $z$; thus, we have the separation
	\begin{equation}
	U(x,y,z)=u(x)+p_y y+p_z z,
	\end{equation}
	with $u(x)$ satisfying 
	\begin{equation}
	(u'(x))^2+(p_y-\rho_1 x)^2+p_z^2=2E.
	\end{equation}
	For later use let us consider more generally an equation of type
	\begin{equation}
	(u'(x))^2+\alpha_1 x^2+\alpha_2x+\alpha_3 =0
	\end{equation}
	whose solution is 
	\begin{equation}
	\begin{split}\label{u'(x)}
		u(x) ={}& \pm\left[\left(\frac{\alpha_2 ^2}{8\alpha_1 ^{3/2}}- \frac{\alpha_3 }{2\sqrt{\alpha_1 }}\right) \arctan\left(\frac{2 \alpha_1 x + \alpha_2 }{2 \sqrt{\alpha_1 } \sqrt{-\alpha_1 x^2 - \alpha_2 x - \alpha_3 }}\right) \right.+\\
		&+\left. \left(\frac{x}{2} + \frac{\alpha_2 }{4 \alpha_1 }\right) \sqrt{-\alpha_1 x^2 - \alpha_2 x - \alpha_3 }\right]+c_1.
	\end{split}
	\end{equation}
	In particular, the Hamilton's characteristic function reads in our case
	\begin{equation}\label{U konst pole}
	\begin{split}
		U={}&\frac{(p_z^2 -2E)}{2 \rho_1}\arctan\left(\frac{(-\rho_1 x + p_y)}{\sqrt{-\rho_1^2 x^2 + 2 p_y \rho_1 x + a}}\right)-\\
		 &- \frac{(\rho_1 x - p_y)}{2\rho_1}\sqrt{-\rho_1^2 x^2 + 2 p_y \rho_1 x + a}+p_y y+p_z z,
	\end{split}
	\end{equation}
where $a=2 E-p_y^2 - p_z^2$.

In the cylindrical coordinates we use the gauge \eqref{konst pole schr cyl}, in which the Hamiltonian does not depend on $\phi$ and $Z$, therefore Hamilton's characteristic function 
	\begin{equation}
		U(r,\phi,Z)=u(r)+p_\phi \phi+p_Z Z
	\end{equation}
reduces the Hamilton-Jacobi equation to
\begin{equation}
	\frac{1}{2}\left[u'(r)^2+\frac{(2 p_\phi-\rho_1 r^2)^2}{4r^2}+p_Z^2\right]=E.
\end{equation}
Let us again consider more general form of the equation for later use, namely
\begin{equation}
	(u'(r))^2=\frac{a r^4+b r^2+c}{r^2}.
\end{equation}
Assuming we take the square of a positive real number, the result to the last equation is of the following type ($v=r^2$)
\begin{equation}
\begin{split}\label{int}
u(v)=\int \!{\frac {\sqrt {a{v}^{2}+bv+c}}{v}} {\rm d}v=&\sqrt {a{v}^{2}+bv
	+c}+{\frac {b}{2\sqrt {a}}\ln \left( {\frac {b+2av}{2\sqrt {a}}}+
	\sqrt {a{v}^{2}+bv+c} \right)}\\&-\sqrt {c}\arctan \left(\frac {bv+2 c}{2
		\sqrt {c}\sqrt {a{v}^{2}+bv+c}} \right)+C.
\end{split}
\end{equation}
The integration constant can absorb the purely imaginary constants arising from the possibly negative value of $a$ in the second term, so we can use this form of the solution in that case as well with $a$ replaced by $|a|$. If $c<0$, we must change the third term to argtanh and $c$ to $|c|$. The explicit expression for the solution of the Hamilton-Jacobi equation  	becomes too complicated for practical use and we have already calculated the trajectories. Thus, we do not present it here.

\subsection{Case $\mathrm{rank}(M)=2$ subcase 2b 1.1)}\label{sec:case 2b 1.1)}
This case corresponds to
\begin{equation}
	\tau(\phi)=\tau_0,\quad \sigma(r)=\frac{\tau_0}{r^2}+\sigma_0,\quad \mu(Z)=0,\quad \psi'(\phi)\neq 0,
\end{equation} 
which yields the following magnetic field
\begin{equation}
B^\phi = 0, \quad B^r = 0, \quad B^Z = -\rho'(r)-\frac{\psi''(\phi)+\psi(\phi)}{2 r^2}
\end{equation}
and the separable scalar potential with vanishing $Z$ component $W(r,\phi,Z)= W_{12}(r,\phi)\equiv W(r,\phi)$ satisfying
eq.~\eqref{mcsween1} and eq.~\eqref{mcsween2}.


Recalling eq.~\eqref{k4k5}, we have only one non-trivial compatibility equation, namely eq.~\eqref{komp 1}. Differentiating it twice with respect to $r$ to eliminate $\psi(\phi)$, we get an equation for $\rho(r)$
	\begin{equation}\label{rho eq}
		(k_1\cos(\phi) + k_2\sin(\phi))[ r^3 \rho^{(4)}(r)+5 r^2 \rho'''(r) +2 r \rho''(r) - 2 \rho'(r)]=0.
	\end{equation}
Because the vanishing of the constants $k_1,$ $k_2$ would mean no additional integrals of motion (we would the second order cylindrical integrals or their reduced first order form only), we need $\rho(r)$ of the form
\begin{equation}\label{rho}
	\rho(r)=\rho_3 \ln(r)+\rho_2 r^2 +\frac{\rho_1}{r}+\rho_0.
\end{equation}
	Using this in eq.~\eqref{komp 1} once differentiated with respect to $r$, we obtain
\begin{equation}\label{psi}
	k_6(\psi'''(\phi)+\psi'(\phi))-2 \rho_3(k_2\sin(\phi) +k_1 \cos(\phi))=0,
\end{equation}
which splits the considerations into two subcases: $k_6\neq 0$ and $k_6=0$.
\begin{enumerate}[label=\Roman*.]
	\item\label{k6 nenula 2b} $k_6\neq 0$: In this subcase we solve eq.~\eqref{psi} for $\psi(\phi)$ and get
	\begin{equation}
	\psi(\phi) =-\frac{1}{k_6} [((k_1\phi+k_2)\rho_3+\psi_2 k_6)\cos(\phi)+((k_2\phi-k_1)\rho_3-\psi_1 k_6)\sin(\phi)]+\psi_3.	\end{equation}
	Using this in eq.~\eqref{komp 1} (without differentiation), we get an algebraic equation with only one solution (excluding $k_1=k_2=0$): $\psi_3=\rho_1$ and $\rho_3=0$, i.e.
	\begin{equation}\label{psirho}
		\psi(\phi) =-\psi_2 \cos(\phi)+\psi_1 \sin(\phi)+\rho_1,\quad \rho(r)=\frac{\rho_2 r^3 + \rho_1}{r}+\rho_0,
	\end{equation}
	so we obtained the constant magnetic field
	\begin{equation}
		B^r=0,\quad B^\phi=0,\quad B^Z=-\rho_2 r.
	\end{equation}
	Translating this into the Cartesian coordinates using eq.~\eqref{transformB}, it reads
	\begin{equation}
		B^x=0,\quad B^y=0,\quad B^z=-\rho_2.
	\end{equation}
	We note that $\psi(\phi)$ from eq.~\eqref{psirho} implies vanishing of the quantum correction, so the quantum version of the system is identical to the classical version in this subcase.
	
	Having solved the compatibility conditions, we solve the original equations \eqref{ord1 lin} and \eqref{ord0 lin} to obtain
	\begin{align}
		W(r, \phi) &= W\left(\frac{k_6 r^2}{2} +k_2 r \cos(\phi) - k_1 r \sin(\phi)\right),\\
		m(r, \phi, Z) &= \rho_2\left(k_2 r\cos(\phi)-k_1 r\sin(\phi) + \frac{k_6 r^2}{2}\right).
	\end{align}
	(We set the integration constant to 0 in $m$ so that we get $\tilde{X}_1=p_Z^A$ in accordance with eq.~\eqref{IP konst pole} in Subsection~\ref{sec:case 2a lin}.)
	
	The potential must, however, satisfy eq.~\eqref{mcsween1} and eq.~\eqref{mcsween2}, where the quantum correction vanishes. 
	Inserting the obtained form of the scalar potential into the first of them, we get
	\begin{multline}
	W' \cdot [(k_2 \rho_2 \sin(\phi) + k_1 \rho_2 \cos(\phi)) r^3 + ((k_2 \rho_0 - k_6\psi_2) \sin(\phi)+\\+ (k_1 \rho_0 - k_6 \psi_1)\cos(\phi)) r + k_1 \psi_2-k_2\psi_1]=0.
	\end{multline}
	If we assume that the scalar potential $W$ is not constant, the terms in square bracket must vanish separately for all powers of $r$. The $r^3$ term in the bracket implies either $k_1=k_2=0$, which means only 2 reduced cylindrical integrals, or $\rho_2=0$ and vanishing magnetic field. Thus, the only interesting case for us is the constant scalar potential
	\begin{equation}
		 W(r,\phi)= W_0.
	\end{equation}
	Equation \eqref{mcsween2} is now also satisfied. Thus, we have arrived at the system \eqref{IP konst pole} and we refer the reader to Subsection~\ref{sec:case 2a lin}.
	
	\item\label{k6 nula 2b} $k_6=0$: With this assumption eq.~\eqref{rho} implies $\rho_3=0$ and together with eq.~\eqref{komp 1} we get 
	\begin{align}
		\rho(r)&=\frac{\rho_2 r^3 + \rho_1}{r}+\rho_0,\label{rho 1.1}\\
		\begin{split}
		\psi(\phi) &=\frac{c_1\cos(2\phi)+c_2\sin(2\phi)+c_3}{k_2 \cos(\phi)-k_1 \sin(\phi)}+\rho_1.
		\end{split}
	\end{align}
	The corresponding magnetic field $B$ reads
	\begin{equation}
	B^r=0,\quad B^\phi=0,\quad B^Z=-\rho_2 r+\frac{\xi}{r^2 (k_1 \sin(\phi)- k_2 \cos(\phi))^3},
	\end{equation}
	where $\xi=-(c_1 + c_3) k_1^2 - 2 c_2 k_1 k_2 + k_2^2 (c_1 - c_3)$.
	Translating this into the Cartesian coordinates using eq.~\eqref{transformB}, it reads
	\begin{equation}
	B^x=0,\quad B^y=0,\quad B^z=-\rho_2+\frac{\xi}{(k_1y- k_2x)^3}.
	\end{equation}
	
	In order to simplify the equations, we use the formula
	\begin{equation}\label{k2=1}
		k_1\sin(\phi)-k_2\cos(\phi)=\sqrt{k_1^2+k_2^2}\cos(\phi+\phi_0),\quad \cos(\phi_0)=-\frac{k_2}{\sqrt{k_1^2+k_2^2}},
	\end{equation}
	rotate the coordinate system so that $\phi_0=0$	and redefine the constants $c_i$ in $\psi(\phi)$ to get 
	\begin{equation}\label{psi spec}
		\psi(\phi) =\frac{c_1\cos(2\phi)+c_2\sin(2\phi)+c_3}{\cos(\phi)}+\rho_1.
	\end{equation}
	The corresponding magnetic field $B$ reads in the cylindrical coordinates
	\begin{equation}
	B^r=0,\quad B^\phi=0,\quad B^Z=-\rho_2 r+\frac{c_1-c_3}{r^2 (\cos(\phi))^3}
	\end{equation}
	and in the Cartesian coordinates
	\begin{equation}
	B^x=0,\quad B^y=0,\quad B^z=-\rho_2+\frac{c_1-c_3}{x^3}.
	\end{equation}
	Having solved the compatibility conditions, we solve the original equations \eqref{ord1 lin} and \eqref{ord0 lin} to obtain 
	\begin{align}
		W ={}& W(r \cos(\phi))= W(x),\\
		m={}& \rho_2 r-\frac{c_1-c_3}{2 r^2(\cos(\phi))^2}=
		\rho_2 \sqrt{x^2+y^2}-\frac{c_1-c_3}{2x^2}.
	\end{align}
	(We omit the constant of integration, which is only added to the integral of motion.)
	
	We need to satisfy 2 more equations to ensure integrability, namely eq.~\eqref{mcsween1} and eq.~\eqref{mcsween2}.
	The first of them is
		\begin{equation}\label{rce pro W}
		\begin{split}
		&\cos(\phi)^3 r^3\left[((\rho_2 r^3+\rho_0 r)\cos(\phi)+2(c_1-c_3))\sin(\phi)-2 c_2\cos(\phi)\right]W'(r \cos(\phi))\\
		&+\tfrac{3}{2}\hbar^2(c_1-c_3)\sin(\phi)=0.
		\end{split}\raisetag{\baselineskip}
		\end{equation}
		(We differentiate $W$ with respect to its argument, i.e. $r\cos(\phi)$.)
		
		In the classical case, $\hbar\to 0$, the conditions for non-vanishing magnetic field $c_1\neq c_3$ and $\rho_2\neq 0$ imply constant scalar potential $W$.
		
		In the quantum case, there is another possibility in addition to $W=W_0$: If $c_1\neq c_3$, we must have 
		\begin{equation}
			\rho_2=0,\quad \rho_0=0,\quad c_2=0,\quad W=-\frac{3\hbar^2}{4(r\cos(\phi))^2}=-\frac{3\hbar^2}{4 x^2}.
		\end{equation}
	Inserting this potential $W$ into eq.~\eqref{mcsween2} with $\rho(r)$ from eq.~\eqref{rho 1.1} and $\psi$ from eq.~\eqref{psi spec}, the numerator becomes 
	\begin{equation}\label{mcsween W0}
	(c_1 - c_3) [(\rho_2 r^3+\rho_0 r) \sin(\phi) \cos(\phi)+ 2 (c_1-c_3) \sin(\phi) - 2 c_2 \cos(\phi)]=0.
	\end{equation}
	Thus, we have a contradiction due to $c_1\neq c_3$ and the potential must be constant in quantum mechanics as well.
	
If we insert constant potential $W=W_0$ into eq.~\eqref{mcsween2}, it is satisfied if and only if $c_1=c_3$, because we obtain the same numerator as in eq.~\eqref{mcsween W0}. The result is, therefore, that both the magnetic field $B^z$ and scalar potential $W$ must be constant. We have again the well-known maximally superintegrable system from \cite{Landau,Marchesiello2015}, see Subsection~\ref{sec:case 2a lin}.
\end{enumerate} 

\subsection{Case $\mathrm{rank}(M)=2$ subcase 2b 1.2)}\label{sec:2b 1.2}
This case means 
\begin{equation}
	\psi'(\phi)\neq 0,\quad \mu(Z)=0,\quad \rho(r)=\frac{\rho_0}{r},\quad \tau(\phi)=\tau_0+\frac{\tau_1}{(\psi(\phi)-\rho_0)^2},\quad \sigma(r)=\frac{\tau_0}{r^2}+\sigma_0,
\end{equation}
thus, the following magnetic field
\begin{equation}
	B^r(r, \phi, Z) = -\frac{\tau_1\beta'(\phi)}{r^2 \beta(\phi)^3},\quad B^Z(r, \phi, Z) = -\frac{\beta(\phi) + \beta''(\phi)}{2 r^3},\quad B^\phi(r, \phi, Z) = \frac{\tau_1}{r^3 \beta(\phi)^2},
\end{equation}
and the scalar potential \eqref{12Wsolved2} 
\begin{equation}
\begin{split}
W ={}& \frac{W_0}{r^2 \beta(\phi)^2}+\frac{\beta(\phi) \beta''(\phi)+\beta'(\phi)^2+\frac{f_1}{4}
	-\frac{\tau_1^2}{\beta(\phi)^4}+2 \beta(\phi)^2}{8r^4}+\\
&+\frac{\hbar^2(2\beta(\phi) \beta''(\phi) + \beta(\phi)^2- \beta'(\phi)^2)}{8r^2 \beta(\phi)^2},
\end{split}
\end{equation}
where $\beta(\phi)$ is the auxiliary function $\psi(\phi)$ after a shift
\begin{equation}
	\beta(\phi)=\psi(\phi)-\rho_0.
\end{equation}
The function $\beta(\phi)$ must satisfy eq.~\eqref{betasimple}
or equivalently its reduced form \eqref{betafirstorder}.
Using this equation, we can simplify the potential from eq.~\eqref{Wbeta} and obtain the potential from eq.~\eqref{12W}, namely
\begin{equation}\label{Wbeta}
W = \frac{W_0}{r^2 \beta(\phi)^2}-\frac{(4 \tau_1^2 + \beta_2)}{32 r^4 \beta(\phi)^4}
+\hbar^2\frac{f_1 \beta(\phi)^4 -12 \beta_1 \beta(\phi)^2 - 5 \beta_2}{32 r^2 \beta(\phi)^6}.
\end{equation}
where $\beta_1$ and $\beta_2$ are integration constants in the reduced equation \eqref{betafirstorder}.

Let us continue solving equations \eqref{komp 1}--\eqref{komp 3}, taking eq.~\eqref{k4k5} into account. Differentiating eq.~\eqref{komp 2} with respect to $r$, we obtain (multiplied by $\beta(\phi)^3$, which must be non-zero)
\begin{equation}\label{tau beta}
	k_6\tau_1 \beta'(\phi)=0.
\end{equation}
Thus, we split our considerations into 4 subcases.
\begin{enumerate}[label=\Roman*.]
	\item $\tau_1=0$, $k_6\neq 0$:
	Differentiating eq.~\eqref{komp 1} with respect to $r$, we obtain
	\begin{equation}
		k_6(\beta'(\phi)+\beta'''(\phi))=0.
	\end{equation}
	Assumption $k_6\neq 0$ leads to 
	\begin{equation}
		\beta(\phi)=c_1\cos(\phi)+c_2\sin(\phi)+c_3,
	\end{equation}
	but the constant $c_3$ must vanish to satisfy eq.~\eqref{komp 1}, implying no magnetic field.
	\item $\tau_1=0$, $k_6 = 0$:
	From the last remaining compatibility equation \eqref{komp 1} we obtain $\beta(\phi)$ of the form
	\begin{equation}\label{beta}
	\beta(\phi) = \frac{a_1 + a_2\sin(2\phi) + a_3 \cos(2 \phi)}{k_1 \sin(\phi) -k_2 \cos(\phi)},
	\end{equation}
	leading to a system with non-vanishing magnetic field in the $z$-direction only
	\begin{equation}
	B^Z(r, \phi, Z) = -\frac{(a_1 + a_3) k_1^2 + 2 a_2 k_2 k_1 + k_2^2 (a_1 - a_3)}{r^2 (k_1\sin(\phi)- k_2 \cos(\phi))^3},
	\end{equation}
	where $a_i$ are constants. Transforming the magnetic field into the Cartesian coordinates using eq.~\eqref{transformB}, we get a somewhat simpler expression
	\begin{equation}
	B^z(x,y,z) = -\frac{(a_1 + a_3) k_1^2 + 2 a_2 k_2 k_1 + k_2^2 (a_1 - a_3)}{(k_1 y-k_2 x)^3}.
	\end{equation}
	Without loss of generality, we can simplify the previous expressions by using eq.~\eqref{k2=1}, choosing our coordinate system and redefining the constants $a_1,$ $a_2$ and $a_3$ so that $k_1=1$, $k_2=0$, i.e.
	\begin{equation}\label{beta spec}
		\beta(\phi) = \frac{a_1 + a_2\sin(2\phi) + a_3 \cos(2 \phi)}{\sin(\phi)}.
	\end{equation}
	The magnetic field simplifies to
	\begin{equation}
	B^Z(r, \phi, Z) = -\frac{(a_1 + a_3)}{r^2 (\sin(\phi))^3},\quad B^z(x,y,z) = -\frac{(a_1 + a_3)}{y^3},
	\end{equation}
	with all other components vanishing both in the cylindrical and Cartesian coordinates.
	
	We have to check if $\beta(\phi)$ of the form eq.~\eqref{beta spec} satisfies eq.~\eqref{betasimple}. For that to be the case the constants must satisfy (omitting the trivial $a_1=a_2=a_3=0$)
	\begin{equation}
		a_1 = -a_3, \quad f_1 = -16 (a_2^2 +a_3^2),
	\end{equation}
	which, however, implies vanishing of the magnetic field. This subcase, therefore, does not admit any superintegrable system with additional first order integrals and non-vanishing magnetic field.

		\item $\tau_1\neq 0$, $k_6\neq0$: Here equation \eqref{tau beta} implies $\beta(\phi)=\gamma=const$, which satisfies eq.~\eqref{betasimple} trivially. However, eq.~\eqref{komp 2} in this case reads
		\begin{equation}
		\tau_1(k_1\cos(\phi) +k_2 \sin(\phi))=0, 
		\end{equation}
		which (recall $\tau_1\neq 0$) implies no additional integral in this case.

			\item $\tau_1\neq 0$, $k_6=0$:\label{system 2b 1.2} From eq.~\eqref{komp 1} we obtain
				\begin{equation}\label{beta2a}
			\beta(\phi)=c_1(k_1\sin(\phi)-k_2\cos(\phi))
		\end{equation}
		Substituting it into the differential equation \eqref{betasimple}, we obtain
		\begin{equation}
			c_1\left(4(k_1^2 + k_2^2)c_1^2 + f_1\right)(k_1\sin(\phi) +k_2 \cos(\phi))=0,
		\end{equation}
		so we need
		\begin{equation}
			c_1=\pm\sqrt{\frac{-f_1}{4(k_1^2 + k_2^2)}},\quad f_1<0,
		\end{equation}
		because vanishing $\beta(\phi)$ would imply division by 0.
		
		We note that the quantum correction in eq.~\eqref{Wbeta} vanishes for $\beta(\phi)$ of the form \eqref{beta2a}. The system defined by the following magnetic field $B$ and scalar potential $W$ is, therefore, the same in classical and quantum mechanics.
		\begin{align}
		W &=-\left(\frac{2\tau_1^2 \left(k_2^2 +k_1^2 \right)^2} {\left(k_2 \cos (\phi)-k_1\sin (\phi) \right)^4 f_1^2 r^4} +\frac{4 W_0 \left(k_2^2 +k_1^2 \right)}{\left(k_2 \cos (\phi) -k_1\sin (\phi)\right)^2 f_1r^2}\right), \raisetag{10pt}\\
	B^r&=-{\frac{4 \left(k_2 \sin (\phi) +k_1\cos (\phi) \right) \left(k_2^2 +k_1^2 \right) \tau_1}
	{\left(k_2 \cos (\phi) -k_1\sin (\phi)\right)^3 r^2 f_1}},\\
	 	{B^\phi} &=-{\frac{4 \left(k_2^2 +k_1^2 \right) \tau_1}{r^3 \left(k_2\cos (\phi) -k_1\sin \left(\phi\right) \right)^2 f_1}},\quad{B^Z} =0.
		\end{align}
		We note that we can write 
		\begin{equation}
		\frac{k_2 \cos (\phi) -
			k_1\sin (\phi)}{\sqrt{k_2^2+k_1^2}}= \cos(\phi_0)\cos(\phi)-\sin(\phi_0)\sin(\phi)=\cos(\phi+\phi_0),
		\end{equation}
		which corresponds to the rotational symmetry of the system. Without loss of generality, we can rotate the coordinate system so that $\phi_0=\frac{3\pi}{2}$, which corresponds to $k_2=0,$ $k_1=1$. The Cartesian form of the magnetic field and the scalar potential with the adjusted coordinate axis reads
		\begin{align}\label{WB netriv}
		W =-4\left(\frac{\tau_1^2}{2 f_1^2 y^4}	+\frac{W_0 }{f_1 y^2}\right),\quad
		B^x={\frac{4 \tau_1}{f_1 y^3}},	\quad {B^y} =0,\quad B^z=0
		\end{align}
		The corresponding first order integrals of motion are 
		\begin{equation}\label{px pz}
		\tilde{X}_2=p_z^A+{\frac{2\tau_1}{f_1 y^2}}, \quad Y_1=p_x^A.
		\end{equation}
		The other cylindrical integral $X_1$ is of the second order and reads
		\begin{equation}
		X_1=\left(x p_y^A-y p_x^A\right)^2 -{\frac{4\tau_1 \left({x}^2 +y^2 \right)}{f_1 y^2} \left(p_z^A+\frac{2\tau_1}{f_1 y^2}\right)}-{\frac{8 W_0\left({x}^2 +y^2 \right)}{f_1 y^2}},\\
		\end{equation}
		This case in the chosen coordinates corresponds to the Case Ib) or Ic) in \cite{Marchesiello2019} with constants
		\begin{equation}
			a_1=a_2=0,\quad a_3=\frac{2\tau_1}{f_1},\quad b_1=b_2=0,\quad b_3=-\frac{4 W_0}{f_1}.
		\end{equation}
		There is, therefore, another second order integral of motion, namely
		\begin{align}
		X_3&=\left(x p_y^A -y p_x^A \right) p_y^A
		-{\frac{4\tau_1 x}{f_1 y^2}	\left(p_z^A+{\frac{2\tau_1}{f_1 y^2}} \right)}-	{\frac{8 W_0 x}{f_1 y^2}},
		\end{align}
		which in the natural choice of gauge
		\begin{equation}\label{gauge case Ib}
		A_x(x,y,z)=0,\quad A_y(x,y,z)=0,\quad A_z(x,y,z)=-\frac{2\tau_1}{f_1 y^2}
		\end{equation}
		simplifies to
		\begin{align}
		X_3&=\left(x p_y -y p_x \right) p_y	-{\frac{4\tau_1 x}{f_1 y^2}	p_z}-{\frac{8 W_0 x}{f_1 y^2}}.
		\end{align}
		There are, however, only 4 independent integrals of motion, because 
		\begin{equation}
		(\tilde{X}_2^2 + Y_1^2 - 2H)X_1 + X_3^2=-\frac{4}{f_1}(\tilde{X}_2 \tau_1 + 2 W_0)(2H-\tilde{X}_2^2).
		\end{equation}
		The non-vanishing Poisson brackets read
		\begin{align}
		\{X_3,X_1\}_\text{P.B.}&=2 {p_x^A} \left(x p_y^A-y p_x^A\right)^2 -{\frac{16 W_0\left({x}^2 +y^2 \right) {p_x^A}}{f_1 y^2}}-\\
		&-{\frac{8\tau_1 \left({x}^2 +y^2 \right) {p_x^A}}{f_1 y^2} \left({p_z^A} +
			{\frac{2\tau_1}{f_1 y^2}} \right)}=2 Y_1 X_1\nonumber,\\
		\{X_3,Y_1\}_\text{P.B.}&=\left(p_y^A\right)^2-{\frac{4\tau_1}{f_1 y^2}	\left(p_z^A+{\frac{2\tau_{1}}{f_1 y^2}}\right)}-
		{\frac{8 W_0 }{f_1 y^2}}=2H-\tilde{X}_2^2-Y_1^2\label{X5 2b},\\
		\{X_1,Y_1\}_\text{P.B.}	&=2p_y^A \left(x p_y^A-y p_x^A\right) -
		{\frac{8\tau_1 x}{f_1 y^2}\left(p_z^A+{\frac{2\tau_1}{f_1 y^2}} \right)}-\frac{16 W_0 x}{f_1 y^2}=2 X_3,
		\end{align}
		so the Poisson algebra is already closed in the sense that the Poisson bracket of any pair of integrals is a function (polynomial) in previously known integrals, and thus generates no new independent integral. Therefore, the system is not second order maximally superintegrable, as it was found in~\cite{Marchesiello2019}.
		
		Let us now compute the trajectories and see if the bounded ones are closed, which is typical for maximally superintegrable systems.
		The Hamilton's equations corresponding to the gauge-fixed Hamiltonian 
		\begin{equation}
			H=\frac{1}{2}\left((p_x)^2+(p_y)^2+\left(p_z-\frac{2\tau_1}{f_1 y^2}\right)^2\right)-{\frac{4 W_0 f_1 y^2 +2\tau_1^2}{f_1^2 y^4}}\label{Ham}
		\end{equation}
		 read
		\begin{gather}\label{Ham eq}
		\dot{x}=p_x,\quad \dot{y}=p_y,\quad \dot{z}=p_z-\frac{2\tau_1}{f_1 y^2},\\
		\dot{p}_x=0,\quad \dot{p}_y=-\frac{4\tau_1p_z+8 W_0}{f_1 y^3},\quad \dot{p}_z=0.
		\end{gather}
		The equations determining $x$ and $p_x$ correspond to the free motion of the particle in the $x$-direction
		\begin{equation}
		x(t)=p_{x0}t+x_0.
		\end{equation} 
		Because the momentum $p_z$ is conserved, we find that $y(t)$ must satisfy the following ODE
		\begin{equation}\label{y 2}
		\ddot y=-\frac{4\tau_1p_z+8 W_0}{f_1 y^3}.
		\end{equation}
		We note that the equation is not defined at $y(t)=0$.
		If $\tau_1p_z+2 W_0=0$, we have the free motion in this direction as well. We continue with the assumption that it is not the case.
		
		Equation \eqref{y 2} does not depend on the independent variable $t$, so it admits $\dot{y}$ as an integrating factor and (assuming it is non-zero) we obtain 
		\begin{equation}\label{y t}
		\dot{y}(t)=\pm\frac{\sqrt{f_1(-C_1 f_1 y^2 + 4 \tau_1 p_z + 8 W_0)}}{f_1 y},
		\end{equation}
		where the integration constant $C_1$ is real and we assume that the square root is well defined.
		Solving the separable ODE \eqref{y t}, we get
		\begin{equation}\label{y(t)}
		y(t)=\pm\frac{\sqrt{C_1 f_1\left(C_1^2 f_1(t-t_0)^2 - 4 p_z\tau_1 - 8 W_0\right)}}{C_1 f_1}.
		\end{equation}
		We remind the reader that eq.~\eqref{y 2} is not defined for $y(t)=0$, so we must restrict the independent variable $t$ so that this does not occur.
		The integration constant $t_0$ translates the origin of time and $C_1$ depends on the initial values and constants of the system
		\begin{equation}
		C_1=-\frac{f_1 y_0^2 \pm\sqrt {f_1^2 y_0^{4}+16t_0^2 f_1 p_z\tau_1 +32 W_0 t_0^2 f_1}}{2t_0^2 f_1}.
		\end{equation}
		The sign of the square root must be chosen so that the square root in eq.~\eqref{y(t)} is well defined (positive argument) for $t$ in a suitably chosen interval, which also determines the possible values of $t_0$.
		
		The sign of $C_1$ determines the motion of the particle in the $y$ coordinate: If $C_1>0$, then the particle escapes to the infinity. If $C_1<0$, then it falls on the origin $y=0$, where eq.~\eqref{y 2} has a singularity.
		
		To solve the case $C_1=0$, we must return to the eq.~\eqref{y t}, which leads to the singular solution
		\begin{equation}\label{y(t) sing}
		y(t)=\frac{2\sqrt{f_1(t-t_0)\sqrt{f_1 (p_z\tau_1 + 2 W_0)}}}{f_1}.
		\end{equation}
		This case is unbounded or goes to $y=0$ (where eq.~\eqref{y 2} is ill defined) for $f_1>0$ and $f_1<0$, respectively.
		
		We can now solve the remaining equation in eq.~\eqref{Ham eq}, namely 
		\begin{equation}\label{z 1}
		\dot{z}=p_z-\frac{2\tau_1}{f_1 y^2}.
		\end{equation}
		The solution using $y(t)$ from eq.~\eqref{y(t)} is
		\begin{equation}
		z (t) =-\frac{\tau_1 C_1{\rm argtanh} \left({\frac{C_1^2 f_1 \left(t-t_0\right)}
				{2\sqrt {C_1^2 f_1 \left(\tau_1 p_z+2 W_0\right)}}}\right)}{\sqrt {C_1^2 f_1 \left(\tau_1 p_z+2 W_0\right)}} + p_zt+C_2.
		\end{equation}
		The form of $z(t)$ above assumes that the argument of argtanh is real, i.e.
		\begin{equation}
		f_1 \left(\tau_1 p_z+2 W_0\right) >0,
		\end{equation}
		and not equal to $\pm1$ (domain of argtanh).
		On the other hand, if 
		\begin{equation}
		f_1 \left(\tau_1 p_z+ 2 W_0\right) <0,
		\end{equation}
		we can rewrite the argtanh using the identity
		\begin{equation}
		{\rm argtanh}(z)=\i\arctan(\i z), \quad \forall z\in \mathbb{C}\setminus\{\pm 1,\pm\i\}
		\end{equation}
		and cancel the imaginary unit with the one from the denominator. Because
		\begin{equation}
		\lim_{x\to\pm 1}{\rm argtanh}(x)=\pm \infty
		\end{equation} 
		and it is dominant with respect to $p_z t$, $z(t)$ is unbounded in the first case. In the second case, arctan is bounded and $z(t)$ is unbounded if and only if $p_z\neq0$.
		
		Using the singular solution eq.~\eqref{y(t) sing} in eq.~\eqref{z 1}, we get
		\begin{equation}
		z(t) = -\frac{\tau_1\ln(t - t_0)}{2\sqrt{f_1(p_z\tau_1 + 2 W_0)}} + p_z t + C_2,
		\end{equation}
		which is clearly unbounded as $t$ goes to $\infty$ or $t_0$.
		
		The last possibility, $y(t)=p_y^0 t+y_0$, leads to
		\begin{equation}
		z(t)=p_zt+\frac{2\tau_1}{f_1 p_y^0(p_y^0 t+y_0)}+C_3,
		\end{equation}
		which is unbounded if $p_z\neq0$ and singular at $t=0$ if $p_y\neq0$. (If $p_y^0=0$, the second term of $z(t)$ would be $\frac{2\tau_1 t}{f_1 y_0^2}$, i.e. free motion with a modified momentum.)
		
		To sum up, the motion in the $x$-direction is free. The motion in the $y$-direction can be free, unbounded, or bounded with the fall on the singular point $y=0$. The motion in the $z$-direction is always unbounded. Therefore, there are no bounded trajectories and we have no hint of higher order maximal superintegrability.
		
		Let us now turn to the quantum version of the system and its stationary Schr\"odinger equation. 
		Looking again at the integrals \eqref{px pz} and the Hamiltonian \eqref{Ham} in the gauge \eqref{gauge case Ib}, we see that we can write the Hamiltonian as
		\begin{equation}
		H=\frac{1}{2}(X_3^2+X_4^2)+\frac{1}{2}X_5,
		\end{equation}
		where 
		\begin{equation}
		X_5=[X_1,X_4]=\left(p_y^A\right)^2-{\frac{4\tau_1}{f_1 y^2}	\left(p_z^A+{\frac{\tau_1}{f_1 y^2}} \right)}-{\frac{8 W_0 }{f_1 y^2}},
		\end{equation}
		see eq.~\eqref{X5 2b}, which is, however, written in terms of the Poisson bracket. This confirms the separation of the Hamiltonian \eqref{Ham} in the Cartesian coordinates and the stationary Schr\"odinger equation reduces as follows.
		\begin{gather}
		\psi(\vec{x})=\exp\left(\frac{\i}{\hbar}\lambda_1 x\right)\exp\left(\frac{\i}{\hbar}\lambda_3 z\right)g(y),\\
		-\frac{\hbar^2}{2}g''(y)=\left(\frac{4(\tau_1+2 W_0)}{f_1 y^2}+E\right)g(y),\label{reduced schr}
		\end{gather}
		where $\lambda_1$ and $\lambda_3$ are eigenvalues of the momentum operators $p_x$ and $p_z$, respectively, i.e.
		\begin{equation}
		X_3 \psi(\vec{x})=\lambda_3\psi(\vec{x}),\quad X_4\psi(\vec{x})=\lambda_1\psi(\vec{x}).
		\end{equation}
		We substitute $g(y)=\sqrt{y}k(y)$ and obtain (after simplifying)
		\begin{equation}
		y^2k''(y)+yk'(y)+\left[\frac{2Ey^2}{\hbar^2}-\left(\frac{1}{4}-\frac{8a}{\hbar^2 f_1}\right)\right]k(y)=0,
		\end{equation}
		where $a=\tau_1+2 W_0$, which is the Bessel equation \cite[Chapter 9]{AbramowitzStegun}, \cite[Chapter~10]{DLMF} modulo a transformation of the independent variable. Equation~\eqref{reduced schr} can, therefore, be solved in terms of Bessel functions of the first and second kind ${\sl J}(\alpha;x),$ ${\sl Y}(\alpha;x)$:
		\begin{equation}
		g \left( y \right) =c_1 \sqrt {y}{{\sl J}
			\left({\frac{\i\sqrt {32 a - \hbar^2 f_1}}{2\hbar\sqrt {f_1}}};\sqrt {{\frac{2E}{\hbar^2}}}y\right)}+
		c_2 \sqrt {y}{{\sl Y}\left({\frac{\i\sqrt{32 a-\hbar^2 f_1}}{2\hbar\sqrt {f_1}}};\sqrt {{\frac{2E}{\hbar^2}}}y\right)},
		\end{equation}
		
		Let us also have a look at the Schr\"odinger equation in the cylindrical coordinates. The corresponding Hamiltonian reads
		\begin{equation}\label{ham sph}
		\hat{H}=-\frac{\hbar^2}{2}\left(\frac{1}{r}\pd_r(r\pd_r)+\frac{1}{r^2}\pd_{\phi\phi}+\pd_{ZZ}\right)+\frac{2\i\hbar\tau_1}{f_1 r^2 (\sin(\phi))^2}\pd_Z-\frac{4 W_0}{f_1 r^2 (\sin(\phi))^2}.
		\end{equation}
		Because the Hamiltonian does not depend on $Z$, it commutes with $P_Z=-\i\hbar\pd_Z$, so we can separate the $Z$ coordinate. We show that we can separate $r$ and $\phi$ as well. For that we use the \emph{ansatz}
		\begin{equation}
		\psi(r,\phi,Z)=f(r)g(\phi)\exp\left(\frac{\i}{\hbar}\lambda_3 z\right).
		\end{equation}
		Substituting it into the stationary Schr\"odinger equation with the Hamiltonian \eqref{ham sph}, we get
		\begin{equation}
		\begin{split}
		&\frac{\hbar^2}{2}\left[\left(f''(r)+\frac{1}{r} f'(r)\right)g(\phi)+\frac{1}{r^2}f(r)g''(\phi) \right]+\\ &+\left[\frac{2\tau_1\lambda_3+4 W_0}{f_1 r^2(\sin(\phi))^2}
		+\left(E-\frac{\lambda_3^2}{2}\right)\right]f(r)g(\phi)=0.
		\end{split}
		\end{equation}
		Dividing by $\frac{f(r)g(\phi)}{r^2}$, we can separate the terms to get the following equations
		\begin{align}
		f''(r) =&-\frac{f'(r)}{r} +\frac{c+2r^2 (\lambda_3^2 - 2 E)}{r^2 \hbar^2} f(r),\label{f(r)}\\ 
		g''(\phi) =& -\frac{c}{\hbar^2} + \frac{4 g(\phi) (\tau_1 \lambda_3 - 2 W_0)}{\sin(\phi)^2 f_1 \hbar^2}, \label{g(phi)}
		\end{align}
		where $c$ is the separation constant.
		
		Equation \eqref{f(r)} multiplied by $r^2$ is the Bessel equation \cite[Chapter 9]{AbramowitzStegun}, \cite[Section 10.2]{DLMF} (modulo a transformation of independent variable), so the solution in terms of Bessel functions is
		\begin{equation}
		f(r)=c_1{\sl J}\left(\frac{\sqrt{c}}{ \hbar}, \sqrt{2E-\lambda_3^2}{\hbar} r\right) + c_2{\sl Y}\left(\frac{\sqrt{c}}{\hbar}, \sqrt{2E-\lambda_3^2}{\hbar} r\right).
		\end{equation}
		Equation \eqref{g(phi)} can be solved in terms of the hypergeometric function ${}_2F_1(a,b;c;Z)$ \cite[Chapter 15]{AbramowitzStegun}, \cite[Chapter 5]{DLMF}, but the expression is too complicated for any practical use. Either way, we know that the spectrum is continuous due to the plane wave in $Z$-direction.
		
	Let us consider the stationary case of Hamilton-Jacobi equation. Starting in the Cartesian coordinates, we have $p_x$ and $p_z$ as the integrals of motion corresponding to cyclic coordinates in our gauge \eqref{gauge case Ib}. Thus, the Hamilton's characteristic function reads
		\begin{equation}
			U(x,y,z)=p_x x+u(y)+p_z z,
		\end{equation}
	where $u(y)$ is determined by the following reduced equation.
		\begin{equation}
			\frac{1}{2}\left[p_x^2+(u'(y))^2+\left(p_z-\frac{2\tau_1}{f_1 y^2}\right)^2\right]-4\left(\frac{\tau_1^2}{2 f_1^2 y^4}+\frac{W_0}{f_1 y^2}\right)=E.
		\end{equation}
	The resulting quadrature 
		\begin{equation}
			u(y)=\int\frac{\sqrt{f_1 ((2 E-p_x^2 - p_z^2) f_1 y^2+ 4 p_z \tau_1 + 8 W_0)}}{f_1 y}\d y
		\end{equation}
	has the following solution
	\begin{equation}
	u(y)=\sqrt{\frac{ \alpha f_1 y^2+ 4 \beta}{f_1}} - \frac{2 \beta}{\sqrt{\beta f_1}} \ln\left(\frac{\sqrt{\beta f_1} \sqrt{f_1 (\alpha f_1 y^2+ 4 \beta)} + 2\beta f_1}{y}\right)+C,
	\end{equation}
	where $\alpha=2 E-p_x^2 - p_z^2$ and $\beta=p_z \tau_1 + 2 W_0$.
		
	In the cylindrical coordinates with the same gauge \eqref{gauge case Ib} we have only one cyclic coordinate, namely $Z$, so we use the \emph{ansatz}
	\begin{equation}
		U(r,\phi,Z)=v(r)+w(\phi)+p_Z Z.
	\end{equation}
	We separate the equation as follows.
	\begin{align}
	\frac{[(v'(r))^2 + p_Z^2]r^2}{2}- Er^2 =c=-\frac{(w'(\phi))^2}{2}+\frac{2 p_Z \tau_1 + 4 W_0}{f_1 (\sin(\phi))^2}.
	\end{align}
	Solution to the separated equations is
\begin{flalign}
&\begin{aligned}
	v(r)&=\sqrt{\alpha r^2 + 2 c} - \sqrt{2c} \ln\left(\frac{\sqrt{2c\alpha r^2 + 4 c^2} + 2 c}{r}\right) + c_1,
\end{aligned}\\
&	\begin{aligned}
		w(\phi)&=\frac{\sqrt{2 c} \ln\left(\sqrt{2 c^2 f_1^2 \cos(2 \phi)+2 c f_1 \beta} \cos(\phi)-c f_1 \cos(2 \phi)-2 \gamma\right)}{2 \sqrt{\gamma}}+\\
		&+\frac{\sqrt{\gamma} \ln\left(4 \cos(\phi) \sqrt{\gamma c f_1 \cos(2 \phi)+\beta \gamma}+(6\gamma-\beta) \cos(2 \phi)+\beta-\gamma\right)}{{\sqrt{f_1}}}-\\
		&-\frac{2 \sqrt{\gamma} \ln(\sin(\phi))}{\sqrt{f_1}}+c_2,
	\end{aligned}
\end{flalign}
where $\alpha=2E-p_Z^2$, $\beta=4 p_Z \tau_1+8 W_0-c f_1$ and $\gamma=p_Z \tau_1+2 W_0$.
\end{enumerate}

\subsection{Case $\mathrm{rank}(M)=1$ subcase 3a 1a)}\label{sec: 3a 1a}
The assumptions of this case are

\begin{equation}
	\mu(Z)\neq 0,\quad \psi(\phi)= 0,\quad	\rho(r)= 0,\quad \sigma(r)= 0,
\end{equation}
which means 
\begin{equation}
	B^r(r, \phi, Z) = \frac{\tau'(\phi)}{2r^2}-\frac{\mu'(Z) r^2}{2},\ B^\phi(r, \phi, Z) = \frac{\tau(\phi)}{r^3},\ B^Z(r, \phi, Z) = \mu(Z) r.
\end{equation}
We note that $\psi(\phi)=0$ implies vanishing quantum correction, so the classical and quantum systems coincide.


Recalling eq.~\eqref{k4k5}, differentiating eq.~\eqref{komp 2} with respect to $Z$ gives
\begin{equation}
	(k_2 \cos(\phi)-k_1 \sin(\phi))\mu''(Z)=0.
\end{equation}
Thus, in order to have additional integrals of motion, we need $\mu(Z)=\mu_1 Z+\mu_0$. From eq.~\eqref{komp 2} twice differentiated with respect to $r$ we immediately get $\mu_1=0$, so we have
\begin{equation}
	\mu(Z)=\mu_0\neq 0.
\end{equation}
Differentiating eq.~\eqref{komp 2} with respect to $r$ once this time, we obtain the equation which leads to splitting  of further considerations:
\begin{equation}
	k_6\tau'(\phi)=0,
\end{equation}
i.e. we split into $k_6=0$ and $\tau(\phi)=\tau_0$. It is convenient to further split the second subcase into $\tau_0=0$ and $\tau_0\neq 0$.

\begin{enumerate}[label=\Roman*.]
	\item $\tau(\phi)=\tau_0\neq 0$: Using these assumptions in eq.~\eqref{komp 2}, which reads 
	\begin{equation}
		\tau_0(k_2 \cos(\phi)+k_1 \sin(\phi))=0,
	\end{equation}
	implies $k_1=k_2=0$, so the only first order integrals of the motion are
	\begin{equation}
	\tilde{X}_1=p_Z^A-\frac{\tau_0}{2r^2},\quad \tilde{X}_2=p_\phi^A-\frac{\mu_0r^2}{2},
	\end{equation}
	i.e. the reduced cylindrical integrals, so this case is only integrable.
	\item \label{konst pole W(Z)} $\tau(\phi)=0$: This leads to the constant magnetic field in the $z$-direction
	\begin{equation}\label{B3a1a}
	\quad B^r=0,\quad B^\phi=0,\quad B^Z=\mu_0 r,
	\end{equation}
	the scalar potential $W$ of the form of eq.~(111) from \cite{Fournier2019} with redefined $W_1(r)$
	\begin{align}\label{potentialmu}
	W &= W_1(r) + W_3(Z),
	\end{align}
	which must satisfy an additional constraint, namely eq.~\eqref{ord0 lin}
	\begin{equation}
		(k_1\cos(\phi)+k_2 \sin(\phi)) W_1'(r) +k_3 W_3'(Z)=0.
	\end{equation}
	The case $k_1=k_2=0$ implies no additional integrals, we will therefore consider only
	\begin{equation}
		W_1(r) = W_0.
	\end{equation} 
	Subsequently we need $k_3=0$ or $W_3(Z)= W_3$. We can set $W_3=0$ by another redefinition of $W_1(r).$ We have, therefore, 2 superintegrable subcases: The constant magnetic field from eq.~\eqref{B3a1a} with either constant potential 
	\begin{equation}
		W= W_0, \text{ without loss of generality }W=0, 
	\end{equation}
	which has four first order integrals of motion and was already considered, see eq.~\eqref{IP konst pole} (with $\rho_1=-\mu_0$),
	or the potential 
	\begin{equation}
		W= W_3(Z)\equiv W_3(z)
	\end{equation}
	with the first order integrals
	\begin{equation}\label{IP konst WZ}
		Y_1=p_x^A+\mu_0 y,\quad Y_2=p_y^A-\mu_0 x,\quad \tilde{X}_1=L_z^A-\frac{\mu_0 (x^2+y^2)}{2},
	\end{equation}
	complemented by the irreducible second order cylindrical integral
	\begin{equation}\label{X2 3a}
		X_2=\left(p_z^A\right)^2+2 W_3(z).
	\end{equation}
	In both cases we have a relation connecting the integrals
	\begin{equation}\label{H W(Z)}
		H=\frac{1}{2}\left(Y_1^2+Y_2^2+X_2\right)+\mu_0\tilde{X}_1,
	\end{equation}
	where in the first case $X_2=\tilde{X}_2^2$, so these integrals ensure minimal superintegrability only.
	
	
	The second system is Case B in \cite{Marchesiello2017}, where it was shown that this system is second order maximally superintegrable for 2 special forms of the scalar potential $W$:
	\begin{align}\label{WZ special}
		W_3(Z)=\frac{c}{z^2}+\frac{\mu_0^2 z^2}{8},\quad W_3(Z)=\frac{\mu_0^2}{2}z^2,
	\end{align}
	where $c\geq0$ is a constant. (The remaining values $c<0$ would allow fall on the $z=0$ plane, where the potential is ill-defined.)
	In \cite{Marchesiello2019} it was found that the system can be reduced to a 2D system with Hamiltonian of the form (with a permutation of coordinates)
	\begin{equation}\label{Kamiltonian}
		K(\vec{X},\vec{P})=\frac{1}{2}\left(P_1^2+P_2^2\right)+\frac{\mu_0^2}{2}Y^2+W(Y).
	\end{equation}
	Seen as a 3D system, it has 2 cyclic coordinates $Z,$ $P_3$, and is therefore maximally superintegrable if and only if the 2D system with Hamiltonian \eqref{Kamiltonian} is superintegrable. Note that the Hamiltonian \eqref{Kamiltonian} corresponds to a system without magnetic field, which is a well-studied problem. For superintegrable cases of order at most 3 see \cite{Miller2013}. The quadratic cases are the potentials from eq.~\eqref{WZ special}, where the second potential can be modified by a translation in $z$ and a constant shift of the potential (because $\mu_0 \neq 0$) to become
	\begin{equation}
		W_3(Z)=\frac{\mu_0^2}{2}z^2+c z.
	\end{equation}
	For some higher order systems see \cite{Marchesiello2019} and references therein.
	
	Let us solve the Hamiltonian equations of motion and the stationary Schr\"odinger equation for the cases in \cite{Marchesiello2017}, i.e. with scalar potential $W$ from eq.~\eqref{WZ special}. For that we choose the gauge 
	\begin{equation}\label{gauge WZ}
	\vec{A}(\vec{x})=(0,\mu_0 x, 0).
	\end{equation}
	For the first potential in eq.~\eqref{WZ special} the following independent second order integral of motion was found
	\begin{equation}\label{X5}
		X_5=\left(L_x^A\right)^2+\left(L_y^A\right)^2-\mu_0 z^2 L_z^A+\frac{\mu_0^2}{4}z^2(x^2+y^2)+\frac{2c}{z^2}(x^2+y^2).
	\end{equation}
	The equations of motion in the gauge~\eqref{gauge WZ} read \cite{Marchesiello2017}
	\begin{gather}
		\dot{x}=p_x,\quad \dot{y}=p_y+\mu_0 x,\quad \dot{z}=p_z,\\
		\dot{p}_x=-\mu_0(p_y+\mu_0 x),\quad \dot{p}_y=0,\quad \dot{p}_z=-\frac{1}{4}\left(\mu_0^2z-\frac{8c}{z^3}\right).
	\end{gather}
	The explicit trajectories are 
	\begin{align}
		x(t)=& \frac{p_x^0}{\mu_0}\sin(\mu_0 t)+\left(\frac{p_y^0}{\mu_0}+x_0\right)\cos(\mu_0 t)-\frac{p_y^0}{\mu_0},\\
		y(t)=&\left(\frac{p_y^0}{\mu_0}+x_0\right)\sin(\mu_0 t)-\frac{p_x^0}{\mu_0}\cos(\mu_0 t)+\frac{p_x^0}{\mu_0}+y_0,\\
		z(t)=& \frac{\sqrt{\mu_0 ^2 z_0^4 (\cos(\mu_0 t) + 1) + 4 \mu_0 p_z^0 z_0^3 \sin(\mu_0 t) + 4 \left((p_Z^0)^2 z_0^2 + 2 c\right) (1 - \cos(\mu_0 t))}}{\sqrt{2} \mu_0 z_0}
	\end{align}
	for $c> 0$, otherwise ($c=0$) the potential is the harmonic oscillator and we get ($x(t)$ and $y(t)$ remain the same)
\begin{equation}
	z(t)=2\left(\frac{p_z^0}{\mu_0}\sin\left(\frac{\mu_0}{2} t\right)+\frac{z_0}{2}\cos\left(\frac{\mu_0}{2}t\right)\right).
\end{equation}
In both cases the trajectories are periodic and bounded (in the $c>0$ case if $z_0\neq 0$, so that the solution is well defined). 

	For the second potential in eq.~\eqref{WZ special}, the fifth independent integral reads \cite{Marchesiello2017}
	\begin{equation}
		X_5=p_x^A L_y^A-p_y^A L_x^A-\mu_0 z L_z^A.
	\end{equation}
	The equations of motion in the gauge~\eqref{gauge WZ} read
	\begin{gather}
	\dot{x}=p_x,\quad \dot{y}=p_y+\mu_0 x,\quad \dot{z}=p_,\\
	\dot{p}_x=-\mu_0(p_y+\mu_0 x),\quad \dot{p}_y=0,\quad \dot{p}_z=-\mu_0^2 z.
	\end{gather}
	The explicit trajectories are 
	\begin{align}
	x(t)=&{} \frac{p_y^0}{\mu_0}\sin(\mu_0 t)+\left(\frac{p_y^0}{\mu_0}+x_0\right)\cos(\mu_0 t)-\frac{p_y^0}{\mu_0},\\
	y(t)=&{}\left(\frac{p_y^0}{\mu_0}+x_0\right)\sin(\mu_0 t)-\frac{p_y^0}{\mu_0}\cos(\mu_0 t)+\frac{p_x^0}{\mu_0}+y_0,\\
	z(t)=&{}\frac{p_z^0}{\mu_0}\sin\left(\mu_0 t\right)+z_0\cos\left(\mu_0 t\right).
	\end{align}
	
	Concerning the Schr\"odinger equation, we first note that it is separable even in the general minimally superintegrable case with $W_3(Z)$. For that we choose the gauge \eqref{gauge WZ}, use the commutation of integrals $Y_2,$ $X_2$ and consider the wave function $\psi(\vec{x})$ to be their (generalized) eigenfunction
	\begin{equation}
	Y_2\psi(\vec{x})=\lambda_2\psi(\vec{x}),\quad X_2\psi(\vec{x})=\lambda_3\psi(\vec{x}).
	\end{equation} 
	We write $\psi(\vec{x})$ as
	\begin{equation}\label{psi WZ}
	\psi(\vec{x})=f(x) \exp\left(\frac{\i}{\hbar}\lambda_2 y\right)g(z),
	\end{equation}
	where $g(z)$ is (generalized) eigenfunction of $X_2$ satisfying
	\begin{equation}\label{spektr WZ}
	-\hbar^2g''(z)+2W_3(z)g(z)=\lambda_3 g(z)
	\end{equation}
	(to be determined for the specific $W_3(z)$) and $f(x)$ satisfies the reduced Schr\"odinger equation
	\begin{equation}
	\hbar^2 f''(x)=\left[\left(\mu_0 x+\lambda_2\right)^2+\lambda_3-2E\right]f(x).
	\end{equation}
	The last equation is again the equation for 1D harmonic oscillator with energy $E-\frac{\lambda_3}{2}$, angular frequency $\omega=\mu_0$ with the centre of force $x=-\frac{\lambda_2}{\mu_0}$. The corresponding eigenfunctions are
	\begin{equation}
	f_{n}(x)=K_n H_n\left(\sqrt{\frac{\mu_0}{\hbar}}\left(x+\frac{\lambda_2}{\mu_0}\right)\right)\exp\left(-\frac{\mu_0}{\hbar}\left(x+\frac{\lambda_2}{\mu_0}\right)^2\right),
	\end{equation}
	with $K_n$ the normalization constant.
	
	The spectrum of the Hamiltonian, 
	\begin{equation}
		E=\hbar\frac{\mu_0}{2}(2n+1)+\frac{\lambda_3}{2},
	\end{equation}
	is continuous regardless of $\lambda_3$, which must be determined from eq.~\eqref{spektr WZ}, because $\psi(\vec{x})$ from eq.~\eqref{psi WZ} is in not normalizable.
	

	The solution to eq.~\eqref{spektr WZ} with the first potential in eq.~\eqref{WZ special} obtained by Maple\texttrademark~is in terms of Whittaker functions $M_{\mu,\nu}(z),$ $W_{\mu,\nu}(z)$ \cite[Chapter 13]{AbramowitzStegun}, \cite[Section 13.14]{DLMF}
	\begin{equation}
	g(z) =\frac{1}{\sqrt{z}} \left[c_1 M_{\frac{\lambda_3}{\mu_0 \hbar}, \frac{\sqrt{\hbar^2 + 8 c}}{4 \hbar}} \left(\frac{\mu_0 z^2}{2\hbar} \right) + c_2 W_{\frac{\lambda_3}{\mu_0 \hbar}, \frac{\sqrt{h^2 + 8 c}}{4 \hbar}}\left(\frac{\mu_0 z^2}{2 \hbar}\right)\right]
	\end{equation}
	and if the first arguments in them are not negative integers or 0, it can be rewritten in terms of confluent hypergeometric function ${}_1F_1(a,b,z)$ (Kummer $M(a,b,z)$) \cite[Chapter 13]{AbramowitzStegun}, \cite[Section 13.2]{DLMF} as follows
	\begin{align}
	\begin{aligned}
		g(z)&=\tilde{c}_1\exp\left(-\frac{\mu_0 z^2}{4 \hbar}\right) z^{1+\frac{a}{2\hbar}}
	{}_1F_1\left(\frac{1}{2}-\frac{4\lambda-a}{4\mu_0\hbar}, 1+\frac{a}{2\hbar}, \frac{\mu_0 z^2}{2 \hbar}\right)+\\
	&+\tilde{c}_2\exp\left(-\frac{\mu_0 z^2}{4 \hbar}\right)z^{1-\frac{a}{2\hbar}}
	{}_1F_1\left(\frac{1}{2}-\frac{4\lambda+a}{4\mu_0\hbar}, 1-\frac{a}{2\hbar}, \frac{\mu_0 z^2}{2 \hbar}\right),
	\end{aligned}
	\end{align}
	where $a=\sqrt{\hbar^2+8c}.$
	(For $c=0$ reduces to harmonic oscillator with $\omega=\frac{\mu_0}{2}$.)
	
	The solution of eq.~\eqref{spektr WZ} with the second potential in eq.~\eqref{WZ special} is again the harmonic oscillator, this time without shift, whose solutions are
	\begin{equation}
	g_n(z)=K_n H_n\left(\sqrt{\frac{\mu_0}{\hbar}}z\right)\exp\left(-\frac{\mu_0}{2\hbar} z^2\right),
	\end{equation}
	where $K_n$ is the normalization constant, so for the second potential in eq.~\eqref{WZ special} the spectrum of the Hamiltonian is
	\begin{equation}
		E=\hbar\mu_0(2n+1).
	\end{equation}
	It is nevertheless continuous, because $\psi(\vec{x})$ from eq.~\eqref{psi WZ} is not normalizable.
	
	Let us see if the systems separate in the cylindrical coordinates as well. 
	\begin{equation}
	\begin{split}
	\hat{H}={}&-\frac{\hbar^2}{2}\left(\frac{1}{r}\pd_r(r\pd_r)+\frac{1}{r^2}\pd_{\phi\phi}+\pd_{ZZ}\right)-\\
	&-\i\hbar\rho_1 r \cos(\phi)\left(\sin(\phi)\pd_r+\frac{\cos(\phi)}{r}\pd_\phi\right)+\frac{\rho_1}{2} r^2(\cos(\phi))^2+W(Z).
	\end{split}
	\end{equation}
	It is clear that we can separate the $Z$ coordinate from the $r$ and $\phi$, the equation for the $z$ coordinate is \eqref{spektr WZ}. The remaining part of the separation is the same as in the case with constant magnetic field, see eq.~\eqref{sep r phi}, where we have shown that it is impossible to separate $r$ and $\phi$ in this gauge. However, it can be separated in another choice of gauge, see the text under eq.~\eqref{konst pole schr cyl}. 
	
	Stationary Hamilton-Jacobi equation in the Cartesian coordinates
	\begin{equation}
		\frac{1}{2}\left[\left(\pderA{U}{x}\right)^2+\left(\pderA{U}{y}+\mu_0 x\right)^2+\left(\pderA{U}{z}\right)^2\right]+W_3(z)=E
	\end{equation}
	separates because the coordinate $y$ is cyclic. After the $z$-dependence is solved, $\pd_z U=v'(z)=\sqrt{2(\lambda_3-W_3(z))}$, the $x$-dependence is a quadrature of the form \eqref{U konst pole}, and thus the result is
	\begin{equation}
	\begin{split}
		u(x)={}&\frac{E - \lambda_3}{\mu_0} \arctan\left(\frac{\mu_0 x + p_y}{\sqrt{-\mu_0^2 x^2 - 2 \mu_0 p_y x - p_y^2 + 2 E - 2 \lambda_3}}\right) +\\
		&+\frac{(\mu_0 x + p_y)}{2 \mu_0} \sqrt{-\mu_0^2 x^2 - 2 \mu_0 p_y x - p_y^2 + 2 E - 2 \lambda_3}+C_1
	\end{split}
	\end{equation}
	The solution for the first potential in~\eqref{WZ special} is of the type in eq.~\eqref{int} and reads (assuming positive argument in the square root)
\begin{equation}
\begin{split}\label{U(z)}
	v(z)={}&-\frac{\sqrt{-\mu_0 z^4 + 8 \lambda_3 z^2 - 8 c}}{4} - \frac{\sqrt{2c}}{2} \arctan \left(\frac{\sqrt{2}\lambda_3+\sqrt{8}c}{\sqrt{c}\sqrt{-\mu_0 z^4 + 8 \lambda_3 z^2- 8 c}}\right)-\\ &-\frac{\lambda_3}{\sqrt{-\mu_0}} \ln\left(\frac{4 \lambda_3 - \mu_0 z^2}{\sqrt{-\mu_0}}+\sqrt{-\mu_0 z^4 + 8 \lambda_3 z^2 - 8 c}\right).
\end{split}
\end{equation}

For the second potential in~\eqref{WZ special} it reads (with the same assumption)
\begin{equation}
\begin{split}
v(z)&=\frac{z \sqrt{-\mu_0 z^2 + 2 \lambda_3}}{2} + \frac{\lambda_3}{\sqrt{\mu_0}} \arctan\left(\frac{\sqrt{\mu_0} z}{\sqrt{-\mu_0 z^2 + 2 \lambda_3}}\right).
\end{split}
\end{equation}
If $\mu_0<0$, we can write the second term in terms of argtanh and $|c|$.

We try the same thing in the cylindrical coordinates as well. We use the other gauge \eqref{konst pole schr cyl} with $\mu_0=-\rho_1$, so the equation reads
\begin{equation}
	\frac{1}{2}\left[\left(\pderA{U}{r}\right)^2+\frac{1}{r^2}\left(\pderA{U}{\phi}+\frac{\mu_0 r^2}{2}\right)^2+\left(\pderA{U}{Z}\right)^2\right]+W_3(Z)=E.
\end{equation}
Because $z\equiv Z$, we can separate it as in the Cartesian coordinates with the same results. Our choice of gauge ensures $\tilde{X}_1=p_\phi$ ($\phi$ is cyclic), so the $\phi$-dependent part of $U$ reads $p_\phi \phi$. The $r$ dependence is determined form
\begin{equation}
	\frac{w'(r)^2}{2} + \frac{(\mu_0 r^2 + 2 p_\phi)^2}{8 r^2} + \lambda_3 = E.
\end{equation} 
The solution is of the same type as in \eqref{U(z)} with different constants.
	
	\item $k_6=0$, $\tau(\phi)\neq 0$. Assuming that $k_1,$ $k_2$ do not both vanish, we obtain 
	\begin{equation}
		\tau(\phi)=\frac{\tau_1}{(k_2\cos(\phi)-k_1\sin(\phi))^2}.
	\end{equation}
	We will use the form of the scalar potential from \cite{Fournier2019}
	\begin{align}\label{Wmutau}
		W(r,\phi,Z) &= -\frac{1}{4r^2}T''(\phi)M(Z) - \frac{1}{4}T(\phi) M''(Z) - \frac{r^2}{8}M'(Z)^2- \frac{1}{8r^4}T'(\phi)^2 + \nonumber \\
		&+ W_1(r) - \frac{1}{r^2}\left[\frac{C_1}{8} \left(T(\phi)\right)^2-w_0 T(\phi)\right] -\frac{C_2}{8} \left(M(Z)\right)^2+w_0 M(Z),
	\end{align}
	where $M'(Z)=\mu(Z)$ and $T'(\phi)=\tau(\phi)$, which in our case means (assuming $k_1\neq 0$)
	\begin{equation}
		M(Z)=\mu_0 Z+m_0,\quad T(\phi)= -\frac{\tau_1}{k_1(k_1\tan(\phi) - k_2)} + t_0.
	\end{equation}
	Inserting the scalar potential \eqref{Wmutau} into the constraint \eqref{ord0 lin}
	\begin{equation}\label{constr}
		\frac{1}{r}[(k_1\cos(\phi)+k_2 \sin(\phi))r \pd_r W -(k_2 \cos(\phi)-k_1 \sin(\phi)-k_6 r)\pd_\phi W + k_3 r \pd_Z W]=0.
	\end{equation}
	and differentiating with respect to $Z$, we obtain 
	\begin{equation}
		\begin{split}
		\mu_0[C_2 k_3 r^3((k_2^3-3 k_1^2 k_2)\cos(\phi)^3 + k_1 (k_1^2 - 3 k_2^2)\cos(\phi)^2\sin(\phi)+\\
	+ 3 k_2 k_1^2\cos(\phi) - k_1^3\sin(\phi))+2\tau_1(k_1^2+k_2^2)]=0
		\end{split}
	\end{equation}
	To satisfy the equation for all $\phi$, each term must vanish on its own. However, that is not the case for the last term: the assumptions of our case are $\tau_1\neq 0$ and $\mu_0\neq0$, and $k_1,$ $k_2$ must be real, for otherwise we would have complex magnetic field 
	\begin{equation}
		B^\phi(r, \phi, Z) = \frac{\tau(\phi)}{r^3}=\frac{\tau_1}{r^3 (k_2\cos(\phi)-k_1\sin(\phi))^2}.
	\end{equation}
	If $k_1=0$, the function $T(\phi)$ changes to
	\begin{equation}
		T(\phi) = \tau_1\frac{\tan(\phi)}{k_2^2} + t_0
	\end{equation}
	and the differentiated eq.~\eqref{constr} reads
	\begin{equation}
		\mu_0(C_2 k_2 k_3 \mu_0 r^3\cos(\phi)^3 + 2\tau_1)=0,
	\end{equation}
	which has the same consequences. Because the potential from eq.~\eqref{Wmutau} cannot be (under our assumptions) constant, see the first and the last term on the first line, we conclude that the considered subcase $k_6=0$, $\tau(\phi)\neq 0$ does not have consistent scalar potential $W$.
%
\end{enumerate}

\subsection{Case $\mathrm{rank}(M)=1$ subcase 3a 1b)}
The assumptions of this case are
\begin{equation}
\mu(Z)\neq 0,\quad \rho(r)\neq 0,\quad \psi(\phi)= 0,\quad \sigma(r)= 0, \quad	\tau(\phi)= 0,
\end{equation}
which means 
\begin{align}\label{Wrztau0sol}
\begin{aligned}
W &= W_1(r) - \frac{r^2}{8}\mu(Z)^2 + \frac{1}{4}\rho(r)\mu(Z) + W_3(Z),\\
B^r &= - \frac{r^2}{2}\mu'(Z), \quad B^\phi =0, \quad B^Z = r \mu(Z) - \frac{1}{2}\rho'(r).
\end{aligned}
\end{align}


Recalling eq.~\eqref{k4k5}, equation \eqref{komp 2} implies $\mu(Z)=\mu_0$ and subsequently from eq.~\eqref{komp 1} follows $\rho(r) = \rho_1 r^2 +\rho_0$ with $\rho_1\neq \mu_0$ (otherwise the magnetic field vanishes). Due to the last inequality and the form of $B^Z$ in eq.~\eqref{Wrztau0sol} we can set $\rho_1=0$ by redefinition of $\mu_0$, so we obtain the following two superintegrable systems with constant magnetic field:

The magnetic field is the same in both cases, namely
\begin{equation}
	B^r(r, \phi, Z) = 0,\quad B^\phi(r, \phi, Z) = 0, \quad B^Z(r, \phi, Z) = r\mu_0,
\end{equation}
which in the Cartesian coordinates reads
\begin{equation}
	B^x(x,y,z)=0,\quad B^y(x,y,z)=0,\quad B^z(x,y,z)=\mu_0.
\end{equation}
The systems are distinguished by their scalar potentials, which are determined from equation 
\begin{equation}\label{1}
	(k_2 \sin(\phi) +k_1\cos(\phi))\pd_r W +k_3\pd_Z W=0.
\end{equation}
Due to the form of the potential eq.~\eqref{Wrztau0sol} we have 2 possibilities with additional integrals of motion: the constant scalar potential $W=W_0$ with 4 first order integrals from eq.~\eqref{IP konst pole} (with $\rho_1=-\mu_0$) and
$W= W_3(Z)$ with $k_3=0$, i.e. with the integrals $Y_1,Y_2,\tilde{X}_1$ from eq.~\eqref{IP konst pole} and the not-reduced integral 
	\begin{equation}
X_2=\left(p_Z^A\right)^2+2 W_3(Z).
\end{equation}
(The other solutions of eq.~\eqref{1} $W= W(r)$ and $W= W(r,Z)$ admit no additional integrals, only one or both cylindrical integrals reduce to the first order integrals.)

The system with constant potential is well known, see e.g. \cite{Landau}, \cite{Marchesiello2015} and Subsection~\ref{sec:case 2a lin}. It is in fact maximally superintegrable with the integral from eq.~\eqref{paty konst pole} (with $\rho_1=-\mu_0$). The second system was considered in \cite{Marchesiello2017}. For more details see the text around eq.~\eqref{WZ special}, where the same system was considered (with $\rho_1=-\mu_0$).

\subsection{Case $\mathrm{rank}(M)=1$ subcase 3a 2)}
The assumptions of this system are
\begin{equation}
	\tau(\phi)\neq 0,\quad \psi(\phi)= 0,\quad	\mu(Z)=0,\quad \rho(r)= 0
\end{equation}
and lead to the following magnetic field $B$ and scalar potential $W$
\begin{align}\label{Wrzmu0sol}
\begin{split}
W &= W_1(r) - \frac{1}{8r^4}\tau(\phi)^2 + \frac{1}{4r^2}\tau(\phi)\sigma(r) + \frac{1}{r^2} W_2(\phi),\\ 
B^r &= \frac{1}{2r^2}\tau'(\phi), \quad B^\phi = \frac{1}{r^3}\tau(\phi) + \frac{1}{2}\sigma'(r), \quad B^Z = 0.
\end{split}
\end{align}


First, we recall eq.~\eqref{k4k5}, i.e. $k_4=k_5=0$. We also assume that $k_1$, $k_2$ do not both vanish, because otherwise we would have the two cylindrical integrals $X_1,X_2$ only. Equations \eqref{komp 2} and \eqref{komp 3} differentiated twice with respect to $r$ give us the form of $\sigma(r)$:
\begin{equation}
	\sigma(r) = \frac{\sigma_1 r + \sigma_2}{r^2}+\sigma_0.
\end{equation}
Inserting this result, eq.~\eqref{komp 2} once differentiated with respect to $r$ 
\begin{equation}\label{split}
	\sigma_1(k_2 \sin(\phi)+ k_1\cos(\phi))+k_6 \tau'(\phi)=0
\end{equation}
leads to splitting of our considerations into 2 cases, $k_6=0$ and $k_6\neq 0$.

\begin{enumerate}[label=\Roman*.]
	\item $k_6=0$: In this case eq.~\eqref{split} implies $\sigma_1=0$. Subsequent solving of eq.~\eqref{komp 2} and eq.~\eqref{komp 3} for $\tau(\phi)$ yields
	\begin{equation}
	\tau(\phi)=\sigma_2-\frac{\tau_1}{(k_1\sin(\phi)-k_2\cos(\phi))^2}.
	\end{equation}
	The corresponding magnetic field reads
	\begin{equation}\
		B^r = \frac{\tau_1(k_1 \cos(\phi)+k_2 \sin(\phi))}{r^2(k_1\sin(\phi)-k_2\cos(\phi))^3}, \quad B^\phi = \frac{-\tau_1}{r^3(k_1\sin(\phi)-k_2\cos(\phi))^2},\quad B^Z = 0.
	\end{equation}
	Without loss on generality,	we simplify the following analysis using the formula $k_1\sin(\phi)-k_2\cos(\phi)=\sqrt{k_1^2+k_2^2}\cos(\phi+\phi_0)$ and choosing the coordinates (rotating the system) so that $\phi_0=\frac{3}{2}\pi$, i.e. $k_1=0$, and redefine $\tau_1$ so that $k_2=1$.
	
	Writing the simplified magnetic field in the Cartesian coordinates, we get
	\begin{equation}
		B^x= 0, \quad B^y =-\frac{\tau_1} {x^3},\quad B^z = 0.
	\end{equation}
	The compatibility equations are solved, and we continue with eq.~\eqref{ord0 lin} to obtain the form of the scalar potential $W$. The solution in this case is
	\begin{equation}
		W(r,\phi)= W(r\cos(\phi)).
	\end{equation} 
	Comparing this result with the form of the scalar potential in eq.~\eqref{Wrzmu0sol}, the final form of $W$ in the cylindrical coordinates is
	\begin{equation}
		W =\frac{W_0+\sigma_0\tau_1}{4r^2(\cos(\phi))^2} - \frac{\tau_1^2}{8 r^4(\cos(\phi))^4}
	\end{equation}
	and in the Cartesian coordinates reads
	\begin{equation}
		W =\frac{W_0+\sigma_0\tau_1}{4 x^2} - \frac{\tau_1^2}{8 x^4}.
	\end{equation}
	The corresponding integrals of motion are (in the Cartesian coordinates)
	\begin{equation}
		\tilde{X}_2= p_z^A+\frac{\tau_1}{x^2},\quad Y_1=p_y^A
	\end{equation}
	and the not-reduced cylindrical integral 
	\begin{equation}
	X_1=\left(p_\phi^A\right)^2+\sigma_2-\frac{\tau_1}{(\cos(\phi))^2}=(L_z^A)^2+\sigma_2-\frac{\tau_1}{x^2}.
	\end{equation}
	This system was already analysed in Subsection~\ref{sec:2b 1.2} subcase~\ref{system 2b 1.2}, where we had different definitions of constants and a rotated coordinate system.
	\item $k_6\neq 0$. This time the eq.~\eqref{split} can be solved for $\tau(\phi)$:
	\begin{equation}
		\tau(\phi) = -\frac{\sigma_1(k_1\sin(\phi) - k_2\cos(\phi))}{k_6} + \tau_0.
	\end{equation}
	Inserting this into eq.~\eqref{komp 2}, we get
	\begin{multline}
		18 k_1 k_2 \sigma_1\cos(\phi)^2- [9\sigma_1(k_1^2 - k_2^2)\sin(\phi) - 6 k_1 k_6(\tau_0 - \sigma_2)]\cos(\phi) +\\+ k_2[(6 k_6(\tau_0 - \sigma_2)\sin(\phi)) - 9 k_1\sigma_1]=0.
	\end{multline}
	This must be satisfied for all $\phi$, which implies either $k_1=k_2=0$, i.e. no additional integrals, or $\tau_0=\sigma_2$ and $\sigma_1=0$, which means zero magnetic field. This subcase, therefore, does not contain anything interesting for us.
\end{enumerate}

\subsection{Case $\mathrm{rank}(M)=1$ subcase 3a 3)}
The assumptions of the system 
\begin{equation}
	\rho(r)\neq 0,\quad \psi(\phi)=0,\quad \mu(Z)=0,\quad \tau(\phi) = 0,\quad \sigma(r)=0
\end{equation}
lead to the following magnetic field $B$ and scalar potential $W$
\begin{align}\label{r12dpolar}
W &= W_1(r) + W_3(Z), \quad
B^r = 0, \quad B^\phi =0, \quad B^Z = - \frac{1}{2}\rho'(r).
\end{align}


Taking eq.~\eqref{k4k5} into account, the last remaining equation \eqref{komp 1}, namely 
\begin{equation}
	r^2(\rho'(r)-r\rho''(r))(k_2\sin(\phi) +k_1 \cos(\phi))=0,
\end{equation}
implies $\rho(r)=\rho_1 r^2+\rho_0$. (We exclude $k_1=k_2=0$, which leads to no additional integral to the 2 cylindrical ones.)

We have, therefore, again the system with constant magnetic field
\begin{equation}\label{B 3a 3}
	B^r=0,\quad B^\phi=0,\quad B^Z=-\rho_1 r, 
\end{equation}
which in the Cartesian coordinates reads
\begin{equation}
	B^x=0,\quad B^y=0,\quad B^z=-\rho_1.
\end{equation}
We next determine the scalar potential $W$. The zeroth order equation \eqref{ord0 lin} with the potential from eq.~\eqref{r12dpolar} reduces to
\begin{equation}
	(k_2\sin(\phi)+k_1\cos(\phi)) W_1'(r)+k_3 W_3'(Z)=0.
\end{equation}
Assumption of additional integrals, i.e. at least one of the constants $k_1,$ $k_2$ non-vanishing, implies $W_1(r)=w_0$, which we can absorb into $W_3(Z)$ by redefinition. 

Therefore, we have two possible subcases, both with the constant magnetic field eq.~\eqref{B 3a 3}:
The first subcase has constant scalar potential,
\begin{equation}
	W= W_0,
\end{equation}
with 4 first order integrals from eq.~\eqref{IP konst pole},
and the second subcase has the potential
\begin{equation}
	W= W_3(Z),
\end{equation}
the first order integrals $Y_1,$ $Y_2,$ $\tilde{X}_1$ from eq.~\eqref{IP konst pole} and the not-reduced cylindrical integral
\begin{equation}
	X_2=\left(p_Z^A\right)^2+2 W_3(Z).
\end{equation}
The first system was considered in \cite{Landau}, \cite{Marchesiello2015}, see the text around eq.~\eqref{IP konst pole}, and the second in \cite{Marchesiello2017}. For more details on the latter system see the text around eq.~\eqref{WZ special}.

\subsection{Case $\mathrm{rank}(M)=1$ subcase 3a 4)}
This system is defined by 
\begin{equation}
	\mu(Z)=0,\quad\tau(\phi) = 0,\quad\rho(r)=0,\quad\sigma(r)\neq 0, 
\end{equation}
so the magnetic field $B$ and scalar potential $W$ read
\begin{align}\label{Wmtr0sn0}
\begin{split}
	W = W_1(r) + \frac{1}{r^2} W_2(\phi),\quad 
	B^r = 0, \quad B^\phi = \frac{1}{2}\sigma'(r), \quad B^Z = 0.
\end{split}
\end{align}

In this case eq.~\eqref{komp 3} reads 
\begin{equation}
	(- k_2\cos(\phi) + k_1 \sin(\phi)) r^3\sigma'(r)=0,
\end{equation}
which is solved either by $\sigma(r)=\sigma_0$, which implies vanishing magnetic field $B$, or both $k_1$ and $k_2$ must vanish, i.e. no additional integral to the cylindrical ones.

\subsection{Case $\mathrm{rank}(M)=1$ subcase 3b)}
This case assumes 
\begin{equation}
	\psi'(\phi) \neq 0,\quad \mu(Z) = 0,\quad \tau(\phi)=0,\quad \sigma(r)=0
\end{equation}
leading to the magnetic field $B$ and potential $W$ of the form
\begin{align}
	\begin{split}
	W(r,\phi,Z)&= W_{12}(r,\phi)+ W_3(Z),\\
	B^r &= 0, \quad B^\phi = 0, \quad B^Z = -\frac{1}{2 r^2} \left(\rho'(r) r^2+\psi''(\phi) +\psi(\phi) \right)
	\end{split}
\end{align}
and additional equations constraining $W$
\begin{align}\label{eq W b}
	r\psi'(\phi) W_r + \left(r \rho(r) - \psi(\phi) \right) W_\phi+\frac{\hbar^2(\psi'''(\phi)+\psi'(\phi))}{4r^3} = 0, \nonumber \\
	\psi'(\phi) \left(r^3 \rho''(r) - r^2 \rho'(r) + r \rho(r) -3 \psi''(\phi) - 4 \psi(\phi) \right)+ \\
	+ \psi'''(\phi) \left(r \rho(r) - \psi(\phi) \right) - 4 r^5 W_{r \phi} - 8 r^4 W_\phi = 0. \nonumber
\end{align}
This is very similar to the considerations in Subsection~\ref{sec:case 2b 1.1)}, the only important difference being the form of the potential $W,$ which admits dependence on $Z$. Let us therefore continue in the same manner and focus on the differences.


Recalling eq.~\eqref{k4k5} and differentiating eq.~\eqref{komp 1} twice with respect to $r$, we get eq.~\eqref{rho eq} for $\rho(r)$ with solution
\begin{equation}
	\rho(r) =\rho_4 r^2 + \rho_2\ln(r) +\frac{\rho_3}{r}+ \rho_1.
\end{equation}
Inserting it into eq.~\eqref{komp 1} differentiated only once we get
\begin{equation}\label{split2}
	k_6(\psi'''(\phi)+\psi'(\phi))+2 \rho_2(k_4\sin(\phi)+k_5\cos(\phi))=0,
\end{equation}
which splits the considerations into 2 subcases: $k_6\neq0$ and $k_6=0$.

\begin{enumerate}[label=\Roman*.]
	\item $k_6\neq 0$: Solving the compatibility equations in the same way as in Subsection~\ref{sec:case 2b 1.1)} subcase~\ref{k6 nenula 2b}, we get
	\begin{equation}\label{psirho 3b}
	\psi(\phi) =-\psi_2 \cos(\phi)+\psi_1 \sin(\phi)+\rho_1,\quad \rho(r)=\rho_2 r^2+ \frac{\rho_1}{r}+\rho_0.
	\end{equation}
	So we obtained the constant magnetic field
	\begin{equation}
	B^r=0,\quad B^\phi=0,\quad B^Z=-\rho_2 r.
	\end{equation}
	Translating this into the Cartesian coordinates using eq.~\eqref{transformB}, it reads
	\begin{equation}
	B^x=0,\quad B^y=0,\quad B^z=-\rho_2.
	\end{equation}
	We note that $\psi(\phi)$ from eq.~\eqref{psirho 3b} implies vanishing of the quantum correction, so the quantum version of the system is identical to the classical version.
	
	Having solved the compatibility conditions, we solve the original equations \eqref{ord1 lin} to obtain
	\begin{align}
	m(r, \phi, Z) &= \rho_2\left(k_2 r\cos(\phi)-k_1 r\sin(\phi) - \frac{k_6r^2}{2}\right).
	\end{align}
	Equation \eqref{ord0 lin} reads
	\begin{equation}\label{eq}
		(k_1\cos(\phi)+k_2 \sin(\phi))r \pd_r W_{12} -(k_2 \cos(\phi)-k_1 \sin(\phi)+k_6 r)\pd_\phi W_{12} + k_3 r \pd_Z W_3=0.
	\end{equation}
	As in the previous case, we can differentiate the equation to get $k_3 W_3''(Z)=0$, so we further split the considerations.
	\begin{enumerate}[label*=\alph*)]
		\item $k_3=0$\label{1a}: In this subcase we have no constraint on $W_3(Z)$ and the remaining equations determining $W_{12}$ are the same as in Subsection~\ref{sec:case 2b 1.1)} subcase~\ref{k6 nenula 2b}, implying $W_{12}(r, \phi) = W_0$, therefore (absorbing the constant into $W_3(Z)$)
		\begin{equation}
			W(r,\phi,Z)= W_3(Z).
		\end{equation}
		The corresponding integrals of motion are the 3 first order integrals
		\begin{equation}
		Y_1=p_x^A-\rho_2 y,\quad Y_2=p_y^A+\rho_2 x,\quad \tilde{X}_1=L_z^A+\frac{\rho_2 (x^2+y^2)}{2}
		\end{equation}
		and the non-reduced cylindrical integral
		\begin{equation}
			X_2=\left(p_Z^A\right)^2+2 W_3(Z).
		\end{equation}
		The system was considered in \cite{Marchesiello2017}. For more details see the text around eq.~\eqref{WZ special}, where the same system was considered (with $\rho_1=\rho_2$).
		\item $k_3\neq 0$: Here we get 
		\begin{equation}
			W_3(Z)=aZ+b
		\end{equation}
		so eq.~\eqref{eq} reads
		\begin{equation}
		(k_1\cos(\phi)+k_2 \sin(\phi))r \pd_r W_{12} -(k_2 \cos(\phi)-k_1 \sin(\phi)-k_6 r)\pd_\phi W_{12} + k_3 a r=0.
		\end{equation}
		The other equation which we shall solve is the first from eq.~\eqref{eq W b}.
		We assume $a\neq 0$, because if $a=0$ we are back in Subsection~\ref{sec:case 2b 1.1)} subcase~\ref{k6 nenula 2b} with constant scalar potential $W= W_0.$
		Let us analyse the system of equations from the point of view of linear algebra. The extended matrix of the system reads
		\begin{equation}
			\begin{pmatrix}
			r(\psi_2\sin(\phi)+\psi_1\cos(\phi))& r(\rho_2 r^2+\rho_0)+\psi_2\cos(\phi)-\psi_1 \sin(\phi)&0 \\		
			r(k_1\cos(\phi)+k_2\sin(\phi))&-(k_2\cos(\phi)-k_1\sin(\phi)-k_6 r) &-k_3 a r\\
			\end{pmatrix}
		\end{equation}
		The left-hand side $2\times 2$ minor		
		\begin{equation}
		\begin{split}
		-2 (k_1 \psi_2 +k_2 \psi_1)(\cos(\phi))^2+k_1\psi_2+k_2 \psi_1+2(k_1\psi_1-k_2\psi_2)\sin(\phi)\cos(\phi)-\\
		-r[k_1(\rho_2 r^2+\rho_0)-k_6 \psi_1]\cos(\phi)-r[k_2(\rho_2 r^2+\rho_0)-k_6 \psi_2]\sin(\phi)\eqqcolon\text{det}A
		\end{split}
		\end{equation}
		does not vanish for all $r,$ $\phi$ due to our assumptions $\rho_2\neq 0$ (non-zero magnetic field), $k_3\neq 0$, $k_6\neq 0$ and at least one of $k_1,$ $k_2$ non-zero. Therefore, the system has the following unique solution
		\begin{equation}\label{sol1}
			\begin{pmatrix}
			\pd_r W_{12}\\ \pd_\phi W_{12}
			\end{pmatrix}=-\frac{k_3 a}{\text{det}A}
			\begin{pmatrix}
		-[r(\rho_2 r^2+\rho_0)+\psi_2\cos(\phi)-\psi_1 \sin(\phi)]\\ r(\psi_2\sin(\phi)+\psi_1\cos(\phi))
			\end{pmatrix}.
		\end{equation}
		This solution, however, does not satisfy Clairaut compatibility condition $\pd_{r\phi} W_{12}=\pd_{\phi r} W_{12}$ for all $r,$ $\phi$: Taking the $r^6$ term in the numerator of the fraction obtained by inserting the solution \eqref{sol1} in the condition, namely
		\begin{equation}
		a k_3 \rho_2^2(k_1 \sin(\phi) - k_2 \cos(\phi)),
		\end{equation}
		we see that it does not vanish when we take our assumptions into account. Therefore, this case does not allow any solutions with $a\neq 0$, so we are left with the solution in 2b 1.1 subcase~\ref{k6 nenula 2b}
		\begin{equation}
		W= W_0\text{ with }a=0.
		\end{equation}
		For more details about this system see Subsection~\ref{sec:case 2a lin} around eq.~\eqref{paty konst pole}.
	\end{enumerate}

	\item $k_6=0$: With this assumption eq.~\eqref{split2} implies $\rho_3=0$ and together with eq.~\eqref{komp 1} we get 
	\begin{align}
	\rho(r)&=\frac{\rho_2 r^3 + \rho_1}{r}+\rho_0,\\
	\begin{split}
	\psi(\phi) &=\frac{c_1\cos(2\phi)+c_2\sin(2\phi)+c_3}{k_2 \cos(\phi)-k_1 \sin(\phi)}+\rho_1.
	\end{split}
	\end{align}
	The corresponding magnetic field $B$ reads
	\begin{equation}
	B^r=0,\quad B^\phi=0,\quad B^Z=-\rho_2 r+\frac{\xi}{r^2 (k_1 \sin(\phi)- k_2 \cos(\phi))^3},
	\end{equation}
	where $\xi=-(c_1 + c_3) k_1^2 - 2 c_2 k_1 k_2 + k_2^2 (c_1 - c_3)$.
	Translating this into the Cartesian coordinates using eq.~\eqref{transformB}, it reads
	\begin{equation}
	B^x=0,\quad B^y=0,\quad B^z=-\rho_2+\frac{\xi}{(k_1y- k_2x)^3}.
	\end{equation}

	In order to simplify the equations, we use eq.~\eqref{k2=1}, choose the coordinate system so that $\phi_0=0$ and redefine the constants $c_i$ in $\psi(\phi)$ to set $k_2=1$, i.e. from now on 
	\begin{equation}
	\psi(\phi) =\frac{c_1\cos(2\phi)+c_2\sin(2\phi)+c_3}{\cos(\phi)}+\rho_1.
	\end{equation}
	The corresponding magnetic field $B$ reads in the cylindrical coordinates
	\begin{equation}
	B^r=0,\quad B^\phi=0,\quad B^Z=-\rho_2 r+\frac{c_1-c_3}{r^2 (\cos(\phi))^3}
	\end{equation}
	and in the Cartesian coordinates
	\begin{equation}
	B^x=0,\quad B^y=0,\quad B^z=-\rho_2+\frac{c_1-c_3}{x^3}.
	\end{equation}
	
	Equation~\eqref{ord1 lin} is the same as in 2b 1.1 subcase~\ref{k6 nula 2b}, so we have 
	\begin{align}
	m={}& \rho_2 r-\frac{c_1-c_3}{2 r^2(\cos(\phi))^2}+C=\rho_2 \sqrt{x^2+y^2}-\frac{c_1-c_3}{2x^2}+C.
	\end{align}
		
	However, eq.~\eqref{ord0 lin} is different, namely
	\begin{equation}\label{eq W 3b}
	r \sin(\phi)\pd_r W_{12}+\cos(\phi) \pd_\phi W_{12} + k_3 r W_3'(Z)=0,
	\end{equation}
	and the considerations split into 2 subcases, $k_3=0$ and $k_3\neq 0$.
	\begin{enumerate}[label*=\alph*)]
		\item $k_3=0$: In this subcase $W_3(Z)$ drops out of eq.~\eqref{eq W 3b} and remains unconstrained. The remaining equations are the same as in case 2b 1.1, we thus have $W_{12}(r,\phi)= W_0$ with $c_1=c_3$, without loss of generality $W_0=0$, and 
		\begin{equation}
			W= W_3(Z),\quad B^r=0,\quad B^\phi=0,\quad B^Z=-\rho_2.
		\end{equation}
		in both classical and quantum cases. We have, therefore, the same system as in subcase~\ref{1a}, so it in fact allows $k_6\neq 0$, with the corresponding integrals
		\begin{equation}
		Y_1=p_x^A-\rho_2 y,\quad Y_2=p_y^A+\rho_2 x,\quad \tilde{X}_1=L_z^A+\frac{\rho_2 (x^2+y^2)}{2}
		\end{equation}
		and the non-reduced cylindrical integral
		\begin{equation}
		X_2=\left(p_Z^A\right)^2+2 W_3(Z).
		\end{equation}
		The system was considered in \cite{Marchesiello2017}. For more details see the text around eq.~\eqref{WZ special}, where the same system was considered (with $\rho_1=\rho_2$).
		
%
	\item $k_3\neq 0$: 
	Here we get 
	\begin{equation}
	W_3(Z)=aZ+b
	\end{equation}
	so eq.~\eqref{eq} reads
	\begin{equation}
	\sin(\phi)r \pd_r W_{12} -\cos(\phi)\pd_\phi W_{12} + k_3 a r=0.
	\end{equation}
	Another equation we need to consider is the first from eq.~\eqref{eq W b}.
	We assume $a\neq 0$, because if $a=0$ we are back in case 2b 1.1 subcase~\ref{k6 nula 2b} with constant scalar potential $W= W_0.$
	
	Preforming the same linear-algebraic analysis as in case I.b), the following must hold
	\begin{equation}
	\rho_2^2 \cos(\phi)=0.
	\end{equation}
	However, it is satisfied if and only if the magnetic field vanishes. Therefore, this case does not allow any solutions with $a\neq 0$ and we have nothing but the solution from 2b 1.1 subcase~\ref{k6 nenula 2b}
	\begin{equation}
	W= W_0,\quad a=0,\quad c_1=c_3.
	\end{equation}
	For more details about that system see Subsection~\ref{sec:case 2a lin} around eq.~\eqref{paty konst pole}.

	\end{enumerate}
\end{enumerate}

To sum up, we have found 3 first order superintegrable systems with non-trivial magnetic field. 
\begin{enumerate}
	\item The first system has constant magnetic field $B^z=\mu_0$ and vanishing scalar potential $W=0$. This is a well-known system with 4 first order integrals of motion \eqref{IP konst pole}, analysed already in \cite{Landau}. For more details see Subsection~\ref{sec:case 2a lin}, where we cite the non-polynomial fifth integral of motion from \cite{Marchesiello2015}.
	\item The second system is in fact a class of systems with constant magnetic field $B^z=\mu_0$ and the unconstrained potential $W=W(Z)$. It has 3 first order integrals from \eqref{IP konst WZ} and the cylindrical integral $X_2=(p_Z^A)^2+2W(Z)$. The system is reducible to a 2D system without magnetic field and is maximally superintegrable if and only if the 2D system is superintegrable \cite{Marchesiello2019}. Two second order maximally superintegrable versions of $W(Z)$, see eq.~\eqref{WZ special}, are considered in \ref{sec: 3a 1a} subcase~\ref{konst pole W(Z)}
	\item The last system is the only one with more interesting magnetic field
		\begin{align}
	W =-4\left(\frac{\tau_1^2}{2 f_1^2 y^4}	+\frac{W_0 }{f_1 y^2}\right),\quad
	B^x={\frac{4 \tau_1}{f_1 y^3}},	\quad {B^y} =0,\quad B^z=0.
	\end{align}
	It was found in \cite{Marchesiello2019} and considered here in Subsection~\ref{sec:case 2b 1.1)} subcase~\ref{system 2b 1.2} It is only minimally second order superintegrable and the trajectories are unbounded, so we have no information about potential higher order integrals.
\end{enumerate}
As shown in Subsection~\ref{sec:general}, the additional integrals must be $p_x^A+m(\vec{x})$ or $p_y^A+m(\vec{x})$ (or their linear combination, but we can rotate the coordinate system). In all three cases we can choose the gauge so that they are $p_x$ or $p_y$. The same is true for the first order reduced cylindrical integral $\tilde{X}_1$, which in suitable gauge reads $\tilde{X}_1=p_\phi$. This implies that the corresponding Hamilton-Jacobi and Schr\"odinger equations separate in the cylindrical as well as Cartesian coordinates and the equations are solved in the corresponding subsections. The spectrum is continuous in all three cases.

\section{Second order integrals}\label{sec:second ord}
In this section we will study second order superintegrable systems of the cylindrical type. We consider the classical cases only because of the computational complexity introduced by the non-trivial quantum correction \eqref{corr cyl}. 

The third order equations \eqref{ord3} can be solved regardless of the magnetic field with the solution in eq.~\eqref{sol third 1}--\eqref{sol third 6}.
So we have to solve the remaining equations \eqref{ord2}--\eqref{ord0} where we substitute the solution of the third order equations and the magnetic field $B$ and scalar potential~$W$ from one of the integrable cases. Here we encounter computational difficulties, because we have to solve 10 differential equations with 20 arbitrary constants $\alpha_{ij}$ and 4 unknown functions $s^r,$ $s^\phi,$ $s^Z$ and $m$ in addition to the magnetic field $B$ and scalar potential $W$ (simplified by assuming integrability).

Due to the computational complexity we were not able to proceed systematically in the general case. We calculated the Clairaut compatibility conditions for functions $s^r$, $s^\phi$, $s^Z$, $\pd_{ab}s^i=\pd_{ba}s^i$, and tried to solve them by taking an \emph{ansatz} on the unknown functions and constants so that the remaining equation \eqref{ord2}--\eqref{ord0} can be solved by Maple\texttrademark. Despite a lot of effort, we were able to find just 3 systems this way, one of them with an additional first order integral, namely the system in Subsection~\ref{sec:2b 1.2} subcase~\ref{system 2b 1.2} One of the second order systems had the additional integral of the (Laplace)-Runge-Lenz type $R_z=L_x p_y-L_y p_x+\ldots$, so we include it in Subsection~\ref{sec:Runge-Lenz}.

Therefore, we have to make some simplifying assumptions. Based on physical considerations, we assume the additional integrals to take the form $L_x^2+L_y^2+L_z^2+\ldots$ and $L_x p_y-L_y p_x+\ldots$, the latter being the $z$-component of (Laplace)-Runge-Lenz vector (in vanishing magnetic field), which corresponds to choosing $\alpha_{11}=1$, $\alpha_{22}=1$, $\alpha_{33}=1$ and $\alpha_{24}=1$, $\alpha_{15}=-1$ with all other vanishing in the solution \eqref{sol third 1}--\eqref{sol third 6} to the third order equations \eqref{ord3}, respectively. We were able to obtain all classical systems of this type, leaving the quantum versions to a later work because the corrections are complicated.

With these assumptions, which significantly decrease the number of auxiliary functions appearing in the second order equations \eqref{ord2} after substituting the form of magnetic field $B$ and scalar potential $W$ of the considered subcase, we were able to solve equations. The next step was to consider the Clairaut compatibility conditions $\pd_{ab}m(r,\phi,Z)=\pd_{ba}m(r,\phi,Z)$ arising from the first order equations \eqref{ord1}, which restrict values of the constants in solution to the second order equations \eqref{ord2} and the form of scalar potential $W$. In this step, it is necessary to consider only those cases, where vanishing of the constants does not imply vanishing of the magnetic field. The last step was to solve the first and zeroth order equations, namely eq.~\eqref{ord1} and eq.~\eqref{ord0}, which can further restrict the potential.

In what follows, we will not present the detailed calculation for brevity, we only list the found systems and solve the Hamilton's and Hamilton-Jacobi equations for new systems. We will start with the system which does not have the integral of the above stated type (Subsection~\ref{sec:biquadr}), following with the systems with $L^2+\ldots$ in Subsection~\ref{sec:L^2} and finally the systems with $L_x p_y- L_y p_x+\ldots$ in Subsection~\ref{sec:Runge-Lenz}.

\subsection{System with bi-quadratic potential and non-zero $B^\phi$}\label{sec:biquadr}

	This system is defined by
	\begin{align}
{B^r} =0,\quad
	{B^\phi} =-2 \gamma r,\quad B^Z =0,\quad 	W(r)=-{\frac{\gamma^2 r^4}{2}}+\frac{1}{2}\delta r^2.
	\end{align}
	
	
	The corresponding magnetic field in the Cartesian coordinates obtained using eq.~\eqref{transformB} is 
	\begin{align}
	{B^x} (\vec{x}) &=2\gamma y,\quad{B^y}
	(\vec{x}) =-2\gamma x,\quad{B^z} (\vec{x}) =0.
	\end{align}
	The first order integrals of motion read 
	\begin{equation}
		\tilde{X}_1=p_\phi^A,\quad \tilde{X}_2=p_Z^A-\gamma r^2,
	\end{equation}
	followed by the second order ones
	\begin{align}
	X_3&={\frac{\left(2\gamma(p_Z^A-\gamma r^2) +\delta \right) r^4\left(\cos(\phi)\right)^2+
			\left(p_r^A r\cos(\phi)-{p_\phi^A}\sin(\phi)\right)^2}{r^2}},\\
	X_4&={\frac{\left(2 \gamma(p_Z^A-\gamma r^2) +\delta \right) r^4\left(\sin(\phi)\right)^2+
	\left(p_r^A r\sin(\phi)+{p_\phi^A}\cos(\phi)\right)^2}{r^2}},\\
	X_5&=\frac{\left[\left(2 \gamma (p_Z^A-\gamma r^2) +\delta \right) r^4 +\left(p_r^A\right)^2 r^2 -\left(p_\phi^A\right)^2\right]\sin(2\phi) +2 p_\phi^A p_r^A r\cos(2\phi)} {2r^2},\\
	H&=\frac{\left(2\gamma(p_Z^A-\gamma r^2) +\delta \right) r^{4}+
	\left((p_Z^A-\gamma r^2)^2+\left(p_r^A\right)^2 \right) r^2 +\left(p_\phi^A\right)^2}{2r^2}.
	\end{align}
	We choose the gauge as follows
	\begin{equation}\label{gauge1}
	{A_r} \left({r,\phi,Z} \right) =0,\quad {A_\phi} \left({r,\phi,Z} \right) =0,\quad A_Z \left({r,\phi,Z} \right) ={{\gamma r^2}}
	\end{equation}
	and the gauge fixed integrals of motion become
	\begin{align}
	\tilde{X}_1&=p_\phi,\quad \tilde{X}_2=p_Z,\label{pphi,pZ}\\
	X_3&={\frac{\left(2 \gamma p_Z +\delta \right) r^4\left(\cos(\phi)\right)^2+
			\left(p_r r\cos(\phi)-p_\phi\sin(\phi)\right)^2}{r^2}},\\
	X_4&={\frac{\left(2 \gamma p_Z +\delta \right) r^4\left(\sin(\phi)\right)^2+\left(p_r r\sin(\phi)+ p_\phi\cos(\phi)\right)^2}{r^2}},\\
	X_5&=\frac{\left[\left(2 \gamma p_Z +\delta \right) r^4 +p_r^2 r^2 -p_\phi^2\right]\sin(2\phi) +2 p_\phi p_r r\cos(2\phi)} {2r^2},\\
	H&={\frac{\left(2 \gamma p_Z +\delta \right) r^4+ \left(p_Z^2+p_r^2 \right) r^2 +p_\phi^2}{2 r^2}}.\label{ham biquadr}
	\end{align}
	Let us translate them into the Cartesian coordinates.
	\begin{align}
		\tilde{X}_1&=x p_y-y p_x,\quad \tilde{X}_2=p_z,\\
		X_3&=p_x^2+(2\gamma p_z+\delta)x^2,\\
		X_4&=p_y^2+(2\gamma p_z+\delta)y^2,\\
		X_5&=p_x p_y+(2\gamma p_z+\delta)xy,\\
		H&=\frac{1}{2}\left[p_x^2+p_y^2+p_z^2+(2\gamma p_z+\delta)(x^2+y^2)\right].
	\end{align}
	From this form of the integrals, we see that this system is Case I d) in \cite{Marchesiello2019} with $a_1=\gamma$, $b_1=\frac{\delta}{2}$ and all other constants vanishing. It was shown that this system is second order minimally superintegrable. We confirm the result, because only 4 of the integrals are functionally independent even for $\gamma =0$ and/or $\delta=0$ due to the relations
	\begin{equation}\label{rel}
	X_3+X_4+\tilde{X}_2^2=2 H,\quad	X_3X_4=X_5^2+2\left(\gamma \tilde{X}_2+\frac\delta {2}\right)\tilde{X}_1^2.
	\end{equation}
	Let us note that $X_3-X_4$ is $X_5$ rotated by $\frac{\pi}{2}$ and multiplied by 2,
	\begin{equation}
	X_3-X_4=\frac{\left[\left(2 \gamma p_Z+\delta \right) r^4 +p_r^2 r^2-p_\phi^2\right]\cos(2\phi)
		-2 p_\phi p_r r \sin (2\phi)}{r^2},
	\end{equation}
	The Poisson algebra is closed in the sense that the Poisson brackets of the integrals are functions (polynomials) in previously known integrals, and thus generate no new independent integrals, because the only non-vanishing Poisson brackets read
	\begin{gather}
	\begin{split}
		\{\tilde{X}_1,X_3\}_\text{P.B.}=2 X_5,\quad \{\tilde{X}_1,X_4\}_\text{P.B.}=-2 X_5,\quad \{\tilde{X}_1,X_5\}_\text{P.B.}=X_4- X_3,\\ \{X_3,X_5\}_\text{P.B.}=4 \gamma\tilde{X}_1 \tilde{X}_2+2\delta \tilde{X}_1,\quad\{X_4,X_5\}_\text{P.B.}=-(4 \gamma \tilde{X}_1 \tilde{X}_2+2\delta \tilde{X}_1).
	\end{split}
	\end{gather}
	
	
	Hamilton's equations of the system in the gauge \eqref{gauge1} are
	\begin{gather}\label{ham eq}
	\dot{r}=p_r,\quad \dot{\phi}=\frac{p_\phi}{r^2},\quad 
	\dot{Z}={\gamma r^2 + p_Z},\\
	\dot{p}_r=\frac{p_\phi^2-\left(2 \gamma p_Z +\delta \right) r^4}{r^3},\quad
	\dot{p}_\phi=0,\quad \dot{p}_Z=0.
	\end{gather}
	
We see that $p_\phi,$ $p_Z$ are constants, as we have already known from eq.~\eqref{pphi,pZ}. We continue with the equation for $r$,
	\begin{equation}\label{r dot dot}
	\ddot{r}={\frac{p_\phi^2-\left(2 \gamma p_Z +\delta \right) r^4}{r^3}},
	\end{equation}
	because it does not depend on the other variables.
	The equation does not explicitly depend on $t$, so we reduce it using the integrating factor $\dot{r}$. Solving the resulting separable ODE, we get
	\begin{equation}\label{r(t) sys1}
	r(t)=\bm{\pm}{\frac{\sqrt{- \beta {C_1} \left(-4\beta(C_1^2 C_2^2 + p_\phi^2)+4 \sqrt{-\beta}C_1^2 C_2{{\rm e}^{-2 \sqrt{-\beta}t}}+C_1^2 {{\rm e}^{-4 \sqrt{-\beta}t}} \right)}} {2\beta{C_1} {{\rm e}^{-\sqrt{-\beta}t}}}}.
	\end{equation}
	The sign must be chosen so that $r(t)\geq 0$. The solution above is valid for $\beta=2 \gamma p_Z +\delta\neq 0$. Although we consider $\gamma\neq 0$ only, the case $\beta=0$ in eq.~\eqref{r dot dot} cannot be excluded and must be solved separately, see the paragraph around eq.~\eqref{r 2} below.
	
	Let us see if all bounded trajectories are closed, which is typical for maximally superintegrable systems \cite{Miller2013}.
	
	We assume that physical quantities must be real functions on the phase space. Because the constants $\gamma,$ $\delta$ appear in the potential and $p_Z$ is a coordinate on phase space, we conclude that $\beta$ must be real, as well as $p_\phi$. The analysis splits into 2 cases, $\beta<0$ and $\beta>0$.
	\begin{enumerate}
		\item If $\beta<0$, the exponential functions in eq.~\eqref{r(t) sys1} are real and positive. Factoring ${\rm e}^{-2\sqrt{-\beta}t}$ from the square root, the expression for $r(t)$ simplifies
		\begin{equation}
		\frac{\sqrt{-\beta {C_1} \left(-4\beta(C_1^2 C_2^2 + p_\phi^2){\rm e}^{2\sqrt{-\beta}t}+4 \sqrt{-\beta}C_1^2 C_2+C_1^2 {{\rm e}^{-2\sqrt{-\beta}t}} \right)}} {-2\beta{C_1}}.
		\end{equation}
		It diverges for $t\to-\infty$, because the constants in the last term must be non-zero. Therefore, the trajectories are not bounded in this case.
		
		\item If $\beta> 0$, the exponentials are complex and can be rewritten in terms of sines and cosines. Using the corresponding \emph{ansatz} in eq.~\eqref{r(t) sys1}
		\begin{equation}
		r(t)=\sqrt{a\sin\left(2\sqrt {\beta}t \right)+b\cos\left(2 \sqrt {\beta}t \right)+c},
		\end{equation} 
		we obtain the following solution
		\begin{equation}\label{r(t)}
		r(t)=\sqrt{a\sin\left(2\sqrt {\beta}t \right)+b\cos\left(2\sqrt {\beta}t \right)+\frac{\sqrt{\beta(\beta a^2+\beta b^2+p_\phi^2)}}{\beta}},
		\end{equation}
		where the constants $a$ and $b$ are determined from the initial conditions and read
		\begin{equation}\label{ab}
			a = \frac{p^r_0 r_0}{\sqrt{\beta}},\quad b = \frac{\beta r_0^4 - r_0^2 (p^r_0)^2 - p_{\phi}^2}{2\beta r_0^2}.
		\end{equation}
		(This form of the solution works over complex numbers as well, which includes the case $\beta<0$.)
		Taking the positive roots, the solution satisfies $r(t)\geq0$ for all $t$ due to
		\begin{equation}
			a\sin\left(2\sqrt {\beta}t \right)+b\cos\left(2\sqrt {\beta}t \right)=\sqrt{a^2+b^2}\cos\left(2\sqrt{\beta}t+t_0\right),
		\end{equation}
		where $t_0$ satisfies $\cos(t_0)=\frac{a}{\sqrt{a^2+b^2}}$, and the estimate
		\begin{equation}
		\frac{\sqrt{\beta(\beta a^2+\beta b^2+p_\phi^2)}}{\beta}\geq\sqrt{a^2+b^2}
		\end{equation}
		valid for all $p_\phi\in \R$, $\beta>0$ and equal if and only if $p^\phi=0$, from which follows
		\begin{equation}
		r(t)^2=\sqrt{a^2+b^2}\left(1+\cos\left(2\sqrt{\beta}t+\phi_0\right)\right)\geq 0.
		\end{equation}
		Therefore, $r(t)$ is bounded and finite in this case. The motion can be periodic only if $p^\phi\neq 0$, for in the opposite case the particle falls on the singularity $r=0$, where eq.~\eqref{r(t) sys1} is not defined. 
	\end{enumerate}
	
	The case $\beta=0$ in eq.~\eqref{r dot dot}, i.e.
	\begin{equation}\label{r 2}
	\ddot{r}=\frac{p_\phi^2}{r^3},
	\end{equation}
	is very similar to the general case. Using the integrating factor $\dot{r}$ again, we get
	\begin{equation}
	r(t)=\pm{\frac{\sqrt {C_1 \left(C_1^2 \left(C_2+t\right)^2 +p_\phi^{2} \right)}}{{C_1}}},
	\end{equation}
	which is clearly unbounded in $t$. (The integration constant $C_1$ cannot be negative.)
	
	So we continue with solving the remaining equations \eqref{ham eq} for the only interesting case $\beta>0$, using the form of $r(t)$ from eq.~\eqref{r(t)}. They are only quadratures and the solutions are
	\begin{align}
	\phi(t)&=\arctan \left({\frac{\left(-b\beta+\sqrt {\beta \left(a^2 \beta+b^2 \beta+p_\phi^2 \right)} \right) 
	\tan \left(\sqrt {\beta}t \right) +a\beta}{p_\phi \sqrt {\beta}}} \right) +C_1,\\
	Z(t)&=\left(\frac{\gamma \sqrt {{a}^2 \beta+{b}^2 \beta+p_\phi^2}}{\sqrt \beta}+p_Z\right) t
	+\frac{\gamma b\sin \left(2 \sqrt {\beta}t \right) -\gamma a\cos \left(2 \sqrt {\beta}t\right) }{2\sqrt\beta}+C_2,
	\end{align}
	where the constants $a$, $b$ in terms of initial conditions are in eq.~\eqref{ab}.
	
	The relationship between the integration constants $C_1,$ $C_2$ and the initial conditions is
	\begin{equation}
	C_1=\phi_0-\arctan\left(\frac{r_0 p_r^0}{p_{\phi}^0\beta}\right),\quad C_2=Z_0+\frac{\gamma r_0 p_r^0\cos(2\sqrt{\beta})}{2\beta}.
	\end{equation}
	We see that $Z(t)$ is bounded if and only if the first term vanishes. Writing it in full, it reads
	\begin{equation}\label{cond}
	\frac{\gamma r_0^2}{2}+p_Z^0+\frac{\gamma[(p_r^0)^2 r_0^2 + (p_\phi^0)^2]}{(4 \gamma p_Z^0 + 2 \delta) r_0^2}=0,
	\end{equation}
	so it is a complicated function of constants of the system $\gamma$, $\delta$ and the initial values.
	
	This system has bounded trajectories only if the initial values and constants $\gamma$ and $\delta$ satisfy eq.~\eqref{cond}. If we exclude this singular possibility, we must conclude that we obtained no information regarding maximal superintegrability of the system from this analysis.

Let us analyse stationary Hamilton-Jacobi equations. Starting in the Cartesian coordinates and gauge \eqref{gauge1}, it reads 
	\begin{equation}
		\frac{1}{2}\left[\left(\pderA{U}{x}\right)^2+\left(\pderA{U}{y}\right)^2+\left(\pderA{U}{z}\right)^2+\left(2\gamma \pderA{U}{z}+\delta\right)(x^2+y^2)\right]=E.
\end{equation}
The $z$-coordinate is cyclic and the other two can be separated into equivalent equations, e.g. the one for $x$ reads
\begin{equation}
	\frac{1}{2}\left(u'(x)\right)^2+\left(\gamma p_z+\frac{\delta}{2}\right)x^2=\lambda_1.
\end{equation}
Its solution is
\begin{equation}
	u(x)=\frac{x}{2} \sqrt{-(2 \gamma p_z + \delta) x^2 + 2 \lambda_1} + \frac{\lambda_1}{\sqrt{2 \gamma p_z + \delta}} \arctan\left(\frac{\sqrt{2 \gamma p_z + \delta} x}{\sqrt{-(2 \gamma p_z + \delta) x^2 + 2 \lambda_1}}\right).
\end{equation}
The result for $y$ is analogous and the separation constants satisfy $E=\lambda_1+\lambda_2+p_z^2$. The value of $E$ therefore determines the appropriate range for $x$ and $y$ through $\lambda_1$ and $\lambda_2$.

The Hamilton-Jacobi equation in the cylindrical coordinates in the gauge \eqref{gauge1} clearly separates because both cylindrical integrals reduce to $\tilde{X}_1=p_\phi$ and $\tilde{X}_2=p_Z$. The only remaining equation to solve is as follows
\begin{equation}\label{rce r}
	(u'(r))^2=2E-p_Z^2-{\frac{\left(2 \gamma p_Z +\delta \right) r^4+p_\phi^2}{r^2}}.
\end{equation}
Assuming positivity of the right-hand side, the solution is
\begin{equation}
\begin{split}
	u(r)&=-\frac{p_\phi}{2} \arctan\left(\frac{(2 E -p_Z^2) r^2-2 p_\phi^2}{2 p_\phi \sqrt{(-2 \gamma p_Z - \delta) r^4 + (2 E-p_Z^2) r^2 - p_\phi^2}}\right)-\\
&- \frac{2E-p_Z^2}{4 \sqrt{2 \gamma p_Z + \delta}} \arctan \left( {\frac {-2(2 \gamma p_Z+ \delta){r}^{2}+2 E-p_Z^{2}}{2\sqrt{2 \gamma p_Z+ \delta}\sqrt {-(2 \gamma p_Z+\delta)r^4 + (p_Z^{2}-2E)r^2-p_\phi^{2}}}} \right)
+\\
& + \frac{ \sqrt{-(2 \gamma p_Z + \delta) r^4 +(2E- p_Z^2) r^2 - p_\phi^2}}{2}.\raisetag{1.3\baselineskip}
\end{split}
\end{equation}
 If $2\gamma p_Z+\delta<0$, the solution works for sufficiently small $r$ and the second term will be argtanh ($-\i\arctan(\i x)={\rm argtanh} (x)$).

\subsection{Systems with additional integral $L^2+\ldots$}\label{sec:L^2}
In this case, there are 2 superintegrable systems with $L^2+\ldots$ as the additional integral plus 2 first order systems already analysed. Both first order systems have constant magnetic field, the first one has vanishing scalar potential (system in Subsection~\ref{sec:case 2a lin}) and the second one $W=W(Z)$ (system in Subsection~\ref{sec: 3a 1a} subcase~\ref{konst pole W(Z)}).

\begin{enumerate}[label=\Roman*.]
	\item \label{rr system} The first system has constant magnetic field
	\begin{equation}
		B^r = 0,\quad B^\phi = 0,\quad B^Z = -\kappa_1 r,\quad W=-\frac{\kappa_1^2 r^2}{8}-\frac{W_0}{2 r^2},
	\end{equation}
	which can be written in the Cartesian coordinates as 
	\begin{equation}
			B^x = 0,\quad B^y = 0 ,\quad B^z = -\kappa_1.
	\end{equation}
	The commuting cylindrical integrals of motion reduce to first order
	\begin{equation}
		\tilde{X}_1= p_\phi^A+\frac{\kappa_1 r^2}{2}=L_z^A+\frac{\kappa_1(x^2+y^2)}{2},\quad \tilde{X}_2=p_Z^A=p_z^A.
	\end{equation}
	The additional integral reads (in the Cartesian coordinates)
	\begin{equation}
		X_3=(L^A)^2+(x^2+y^2+z^2)\left[\kappa_1 L_z^A+\frac{\kappa_1^2 }{4}(x^2+y^2)\right] -\frac{W_0 z^2}{x^2+y^2}.
	\end{equation}
	We can choose the gauge so that $\tilde{X}_1=p_\phi$, namely
	\begin{equation}\label{gauge konst pole}
		A_r=0,\quad A_\phi=-\frac{\kappa_1 r^2}{2},\quad A_Z=0,\quad\text{i.e. }
		\vec{A}(\vec{x})=\left(-\frac{y}{2},\frac{x}{2},0\right).
	\end{equation}
	This system was found in \cite{Marchesiello2018Sph}, where it was shown that it admits another second order integral of motion, namely (in the Cartesian coordinates)
	\begin{equation}
		X_4=\frac{1}{2}\{X_3,\tilde{X}_2\}_{\mathrm{P.B.}}=p_x^A L_y^A-p_y^A L_x^A +\kappa_1 z L_z^A+\frac{\kappa^2}{4}(x^2+y^2)z-\frac{W_0 z}{x^2+y^2}.
	\end{equation}
	This (Laplace)-Runge-Lenz type integral (of the type considered in Subsection~\ref{sec:Runge-Lenz}) is, however, dependent on the previously known integrals due to relation
	\begin{equation}
		X_4^2=2H(X_3-\tilde{X}_1^2+W_0)-\tilde{X}_2^2 X_3-\kappa_1(\tilde{X}_1^3-\tilde{X}_1X_3)+\kappa_1 W_0\tilde{X}_1.
	\end{equation}
	Integrals $\tilde{X}_1$ and $X_3$ Poisson commute.
	Thus, the system is only minimally second order superintegrable.

	We continue with analysing the trajectories. The Hamiltonian equations of motion in the chosen gauge read
	\begin{equation}
		\begin{split}
		\dot{p}_r = \frac{p_\phi^2-W_0}{r^3},\quad \dot{p}_\phi = 0,\quad \dot{p}_Z = 0,\\
		\dot{r} = p_r,\quad \dot{\phi} = \frac{p_\phi}{r^2} - \frac{\kappa_1}{2},\quad \dot{Z} = p_Z.
		\end{split}
	\end{equation}
	
	We have chosen the gauge so that $p_\phi$ and $p_Z$ are constant, so 2 of the equations are no surprise. The free motion in the $z$-direction,
	\begin{equation}
		Z(t)=p_Z^0 t+ Z_0,
	\end{equation}
	 was also clear from the form of the magnetic and electric field. The motion can be periodic only if $p_Z^0=0$.
	 
	We combine the equations for $p_r$ and $r$ to get
	\begin{equation}
		\ddot{r}=\frac{(p_\phi^0)^2-W_0}{r^3},
	\end{equation}
	where $p_\phi^0$ is the initial value of $p_\phi(t)$. We use the integrating factor $\dot{r}\equiv p_r$ (assuming that it is non-zero) and obtain
	\begin{equation}\label{eq nwm}
		\dot{r} = \pm\sqrt{\frac{W_0-(p_\phi^0)^2}{r^2}}.
	\end{equation}
	If $W_0\geq (p_\phi^0)^2$, the solutions to eq.~\eqref{eq nwm} are
	\begin{equation}
		r(t)=\sqrt{\pm 2 t \sqrt{W_0-(p_\phi^0)^2} + r_0^2}.
	\end{equation}
	The plus or minus in the square root corresponds to the sign in eq.~\eqref{eq nwm}, the integration constant $r_0$ must ensure that the square root is well defined for the chosen range of $t$, and only the plus sign of the whole expression is consistent with the definition of $r$.
	
	There are, therefore, 3 possible trajectories: if $W_0> (p_\phi^0)^2$, the negative sign inside the square root leads to the fall on singularity $r=0$ and the positive sign implies the escape to infinity. The last possibility, which corresponds to $W_0= (p_\phi^0)^2$, is the singular solution $r=r_0$.
	
	This means that only the singular case is periodic because solving the remaining equation in eq.~\eqref{eq nwm} yields
	\begin{equation}
		\phi(t)=\left(\frac{p_\phi^0}{r_0^2}-\frac{\kappa_1}{2}\right)t+\phi_0.
	\end{equation}
	The trajectory is therefore a circle of radius $r_0$ in the $z=z_0$ plane.

	We conclude that the analysis gave us little information about maximal superintegrability, because for almost all initial values the trajectory is either unbounded or ends in the singularity.
	
	Let us solve the stationary Hamilton-Jacobi equation in the natural gauge \eqref{gauge konst pole}, where the cylindrical integrals are $\tilde{X}_1=p_\phi$ and $\tilde{X}_2=p_Z$ ($\phi$ and $Z$ are cyclic), so we only need to solve the reduced equation
		\begin{equation}\label{rce I}
		\frac{1}{2}\left[\left(u'(r)\right)^2+\frac{1}{r^2}\left(p_\phi-\frac{\kappa_1 r^2}{2}\right)^2+\left(p_Z\right)^2\right]-\frac{\kappa_1^2 r^2}{8}-\frac{W_0}{2 r^2}=E.
		\end{equation}
	The solution to the stationary Hamilton-Jacobi equation is therefore
\begin{equation}
\begin{split}
	U(r,\phi,Z)&=p_\phi \phi +p_Z Z+\sqrt{a r^2 -b^2} - b \ln\left(\frac{b^2 + b \sqrt{a r^2 - b^2}}{r}\right),
\end{split}
\end{equation}
where $a=(\kappa_1 p_\phi - p_Z^2 + 2 E)$ and $b=\sqrt{W_0-p_\phi^2}$. Here we need $W_0>p_\phi^2$, the other condition $a>0$ then follows from eq.~\eqref{rce I}.
	
	\item\label{L^2 II} The second system is defined by
	\begin{equation}
			\begin{split}
			B^r &= \frac{\kappa_1 r^2}{Z^3}-\kappa_2 r^2 Z,\quad B^\phi= 0,\quad B^Z = r\left(\frac{\kappa_1}{Z^2}+\kappa_2 (r^2+R^2) + \kappa_3  \right),
			\\ W &=\frac{\kappa_3^2}{8}Z^2-\frac{\kappa_2^2}{8}r^2 R^4-\frac{\kappa_1^2 r^2}{8 Z^4}-\frac{\kappa_1 \kappa_2}{4 Z^2}R^4-\frac{\kappa_1 \kappa_3}{4 Z^2}r^2-\frac{\kappa_2 \kappa_3}{4}r^2 R^2+\\			&+\frac{W_1}{r^2} +\frac{W_2}{Z^2}-W_3 R^2,
			\end{split}\raisetag{3\baselineskip}
	\end{equation}
	where $R=\sqrt{r^2+Z^2}=\sqrt{x^2+y^2+z^2}$. The magnetic field in the Cartesian coordinates reads
	\begin{equation}
		B^x=\frac{\kappa_1 x}{z^3}-\kappa_2 x z,\quad B^y=\frac{\kappa_1 y}{z^3}-\kappa_2 y z,\quad B^z=\frac{\kappa_1}{z^2}+\kappa_2(r^2+R^2)+\kappa_3.
	\end{equation}
	One of the cylindrical integrals reduces to the first order integral
	\begin{equation}
		\tilde{X}_1=p_\phi^A -\frac{1}{2}\left(\frac{\kappa_1 r^2}{Z^2}+\kappa_2 r^2 R^2 + \kappa_3 r^2\right).
	\end{equation}
	The second order integrals of motion read
	\begin{align}
	\begin{split}
		X_2&=\left(p_Z^A\right)^2+ \left(\frac{\kappa_1}{Z^2} +\kappa_2 Z^2\right) p^A_\phi -\frac{\kappa_2^2}{2} r^2 Z^4 - \left(\frac{\kappa_2^2 r^4}{2} +\frac{\kappa_2 \kappa_3 r^2}{2}-\frac{\kappa_3^2}{4}+2 W_3\right) Z^2 -\\
	&- \kappa_1 \kappa_2 r^2 - \frac{\kappa_1 \kappa_2 r^4}{2Z^2} - \frac{\kappa_1 \kappa_3 r^2}{2Z^2} +\frac{\kappa_1^2}{4Z^2} +\frac{2 W_2}{Z^2} - \frac{\kappa_1^2 r^2}{2 Z^4},\raisetag{2.25\baselineskip}
	\end{split}\\
	\begin{split}
		X_3&=(Z p_r^A - r p_Z^A)^2 + \frac{Z^2+r^2}{r^2} \left(p^A_\phi\right)^2 -\left(\kappa_2 R^2+\kappa_3\right)R^2 p_\phi^A-\\
		&-\frac{\kappa_1^2}{4Z^4}r^4+\frac{\kappa_2^2}{4} r^2 R^6+ \frac{\kappa_3^2}{4} r^2 R^2 + \frac{\kappa_2\kappa_3}{2}r^2 R^4+\frac{2W_1}{r^2}Z^2+\frac{2W_2}{Z^2}r^2,\\
	\end{split}\\
	\begin{split}
	H&=\frac{1}{2}\left(\left(p_r^A\right)^2+\frac{\left(p_\phi^A\right)^2}{r^2}+\left(p_Z^A\right)^2\right)
		+\frac{\kappa_3^2}{8}Z^2-\frac{\kappa_2^2}{8}r^2 R^4-\frac{\kappa_1^2 r^2}{8 Z^4}-\\
		&-\frac{\kappa_1 \kappa_2}{4 Z^2}R^4-\frac{\kappa_1 \kappa_3}{4 Z^2}r^2-\frac{\kappa_2 \kappa_3}{4}r^2 R^2+\frac{W_1}{r^2} +\frac{W_2}{Z^2}-W_3 R^2.
	\end{split}
	\end{align}
	We omit the Cartesian form of the integrals for brevity as it brings no new insight, we only note that $X_3=(L^A)^2+\ldots$ This system is new to the literature, as far as we know. 
	
	There is only one non-vanishing Poisson bracket among the integrals, namely $X_4:=\{X_2,X_3\}_{\rm P.B.}$, but its square can be related to previous integrals as follows
	\begin{equation}
	\begin{split}
		X_4^2&=32 H X_3 X_2-32 H \tilde{X}_1^2 X_2-16 X_3 X_2^2-	16 \tilde{X}_1(-\kappa_2 \tilde{X}_1^4 + 2 \kappa_2 X_3 \tilde{X}_1^2+\\
		&+ \kappa_3 \tilde{X}_1^2 X_2 - 4\kappa_1 H^2 + 2\kappa_1 H X_2 - \kappa_2 X_3^2 - \kappa_3 X_3 X_2) - 128 W_2 H^2+\\
		&+ 64 \kappa_1 \kappa_3 H \tilde{X}_1^2 + 128 W_2 H X_2 + 4(2\kappa_1 \kappa_2 - \kappa_3^2+ 8W_3) X_3^2+\\
		&+ 8(\kappa_3^2+ 2\kappa_1 \kappa_2 - 8W_3)X_3 \tilde{X}_1^2 + 4(10\kappa_1 \kappa_2  + 8W_3 - \kappa_3^2)\tilde{X}_1^4 - 16\kappa_1 \kappa_3 \tilde{X}_1^2X_2-\\
		&- 32(W_1 + W_2)X_2^2 + 8\tilde{X}_1(16\kappa_3 W_2 H + (\kappa_3^2 \kappa_1  - 2\kappa_1^2 \kappa_2  - 8\kappa_1 W_3)X_3 +\\&
		+ (16\kappa_2 W_2 - \kappa_3^2 \kappa_1  - 4\kappa_1^2 \kappa_2  - 8\kappa_1 W_3)\tilde{X}_1^2 - 8\kappa_3 W_2 X_2) +\\&
		+ 4(2\kappa_1^3 \kappa_2  - \kappa_3^2 \kappa_1^2 + 8 \kappa_1^2 W_3 + 32\kappa_1 \kappa_2 W_1 - 16\kappa_1 \kappa_2 W_2 - 64W_2 W_3)\tilde{X}_1^2+\\
		&+ 32W_1(\kappa_3^2 \kappa_1  - 2\kappa_1^2 \kappa_2  - 8\kappa_1 W_3 + 8\kappa_2 W_2)\tilde{X}_1^2 +\\&+ 64W_1 W_2 (\kappa_3^2 - 2\kappa_1 \kappa_2  - 8W_3). \end{split}\raisetag{\baselineskip}
\end{equation}
	
	We choose the gauge so that $\tilde{X}_1=p_\phi$, namely 
	\begin{equation}\label{gauge L^2 IV}
	A_r=0,\quad A_\phi= \frac{1}{2}\left(\kappa_2 r^2 R^2 + \kappa_3 r^2 + \frac{\kappa_1 r^2}{Z^2}\right),\quad A_Z=0.
	\end{equation}
	The Hamilton's equations are
	\begin{gather}\label{Ham eq L^2 IV} 
		\dot{r} = p_r,\quad	\dot{p}_r =\left(\frac{\kappa_1 \kappa_2}{2} - \frac{\kappa_3^2}{4} - \kappa_2 p_\phi+ 8 W_3\right) r + \frac{p_\phi^2 + 2 W_1}{r^3},\\
		\dot{\phi} = \frac{\kappa_2 r^2}{2} + \frac{\kappa_2 Z^2}{2} + \frac{\kappa_3}{2} + \frac{\kappa_1}{2 Z^2} + \frac{p_\phi}{r^2},\quad \dot{p}_\phi = 0,\\
		\dot{Z} = p_Z,\quad \dot{p}_Z = \left(\frac{\kappa_1 \kappa_2}{2}- \frac{\kappa_3^2}{4} - \kappa_2 p_\phi + 2 W_3\right) Z + \frac{p_\phi \kappa_1 + 2 W_2}{Z^3}.
	\end{gather}
	The fourth equation shows that $p_\phi$ is constant. Taking that into account and combining the equations in the first and third row, respectively, the equations for $\ddot{r}$ and $\ddot{Z}$ are the same modulo some constants, so we will solve only one of them in detail.
	
	The equation for $r$, 
	\begin{equation}
		\ddot{r}=\left(\frac{\kappa_1 \kappa_2}{2} - \frac{\kappa_3^2}{4} - \kappa_2 p_\phi+ 8 W_3\right) r + \frac{p_\phi^2 + 2 W_1}{r^3},
	\end{equation}
	where $p_\phi$ is constant (initial value), is reduced by the integrating factor $\dot{r}$ and the subsequent quadrature gives us the solution of the form
	\begin{equation}
		r(t)=\sqrt{c_1\sin(\omega(t-t_0)) +\sqrt{c_1^2 + \frac{4 C}{\omega^2}}},
	\end{equation}
	where $C=p_\phi^2 + 2 W_1$ and $\omega=\sqrt{2 \kappa_1 \kappa_2 - 4 \kappa_2 p_\phi - \kappa_3^2 + 8 W_3}$.
	The overall sign must be positive due to the definition of $r$. We can assume that $c_1\geq 0$, for otherwise we redefine $t_0$. If $C>0$, the result is a periodic function, otherwise the particle ends in the singularity $r=0$.
	
	The analogous result in the $z$-direction is 
	\begin{equation}
	Z(t)=\pm\sqrt{c_2\sin(\omega(t-t_1)) +\sqrt{c_2^2 + \frac{4 D}{\omega^2}}},
	\end{equation}
	where $D=p_\phi \kappa_1+2 W_2$. This is periodic if and only if $D>0$ because we need a positive number in the square root and we have another singularity at $Z=0$.
	
	The last thing we need to solve is the equation for $\phi$ in \eqref{Ham eq L^2 IV} and the solution reads
	\begin{equation}
	\begin{split}\label{phi 2}
		\phi \left( t \right) &= c_5 + k_1 t + \frac{\kappa_2}{2 \omega^2} \left(\sin(\omega t + c_2) c_1 + c_3 \sin(\omega t + c_4)\right)+\\
		&+ \frac{p_\phi}{\sqrt{C}} \arctan\left(k_2 \tan\left(\frac{\omega t + c_2}{2}\right)\right) + \frac{\kappa_1}{2 \sqrt{D}} \arctan\left(k_3 \tan\left(\frac{\omega t+c_4}{2}\right)\right).
		\end{split}
\end{equation}
where the constant $k_i$ are some complicated functions of the constants in the magnetic field and the scalar
potential, and the initial values, which we omit for readability, except for the important one
\begin{equation}
	k_1=\frac{\kappa_3}{2}+\frac{\kappa_2}{2}\left(\sqrt{c_1^2 + \frac{4 (p_\phi^2 + 2 W_1)}{\omega^2}}+\sqrt{c_1^2 + \frac{4 (p_\phi \kappa_1+2 W_2)}{\omega^2}}\right).
\end{equation}
In order to have periodic trajectories, we need $k_1$ to be commensurable with $\omega=\sqrt{2 \kappa_1 \kappa_2 - 4 \kappa_2 p_\phi - \kappa_3^2 + 8 W_3}$. We see that both of these constants depend on the initial values, so this cannot be a maximally superintegrable system if $\kappa_2\neq0$, it can be at most particularly superintegrable \cite{Turbiner}.

We now briefly consider the case with $\kappa_2=0$ because than neither $\omega$ nor $k_1$ depends on the initial values. Numerical integration of the Hamilton's equations shows that the trajectories close if we add another constraint, namely $W_3=\frac{\kappa_3^2}{8}\left(1-\frac{n^2}{m^2}\right)$, $m,n\in \N$, see a sample trajectory in Figure~\ref{fig:plot1}, i.e. the magnetic field and scalar potential read
\begin{equation}
	\begin{split}\label{L2_IV}
	B^r &= \frac{\kappa_1 r^2}{Z^3},\quad B^\phi=0,\quad B^Z = \frac{\kappa_1 r}{Z^2}+\kappa_3 r,\\
	W&=-\frac{r^2}{8}\left(\frac{\kappa_1}{Z^2} +\kappa_3 \right)^2 + \frac{\kappa_3^2n^2}{8m^2}(r^2+Z^2) + \frac{W_2}{Z^2} +\frac{W_3}{r^2},
	\end{split}
\end{equation}
which in the Cartesian coordinates becomes
\begin{equation}
	B^x=\frac{\kappa_1 x}{z^3},\quad B^y=\frac{\kappa_1 y}{z^3},\quad B^z=\frac{\kappa_1 z}{z^3}+\kappa_3.
\end{equation}
There is no additional first or second order integral of motion and the search for higher order ones is out of the scope of this thesis.

\begin{figure}
	\centering
	\includegraphics[width=0.4\linewidth]{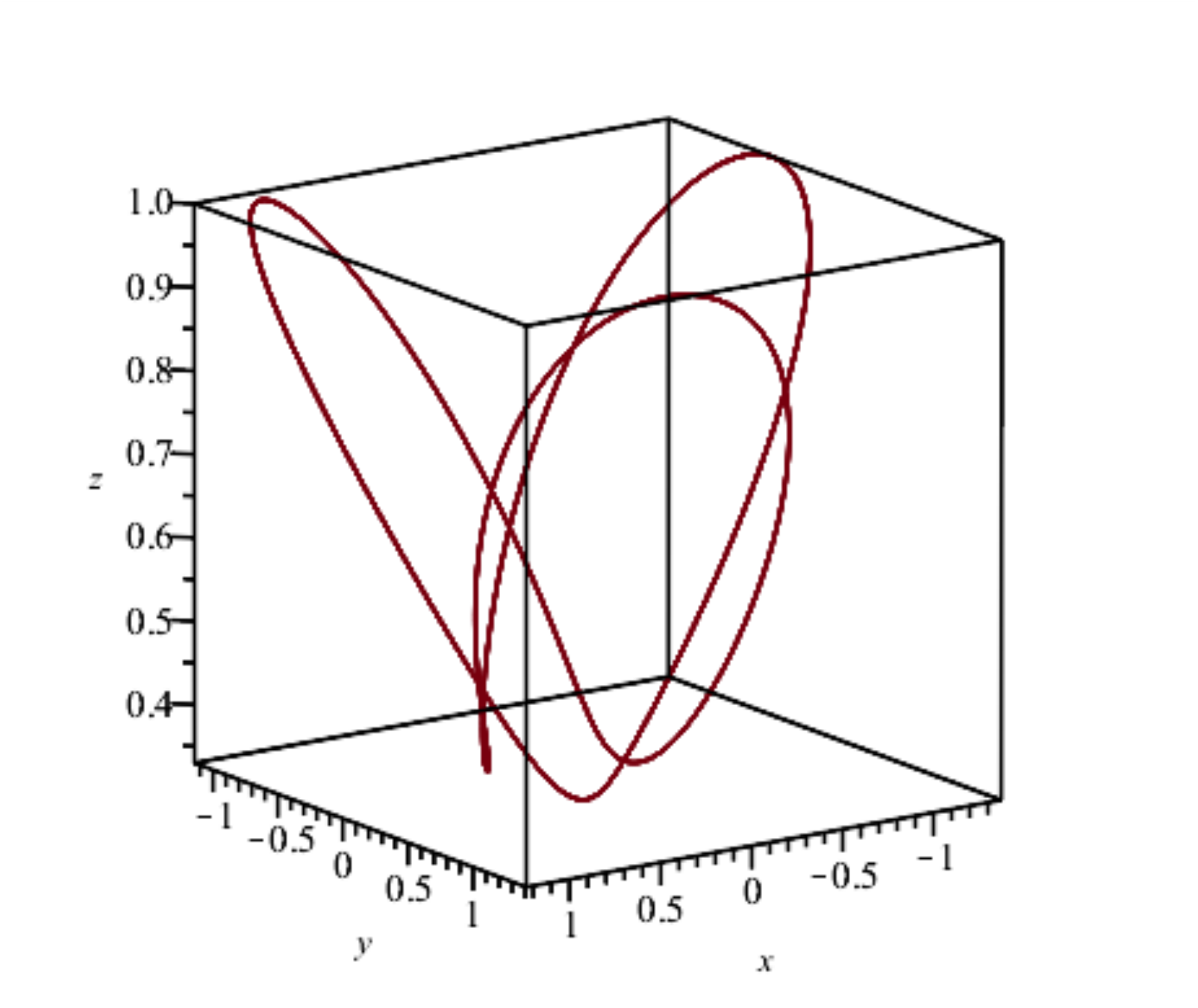}
	\caption{Sample plot of the system in eq.~\eqref{L2_IV} with the initial conditions $[x(0)=1,y(0)=-1,z(0)=1,p_x(0)=1,p_y(0)=0,p_z(0)=0]$ and the value of the constants $[W_1=3,W_2=1,\kappa_1=1,\kappa_3=7,n=3,m=2]$.}
	\label{fig:plot1}
\end{figure}

Let us solve stationary Hamilton-Jacobi equation. In the gauge \eqref{gauge L^2 IV} the coordinate $\phi$ is cyclic, so we can separate it out. The reduced equation separates as follows
	\begin{equation}
	(u'(r))^2 -\frac{\omega^2}{4} r^2 +\frac{C}{r^2}= 2 c=- (v'(Z))^2 +\frac{\omega^2}{4} Z^2+2E-p_\phi \kappa_3-\frac{D}{Z^2}.
	\end{equation}
	where $c$ is the separation constant, $b^2=p_\phi^2 +2 W_2$.
The solutions are
\begin{align}
\begin{split}
	u(r) ={}& \frac{\sqrt{C}}{2} \arctan\left(\frac{2 (c r^2 - C)}{\sqrt{C} \sqrt{\omega^2 r^4 + 8 c r^2 - 4 C}}\right)- \frac{\sqrt{\omega^2 r^4 + 8 c r^2 - 4 C}}{4}+ \\
&	- \frac{c}{\omega} \ln(\omega^2 r^2 + \omega\sqrt{\omega^2 r^4 + 8 c r^2 - 4 C}+ 4 c)+\alpha_1\raisetag{1.25\baselineskip}
\end{split}\\
	\begin{split}
	v(Z) ={}&\frac{\sqrt{D}}{2} \arctan\left(\frac{-(F Z^2 + 2D)}{\sqrt{D} \sqrt{\omega^2 Z^4 - 4F Z^2 - 4 D}}\right)-\frac{\sqrt{\omega^2 Z^4 - 4F Z^2 - 4 D}}{4}+\\
	&+\frac{F}{2\omega} \ln(\omega^2 Z^2 +\omega\sqrt{\omega^2 Z^4 - 4F Z^2 - 4 D}- 2F) +\alpha_2,\raisetag{1.25\baselineskip}
	\end{split}
\end{align}
where we have defined $F=\kappa_3 p_\phi - 2 E + 2 c$.

\end{enumerate}

To conclude, in both cases with additional integral of type $(L^A)+\ldots$ we are able to solve the stationary Hamilton-Jacobi equation by separation of variables because we can choose the gauge so that $\tilde{X}_1=p_\phi$, i.e. we have at least one first order integral. Neither of the systems is second order maximally superintegrable and the Poisson algebra does not generate additional independent integrals. The trajectories are unbounded for system \ref{rr system} for almost all initial values, so we have no hint of additional integrals. The trajectories of system \ref{L^2 II} are in general bounded but not closed, which depends on the initial values, so the system can be at most particularly superintegrable \cite{Turbiner}. Special values of the constants which lead to the magnetic field in eq.~\eqref{L2_IV} imply closed trajectories, so this subcase is a promising candidate for higher order superintegrability.

\subsection{Systems with additional integral $L_x p_y-L_y p_x+\ldots$}\label{sec:Runge-Lenz}
We have already considered two systems with (Laplace)-Runge-Lenz type integral, namely system in Subsection~\ref{sec: 3a 1a} subcase~\ref{konst pole W(Z)} with potential $W(Z)=a Z^2+b Z$ and system~\ref{rr system} in Subsection~\ref{sec:L^2}. Both these systems were found in \cite{BertrandSnobl} (in the first case with $b=0$, which can be assured by a $z$-direction).

Here we analyse 3 additional systems. As far as we know, none of them appeared in the literature in the most general form, although some special cases were already found and they are mentioned in the respective cases.

\begin{enumerate}[label=\Roman*.]

	\item\label{RL II} The first system is defined by 
	\begin{equation}\label{RLII}
	B^r = 0,\quad B^\phi = 0,\quad B^Z = \kappa_1 r - \frac{2 \kappa_2}{r},\quad W =-\frac{\kappa_1^2 r^2}{8}+\kappa_1\kappa_2 \ln(r)- \frac{W_0}{2 r^2}.
	\end{equation}
	The Cartesian version of the magnetic field is
	\begin{equation}
		B^x=0,\quad B^y=0,\quad B^z=\kappa_1-\frac{2\kappa_2}{x^2+y^2}.
	\end{equation}
	The commuting cylindrical integrals reduce to first order integrals
	\begin{equation}
	\tilde{X}_1=p_\phi^A-\frac{\kappa_1 r^2}{2} + 2\kappa_2 \ln(r)=L_z^A-\frac{\kappa_1 (x^2+y^2)}{2} + \kappa_2 \ln\left(x^2+y^2\right)\!,\ \tilde{X}_2=p_Z^A=p_z^A.
	\end{equation}
	The second order integrals are the additional integral and the Hamiltonian
	\begin{align}
	&\begin{aligned}
		X_3={}&r p_r^A p_Z^A -Z \left(\left(p_r^A\right)^2 + \frac{\left(p_Z^A\right)^2}{r^2}\right) - 2 \kappa_2 \phi p_Z^A + \kappa_1 Z p_\phi^A -\\
	&-Z \left(\frac{\kappa_1^2 r^2}{4} - \kappa_1\kappa_2 - \frac{W_0}{r^2}\right),
	\end{aligned}\\
	&\begin{aligned}
		H={}&\frac{1}{2}\left(\left(p_r^A\right)^2+\frac{\left(p_\phi^A\right)^2}{r^2}+\left(p_Z^A\right)^2\right) -\frac{\kappa_1^2 r^2}{8}+\kappa_1\kappa_2 \ln(r)- \frac{W_0}{2 r^2},
	\end{aligned}
	\end{align}
	which in the Cartesian coordinates become
	\begin{align}
	&\begin{split}
	X_3={}&L_x^A p_y^A-L_y^A p_x^A - 2 \kappa_2 \arcsin\left(\frac{y}{\sqrt{x^2+y^2}}\right) p_z^A + \kappa_1 z L_z^A-\\& -z \left(\frac{\kappa_1^2 (x^2+y^2)}{4} - \kappa_1\kappa_2 - \frac{W_0}{x^2+y^2}\right),\\	H={}&\frac{1}{2}\left(\left(p_x^A\right)^2+\left(p_y\right)^2+\left(p_z^A\right)^2\right) -\\
	&-\frac{\kappa_1^2 (x^2+y^2)}{8}+\kappa_1\kappa_2 \ln\left(\sqrt{x^2+y^2}\right)- \frac{W_0}{2 (x^2+y^2)}.
	\end{split}
	\end{align}
	From the form of the magnetic field $B$ and scalar potential $W$ in eq.~\eqref{RLII} we can deduce that the motion separates into free motion in the $z$-direction and motion in the $xy$-plane. Integral $X_3$ is therefore a bit surprising because it connects these independent directions.
	
	As far as we know, the system with $\kappa_2\neq0$ is new to the literature. If $\kappa_2=0$, it is again the system \ref{rr system} from Subsection~\ref{sec:L^2}.
	
	Considering the Poisson algebra, the non-trivial Poisson brackets are
	\begin{equation}
		\{\tilde{X}_1,\tilde{X}_2\}_{\rm P.B.}=\tilde{X}_2^2-2H+\kappa_1 \tilde{X}_1+\kappa_1 W_0,\quad \{\tilde{X}_1,X_3\}_{\rm P.B.}=-2\kappa_2 \tilde{X}_2,
	\end{equation}
	so we do not obtain additional independent integrals.
	
	Let us analyse the Hamiltonian equations of motion. We choose the gauge so that both $\tilde{X}_1$ and $\tilde{X}_2$ are simply momenta
	\begin{equation}\label{gauge RL II}
		A_r=0,\quad A_\phi=\frac{\kappa_1 r^2}{2} - 2\kappa_2 \ln(r),\quad A_Z=0.
	\end{equation}
	The equations of motion then read
	\begin{gather}
		\dot{p}_r =-\frac{(p_\phi-1)\kappa_1^2 r}{2}+\frac{ ((p_\phi-1)\kappa_2+p_\phi) \kappa_1}{r} + \frac{(2\kappa_2 \ln r-p_\phi)^2 - 4 \kappa_2^2 \ln r - W_0}{r^3},\nonumber\\
		\dot{p}_\phi = 0,\quad \dot{p}_Z = 0,\quad \dot{r} = p_r,\quad \dot{\phi} = \frac{2 p_\phi + \kappa_1 r^2 - 4 \kappa_2 \ln r}{2 r^2},\quad \dot{Z} = p_Z.
	\end{gather}
	We have free motion in the $z$-direction, that is
	\begin{equation}
	Z(t)=p_Z^0 t+Z_0.
	\end{equation}
	Because $p_\phi=p_\phi^0$, the equation for $r$ 
	\begin{equation}
		\ddot{r}=-\frac{(p_\phi-1)\kappa_1^2 r}{2}+\frac{ ((p_\phi-1)\kappa_2+p_\phi) \kappa_1}{r} + \frac{(2\kappa_2 \ln r-p_\phi)^2 - 4 \kappa_2^2 \ln r - W_0}{r^3}
	\end{equation}
	is not coupled with $\phi$. We reduce it to the first order ODE by using the integrating factor $\dot{r}$ and obtain the following quadrature
	\begin{equation}
		t=\int^{r(t)}_{r_0}\frac{2 a \d a}{\sqrt{4 c_1 a^2-16 \ln(a)^2 \kappa_2^2 + (\alpha a^2 + 16 \kappa_2 p_\phi) \ln(a) - 2 \kappa_1^2 (p_\phi - 1) a^4 +\beta}},
	\end{equation}
	where $\alpha=8 \kappa_1 ((\kappa_2 + 1) p_\phi - \kappa_2)$ and $\beta=- 4 p_\phi^2 + 8 \kappa_2 p_\phi + 4 W_0$.
	We do not know how to solve the quadrature even in any special cases.
	
		We have no strong hint on maximal superintegrability of the system from the numerical plots of the trajectories in the $xy$-plane for $p_Z=0$ (otherwise the motion is unbounded). Although the trajectory in Figure~\ref{fig:plotRLII1} seems to be closed, if we add momentum $p_x(0)=2$ (note that $p_\phi \equiv L_z$ remains the same), we obtain Figure~\ref{fig:plotRLII2}, where we can see that the trajectory is not closed and with increasing time fills the annulus. Because all our plots showed similar behaviour, we conclude that the system does not have bounded and closed trajectories for most initial values and is therefore not higher order maximally superintegrable.
	
	\begin{figure}
		\centering
		\begin{subfigure}{.5\textwidth}
			\centering
			\includegraphics[width=1\linewidth]{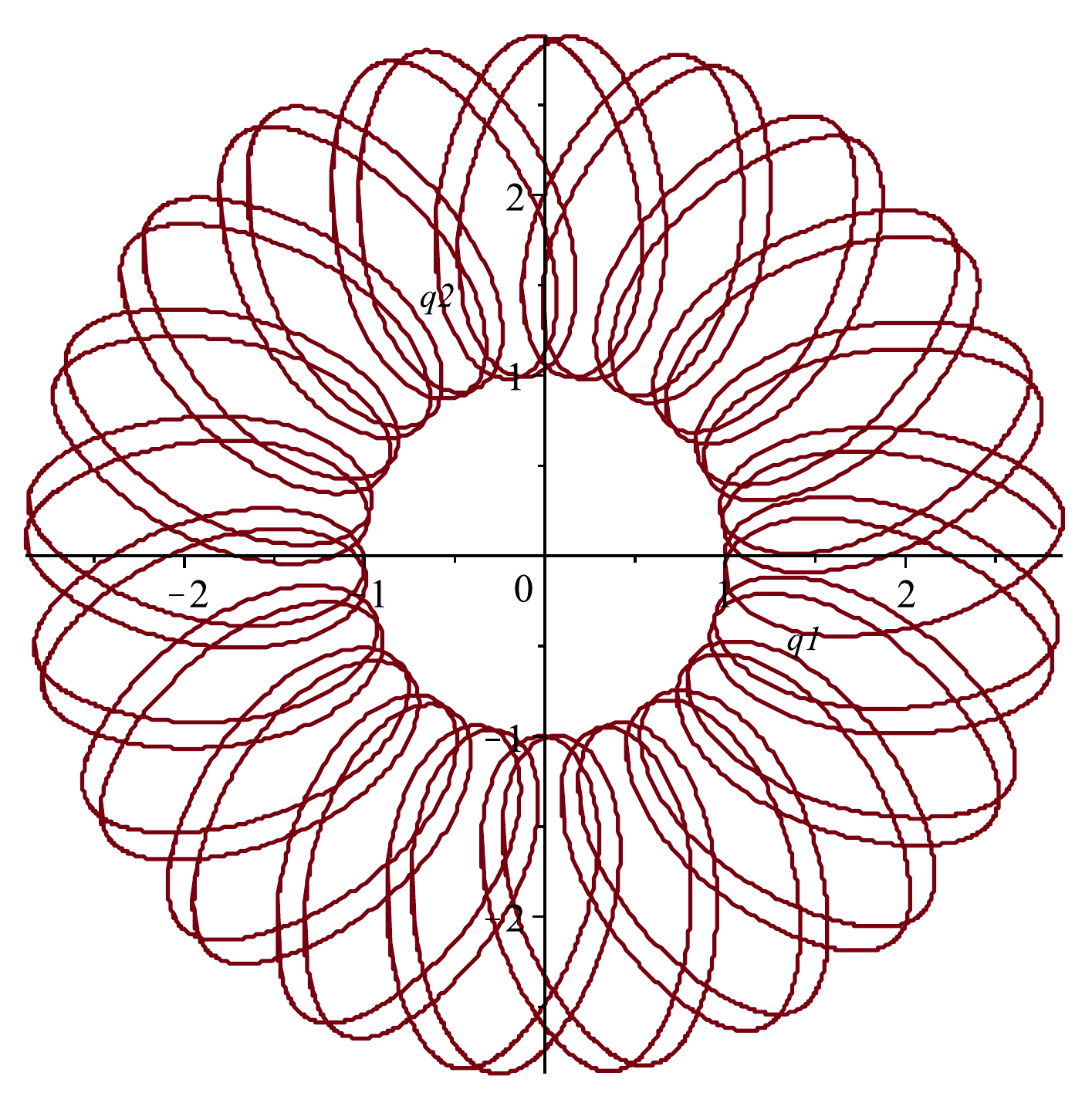}
			\caption{With $p_x(0)=0$}
			\label{fig:plotRLII1}
		\end{subfigure}%
		\begin{subfigure}{.5\textwidth}
			\centering
			\includegraphics[width=1\linewidth]{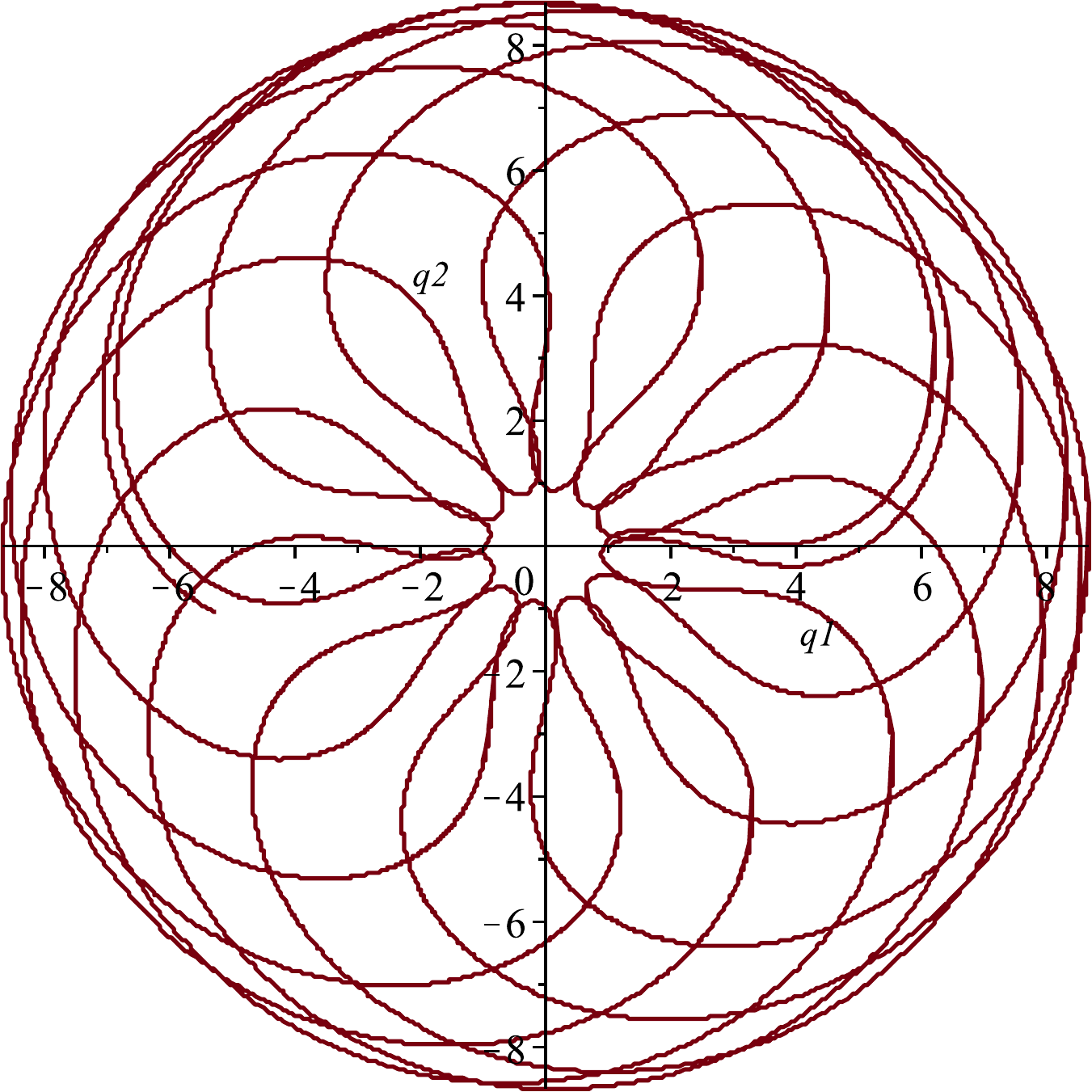}
			\caption{With $p_x(0)=2$}
			\label{fig:plotRLII2}
		\end{subfigure}
		\caption{The $xy$-plane plots of the system eq.~\eqref{RLII} with the initial values $[x(0)=0,y(0) = 0, p_x(0)=0, p_y(0) = 1]$ and the constants $[ \kappa_1 = 1, \kappa_2 = 6, W_0 = -10]$, with $p_x(0)$ according to the subcaption.}
		\label{fig:plotRLII}
	\end{figure}
	
	Considering the stationary Hamilton-Jacobi equation, the choice of gauge \eqref{gauge RL II} leads to separation of cyclic coordinates $\phi$ and $Z$, thus the remaining equation reads
	\begin{equation}
		(u'(r))^2=2 E- p_Z^2 -\kappa_1 p_\phi - \frac{(2 \kappa_2 \ln(r) -p_\phi)^2-W_0}{r^2}.
	\end{equation}
	However, the resulting quadrature
	\begin{equation}
		u(r) = \int_{r_0}^r {\frac{\sqrt{(2 E-p_Z^2-\kappa_1 p_\phi) a^2 -(2 \kappa_2 \ln(r) -p_\phi)^2+W_0}}{a} \d a} 
	\end{equation}
	cannot be solved in terms of known functions as far as we know unless $\kappa_1=0$ and $2E=p_Z^2$. In that case we need $W_0<0$ to balance the remaining bracket ($u(r)$ must be a real function) and the solution is
	\begin{equation}
	\begin{split}
	u(r)={}&\frac{(\ln(r) \sigma_0 + p_\phi) \sqrt{-2 W_0 - (\sigma_0\ln(r) + p_\phi)^2}}{2 \sigma_0} -\\
	&-\frac{W_0}{\sigma_0}\arctan\left(\frac{\sigma_0 \ln(r) + p_\phi}{\sqrt{-2 W_0 - (\sigma_0\ln(r) + p_\phi)^2}}\right).
	\end{split}
	\end{equation}
	Hamilton's characteristic function $U=u(r)+p_\phi \phi+p_z Z$ allows us to calculate 
	\begin{equation}
	r(\phi)=\exp\left(\frac{-p_\phi \pm \sqrt{-\sigma_0^2 (\phi - \phi_0)^2 - 2 W_0}}{\sigma_0}\right).
	\end{equation}
	We must have $W_0\leq0$ and $W_0=0$ is the turning point.

	\item\label{RL III} The second system has the following magnetic field $B$ and scalar potential $W$
	\begin{equation}\label{RLIII}
		B^r = 0,\ B^\phi = 0,\ B^Z = \kappa_1 r,\ W = \frac{(4 W_1 + \kappa_1^2) Z^2}{2} + W_3 Z + \frac{W_1 r^2}{2} - \frac{W_2}{2 r^2}.
	\end{equation}	
	Its magnetic field is constant, as we see from the Cartesian form
	\begin{equation}
		B^x=0,\quad B^y=0,\quad B^z=\kappa_1.
	\end{equation}
	Only $X_1$ reduces to the first order integral
	\begin{equation}
		\tilde{X}_1=p_\phi^A-\frac{\kappa_1 r^2}{2}=L_z^A-\frac{\kappa_1 (x^2+y^2)}{2}.
	\end{equation}
	The second order integrals are
	\begin{align}
		X_2={}&(p_Z^A)^2+(\kappa_1^2 + 4 W_1) Z^2 + 2 W_3 Z,\\
		X_3={}&r p_r^A p_Z^A - Z \left(\frac{\left(p_Z^A\right)^2}{r^2} + \left(p_r^A\right)^2\right) + \left(\kappa_1 Z+\frac{W_3}{\kappa_1} \right) p_\phi^A + \left(W_1 r^2+\frac{W_2}{r^2}\right)Z,\\
		\begin{split}
		H={}&\frac{1}{2}\left(\left(p_r^A\right)^2+\frac{\left(p_\phi^A\right)^2}{r^2}+\left(p_Z^A\right)^2\right)+ \frac{(4 W_1 + \kappa_1^2) Z^2}{2} 
		+ W_3 Z + \frac{W_1 r^2}{2} - \frac{W_2}{2 r^2},
		\end{split}
	\end{align}
	and in the Cartesian coordinates become
		\begin{align}
	X_2={}&(p_z^A)^2+(\kappa_1^2 + 4 W_1) z^2 + 2 W_3 z,\\
	X_3={}&L_x^A p_y^A -L_y^A p_x^A+ \left(\kappa_1 z+\frac{W_3}{\kappa_1} \right) L_z^A + \left(W_1(x^2+y^2) +\frac{W_2}{x^2+y^2}\right)z,\\
	\begin{split}
	H={}&\frac{1}{2}\left(\left(p_x^A\right)^2+\left(p_y^A\right)^2+\left(p_z^A\right)^2\right)+ \frac{(4 W_1 + \kappa_1^2) z^2}{2} +\\
	&+ W_3 z + \frac{W_1 (x^2+y^2)}{2} - \frac{W_2}{2 (x^2+y^2)},
	\end{split}
	\end{align}
	As far as we know, this system is new to the literature if all the constants are non-zero. If $W_1=-\frac{\kappa_1^2}{4}$ and $W_3=0$, we are back to the system \ref{rr system} from Subsection~\ref{sec:L^2} with $p_z^A$, which appeared in \cite{Marchesiello2018Sph} and in \cite{Bertrand}. If $W_1=0$ and $W_2=0$, the system admits $p_x^A+\kappa_1 y$ and $p_y^A-\kappa_1 y$ and was therefore analysed in Subsection~\ref{sec: 3a 1a} subcase~\ref{konst pole W(Z)} and originally found in \cite{Marchesiello2017}. Setting the constants $W_i$ to zero in any other combination does not yield more independent second order integrals.
	
	This time there is only one non-vanishing Poisson bracket in the Poisson algebra
	\begin{equation}
	\begin{split}
		X_4\coloneqq \{X_2,X_3\}_{\rm P.B.}&=2\left(\frac{(p^A_\phi)^2}{r^2} + (p^A_r)^2\right) p^A_Z- \kappa_1 p^A_\phi p^A_Z - \\
		&-\left(W_1 r^2+\frac{ W_2}{r^2}\right)p^A_Z + 2 (\kappa_1^2 Z+ 4 W_1 Z + W_3) r p^A_r.
	\end{split}
	\end{equation}
	This third order integral is not independent due to the following relation
	\begin{equation}
	\begin{split}
		\tfrac{1}{4} X_4^2&=4 X_2 H^2-\left(4 X_2^2+4 \kappa_1 X_1 X_2-4 W_3 X_3+\frac{4 W_3^2 X_1}{\kappa_1}\right)H+2\kappa_9 X_1 X_2^2 +\\
		&+X_2^2+ \left(W_1 \kappa_1^2-2 W_3 X_3-4 W_3 X_1^2+\frac{2 W_3^2}{\kappa_1} X_1+ W_2(\kappa_1^2+4 W_1)\right) X_2-\\
		&-(\kappa_1+4 W_1) X_3^2 +\frac{8 W_1 W_3}{\kappa_1}X_1 X_3-\frac{W_1 W_3^2}{\kappa_1^2}X_1^2+W_1 W_3^2,
	\end{split}\raisetag{1.4 \baselineskip}
	\end{equation}
	so we do not get anything new.
	
	We choose the natural gauge \eqref{gauge konst pole}, which makes $\tilde{X}_1=p_\phi$. The Hamiltonian equations of motion then read
	\begin{gather}\label{rce}
		\dot{p}_r = -\left(\frac{\kappa_1^2}{4} + W_1\right) r + \frac{p_\phi^2 - W_2}{r^3},\quad \dot{p}_\phi = 0,\quad \dot{p}_Z = -\left(\kappa_1^2 + 4 W_1\right) Z - W_3,\\
		\dot{r} = p_r,\quad \dot{\phi} = \frac{p_\phi }{r^2}+ \frac{\kappa_1}{2},\quad \dot{Z} = p_Z
	\end{gather}
	We see that $p_\phi$ is a constant, which corresponds to our choice of gauge.
	The equations for the $z$-direction are independent from the others and can be easily solved
	\begin{equation}
		Z(t)=\frac{p_Z^0}{\omega} \sin\left(\omega t\right) +\left(Z_0+\frac{W_3}{\omega^2}\right) \cos\left(\omega t\right) - \frac{W_3}{\omega^2},
	\end{equation}
	where $\omega=\sqrt{\kappa_1^2 + 4 W_1}$. We assume $\kappa_1^2 + 4 W_1>0$, for otherwise we would write the solution in terms of exponentials and the motion would be unbounded.
	
	We continue with $r$. We reduce the combination of the corresponding equations in eq.~\eqref{rce} to a first order ODE by the integrating factor $\dot{r}$ (assuming it is non-zero), with the result
	\begin{equation}
		\dot{r}^2 = - \frac{\omega^2 r^2}{4} - \frac{p_\phi^2 - W_2}{r^2} +c_1.
	\end{equation}
	Assuming we can take the square root, we solve the quadrature and invert the implicit solution to get
	\begin{equation}
		r(t)=\frac{\sqrt{2 c_1 + 2\sqrt{c_1^2-\omega^2 b^2}[\sin(\omega (t-t_0))]}}{\omega},
	\end{equation}
	where $b=\sqrt{p_\phi^2-W_2}$. The constant $c_1$ must be positive because the amplitude of the second term is smaller and, without loss of generality, we have chosen the plus sign due to freedom to redefine $t_0$.
	
	The last remaining equation from \eqref{rce} gives us 
		\begin{equation}
	\begin{split}\label{phi}
	\phi(t)={}&\frac{p_\phi}{b} \arctan\left(\frac{c_1 \tan\left(\frac{\omega}{2}(t - t_0)\right) - \sqrt{c_1^2-\omega^2 b^2}}{\omega b}\right)-	\frac{\kappa_1}{2}(t - t_0)+c_2,
	\end{split}
	\end{equation}
	 We see that the arctan in the first term is periodic with half the frequency of $r$ and $Z$. 
This implies that the trajectories can be periodic only if $\kappa_1$ is commensurable with $\omega$. It is, however, not sufficient because numerical integration of the Hamilton's equation suggests that they are not closed if $\frac{p_\phi}{b}$ is irrational, see Figure~\ref{fig:RLIII} where $\frac{p_\phi}{b}=\frac{3}{\sqrt{6}}$. For a trajectory where the fraction if rational see Figure~\ref{fig:RLIII2}. This unexpected result is probably because $\phi$ is defined mod $2\pi$. Assuming we can trust the numerical results, this system is not maximally superintegrable because not all bounded trajectories are closed, it can be only particularly maximally superintegrable \cite{Turbiner}.
	 
	\begin{figure}
	\centering
	\includegraphics[width=0.75\linewidth]{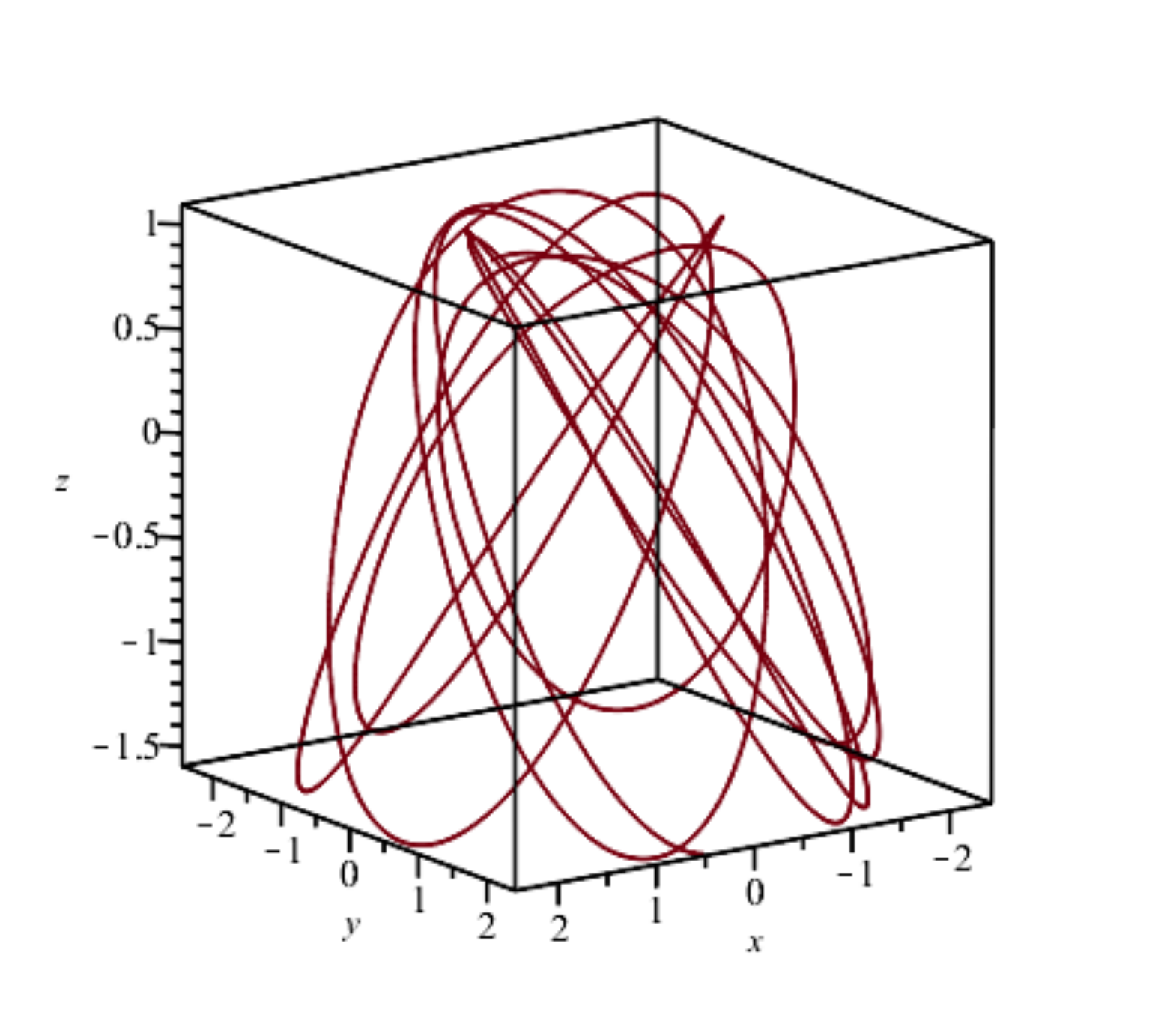}
	\caption{A trajectory for system \eqref{RLIII} with irrational fraction $\frac{p_\phi}{b}=\frac{3}{\sqrt{6}}$. The constants read $W_1 = \frac{3}{4}, W_2 = \frac{135}{16}, W_3 = 1, \kappa_1 = 1$ and the initial values are $x(0)=1,y(0)=0,z(0)=1,p_x(0)=0,p_y(0)=3,p_z(0)=1$.}
	\label{fig:RLIII}
\end{figure}
	\begin{figure}
	\centering
	\includegraphics[width=0.75\linewidth]{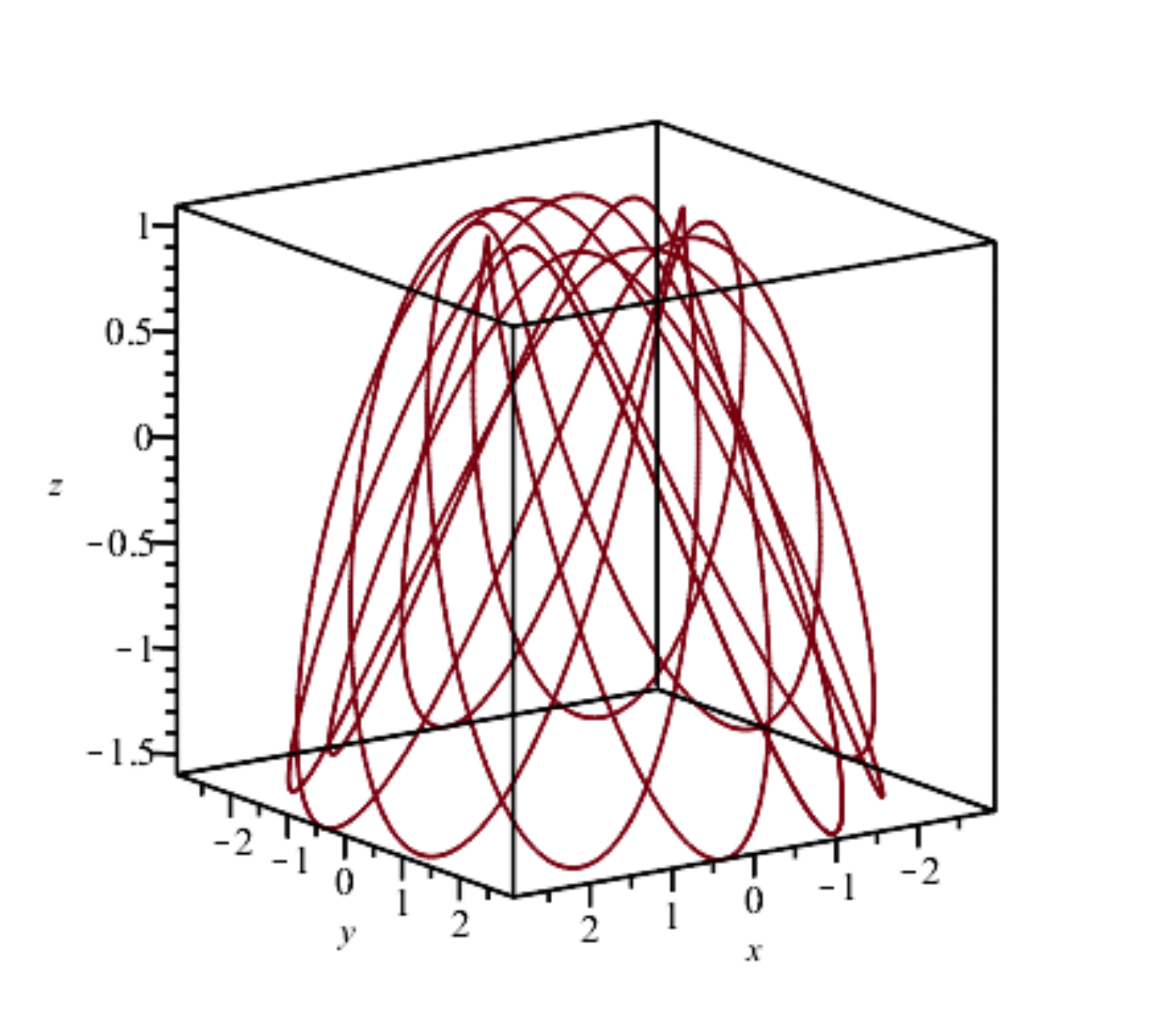}
	\caption{A trajectory for system \eqref{RLIII} with rational fraction $\frac{p_\phi}{b}=\frac{4}{3}$. The constants read $W_1 = \frac{3}{4}, W_2 = -5, W_3 = 1, \kappa_1 = 1$ and the initial values are $x(0)=1,y(0)=0,z(0)=1,p_x(0)=0,p_y(0)=2,p_z(0)=1$.}
	\label{fig:RLIII2}
\end{figure}

	The stationary Hamilton-Jacobi equation separates again. The choice of gauge is clearly \eqref{gauge konst pole}, so that $\phi$ is cyclic. We substitute the \emph{ansatz} $U=u(r)+p_\phi \phi+v(Z)$ and obtain
	\begin{equation}
		u'(r)^2+4\omega^2 r^2 + \frac{b^2}{r^2} = c_1=-v'(Z)^2 -\omega^2 Z^2 - 2 W_3 Z - \kappa_1 p_\phi+2E.
\end{equation} 
The energy $E$ and the separation constant $c$ determine the values range of values of $r$ and $Z$ because $u'(r)$ and $v'(Z)$ must be real. That implies that we can take the square roots in the following solutions if $-b^2=W_2-p_\phi^2>0$:
\begin{align}
\begin{split}
u(r)&=\frac{\sqrt{-b^2}}{2} \ln\left(\frac{2 c_1 r^2 - 2 b^2 + 2 \sqrt{-b^2} \sqrt{-4 \omega^2 r^4 + 2 c_1 r^2 - b^2}}{r^2}\right) +c_3- \\
&-\frac{c_1}{4\omega} \arctan\left(\frac{2 \omega \left(r^2 -\frac{c_1}{4 \omega^2}\right)}{\sqrt{-4 \omega^2 r^4 + 2 c_1 r^2 - b^2}}\right) - \frac{\sqrt{-4 \omega^2 r^4 + 2 c_1 r^2 - b^2}}{2},
\end{split}
\end{align}
\begin{align}
\begin{split}
	v(Z)&={\frac { \left( Z{\omega}^{2}+W_{{3}} \right) \sqrt {-{\omega}^{2}{Z}^{2}-2 W_{{3}}Z+d}}{2{\omega}^{2}}}+\\
	&+{\frac {d{\omega}^{2}+{W_{{3}}}^{2}}{2{\omega}^{3}}\arctan \left( {\frac {Z{\omega}^{2}+W_{{3}} }{\omega \sqrt{-{\omega}^{2}{Z}^{2}-2 W_{{3}}Z+d}}} \right)}+c_4,
\end{split}
\end{align}
where $d =2 E -\kappa_1 p_\phi - 2 c_1$.

	\item\label{RL IV} The last system is defined by 
	\begin{align}
	\begin{split}
		B^r ={}& -\left(\kappa_1 Z + \frac{\kappa_2}{2}\right) r^2,\quad B^\phi = 0, \quad B^Z = \kappa_1 r Z^2 + \frac{\kappa_1 r^3}{2} + \kappa_2 r Z + \kappa_3 r,\\ 
		W = {}&-\frac{\kappa_1^2 r^2 Z^4}{8} - \frac{\kappa_1 \kappa_2 r^2 Z^3}{4} - \left(\frac{\kappa_1^2 r^4}{16} + \frac{\kappa_1 \kappa_3 r^2}{4} + \frac{\kappa_2^2 r^2}{8} - \frac{\kappa_3^2}{2} -2 W_1\right) Z^2 -\\
		&- \left(\frac{\kappa_1 \kappa_2 r^4}{16} + \frac{\kappa_2 \kappa_3 r^2}{4} - W_2\right) Z - \frac{\kappa_1^2 r^6}{128} - \frac{\kappa_1 \kappa_3 r^4}{16} + \frac{W_1 r^2}{2} - \frac{W_3}{2 r^2},
	\end{split}
	\end{align}
which means that the magnetic field in the Cartesian coordinates reads
\begin{equation}
	B^x= -\left(\kappa_1 z + \frac{\kappa_2}{2}\right)x,\ B^y= -\left(\kappa_1 z + \frac{\kappa_2}{2}\right)y,\ B^z=\kappa_1 z^2 + \kappa_2 z+ \frac{\kappa_1 r^2}{2} + \kappa_3.
\end{equation}
We simplify the following expressions by the shift in $Z\to Z-\frac{\kappa_2}{2\kappa_1}$ and redefine the constant $\kappa_3$ so that we have the previous expressions with $\kappa_2=0$. We therefore have
	\begin{align}
\begin{split}
B^r ={}& -\kappa_1 r^2 Z,\quad B^\phi = 0, \quad B^Z = \kappa_1 r Z^2 + \frac{\kappa_1 r^3}{2} + \kappa_3 r,\\ 
W = {}&-\frac{\kappa_1^2 r^2 Z^4}{8} - \left(\frac{\kappa_1^2 r^4}{16} + \frac{\kappa_1 \kappa_3 r^2}{4} - \frac{\kappa_3^2}{2} -2 W_1\right) Z^2 +\\
&+ W_2 Z - \frac{\kappa_1^2 r^6}{128} - \frac{\kappa_1 \kappa_3 r^4}{16} + \frac{W_1 r^2}{2} - \frac{W_3}{2 r^2},
\end{split}
\end{align}
which means that the magnetic field in the Cartesian coordinates reads
\begin{equation}
B^x= -\kappa_1 x z,\quad B^y= -\kappa_1 y z,\quad B^z=\kappa_1 z^2 +\frac{\kappa_1 r^2}{2} + \kappa_3.
\end{equation}
We assume $\kappa_1\neq 0$, for otherwise it reduces to the previous system.

The integrals of motion read
\begin{align}
\begin{split}
\tilde{X}_1&=p_\phi^A-\frac{\kappa_1 r^2 Z^2}{2} - \frac{\kappa_1 r^4}{8} -
\frac{\kappa_3 r^2}{2},\\
	X_2&=\left(p_Z^A\right)^2 +\kappa_1 Z^2 p_\phi^A -\frac{\kappa_1^2 r^2 Z^4}{2} -\left(\frac{\kappa_1^2 r^4}{8} + \frac{\kappa_1 \kappa_3 r^2}{2} -\kappa_3^2 - 4 W_1\right) Z^2 - 2 W_2 Z,
\end{split}\\
\begin{split}
	X_3&=r p_r^A p_Z^A - Z\left(p_r^A\right)^2 - Z\frac{\left(p_\phi^A\right)^2}{r^2} +Z\left(\kappa_1 Z^2 + \frac{\kappa_1 r^2}{2} + \kappa_3\right) p_\phi^A + \frac{W_2}{2} r^2-\\
&- \frac{\kappa_1^2}{64} r^2 Z(3r^2 + 4 Z^2)(r^2 + 4 Z^2)-\kappa_1 \kappa_3 r^2Z\left(\frac{r^2}{4}+\frac{Z^2}{2}\right)
+\left( W_1 r^2 + \frac{W_3}{r^2}\right) Z.
\end{split}
\end{align}
This system is again new to the literature, as far as we know. Setting $\kappa_2=0$ does not yield additional integrals even with $W_1=W_2=0$.

The Poisson algebra has only one non-trivial Poisson bracket
\begin{equation}
	\begin{split}
	X_4&\coloneqq \{X_2,X_3\}_{\rm P.B.}=-2 \left((p^A_r)^2 +\frac{(p^A_\phi)^2}{r^2}\right) p^A_Z- 2 \kappa_1 Z p_\phi^A p_r^A+\\
	&+\left(\kappa_1 (2Z^2 + r^2) + 2\kappa_3\right) p_\phi^A p_Z^A+r \left[\kappa_1^2 r^2 Z^3+\right.\\&\left.
	+\left(\frac{\kappa_1^2 r^4}{4}+ \kappa_1 \kappa_3 r^2 - 2 \kappa_3^2 - 8 W_1\right) Z - 2 W_2\right] p^A_r-\\
	&- \left[\frac{\kappa_1^2 r^2Z^4}{2} + \frac{(\kappa_1^2 r^4 + 2 \kappa_1 \kappa_3 r^2) Z^2}{2}
	+ \frac{3 \kappa_1^2 r^6}{32}+\frac{\kappa_1 \kappa_3 r^4}{2}-2 W_1 r^-\frac{2 W_3}{r^2}\right] p_Z^A.\raisetag{4.5\baselineskip}
	\end{split}
\end{equation}
Thus obtained integral of motion is, however, dependent on the previous integrals due to the following relation
\begin{equation}
	\begin{split}
	X_4^2&=-16 X_2 H^2 +8 (2 X_2^2 +2 \kappa_3 X_1 X_2 + 2 W_2 X_3) H - 4 X_2^3- 8 \kappa_3 X_1 X_2^2+\\
	& + 4[\kappa_1 X_1^3 - \kappa_1 W_3 X_1 - \kappa_3^2 W_3 + 4 W_1 X_1^2 - 4 W_3 W_1 -2 W_2 X_3] X_2+\\
	&+4 (\kappa_1 X_1 + \kappa_3^2 +4 W_1) X_3^2 -8 \kappa_3 W_2 X_1 X_3 + 4 W_2^2 X_1^2 - 4 W_3 W_2^2.
	\end{split}\raisetag{\baselineskip}
\end{equation}

To analyse trajectories we choose the natural gauge, which assures that $\tilde{X}_1=p_\phi$, namely
\begin{equation}\label{gauge RL IV}
A_r=0,\quad A_\phi=\frac{\kappa_1 r^2 Z^2}{2} + \frac{\kappa_1 r^4}{8} +\frac{\kappa_3 r^2}{2},\quad A_Z=0.
\end{equation}
The Hamiltonian equations read
\begin{gather}
	\dot{p}_r =-\frac{\omega^2 r}{4} +\frac{b}{r^3},\quad \dot{p}_\phi = 0,\quad \dot{p}_Z =- \omega^2 Z - W_2,\\
	\dot{r} = p_r,\quad \dot{\phi} = \frac{\kappa_1 r^2}{8} + \frac{\kappa_1 Z^2 + \kappa_3}{2} + \frac{p_\phi}{r^2},\quad \dot{Z} = p_Z,
\end{gather}
where $\omega=\sqrt{\kappa_1 p_\phi + \kappa_3^2 + 4 W_1}$ and $b=p_\phi^2 - W_3$, which are constants due to the second equation ($p_\phi(t)=p_\phi^0$). The equations for $r$ and $Z$ do not depend on other variables. The solution to the equation for $Z$ reads
\begin{equation}
	Z(t) = \frac{p_Z^0}{\omega}\sin(\omega t) + \frac{(Z_0 \omega^2 + W_2)}{\omega^2}\cos(\omega t) -\frac{W_2}{ \omega^2},
\end{equation}
so we see that we need $\omega$ to be real, that is $\kappa_1 p_\phi + \kappa_3^2 + 4 W_1>0$, to have bounded trajectories in the $z$-direction. We continue with this assumption.
 
The equation for $r$, 
\begin{equation}\label{aa}
	\ddot{r}=-\frac{\omega^2 r}{4} +\frac{b}{r^3},
\end{equation}
is reduced by the integrating factor $\dot{r}$ and its solution with standard initial conditions $r(0)=r_0 >0,$ $\dot{r}(0)=\dot{r}_0\equiv p_r^0$ reads
\begin{equation}
	r(t)=\frac{\sqrt{2 c_1+ 2\sqrt{(c_1^2 - 16 b \omega^2 r_0^4)} \sin(\omega t + \theta)}}{2 r_0 \omega},
\end{equation}
where 
\begin{gather}\label{sol}
c_1=\omega^2 r_0^4 + 4 \dot{r}_0^2 r_0^2 + 4 b,\quad
	\theta=\arctan\left(\frac{-\omega^2 r_0^4 + 4 \dot{r}_0^2 r_0^2 + 4 b}{4 \omega \dot{r}_0 r_0^3}\right).
\end{gather}
(It includes the constant solution which we obtain if the inner square root vanishes, i.e. $c_1^2 = 16 b \omega^2 r_0^4$.)
We have chosen the plus sign in front of the sine because we could redefine $\theta$ (shift the time $t$). This solution is periodic if $b>0$, i.e. $\left(p_\phi^0\right)^2>W_3$, otherwise the trajectory ends at the singularity $r=0$. This dependence on angular momenta is reminiscent of Newton's cannonball \cite{Newton}, where high enough velocity perpendicular to the gravity prevents the cannonball from falling back on Earth.

The last unknown is $\phi$, which can now be obtained by a quadrature. The result is a sum of an arctan of a periodic argument with period $\frac{2\pi}{\omega}$, a polynomial in sines and cosine of $\frac{\omega t}{2}$ and the term $A t$, where the constant
\begin{equation}
	A=\frac{\kappa_1 Z_0^2+ 2 \kappa_3}{4}+ \frac{\kappa_1 (2Z_0 W_2 + p_Z^2)}{4 \omega^2}+\frac{c_1^2}{32 \omega^2 r_0^2} +\frac{3\kappa_1 W_2^2}{4 \omega^4}
\end{equation}
depends on the initial values. We conclude that the trajectories can be periodic only if $A$ is commensurable with $\omega$, which is not satisfied for all initial values, the system therefore cannot be maximally superintegrable for general initial conditions. However, it can be so called particularly superintegrable \cite{Turbiner}, which means that there are integrals of motion on some common invariant subspace.

The stationary Hamilton-Jacobi equation is separable again. The azimuthal coordinate $\phi$ is cyclic in our gauge \eqref{gauge RL IV} and the \emph{ansatz} $U=u(r)+p_\phi \phi+v(Z)$ leads to the following separation
\begin{equation}
u'(r)^2 +\alpha_1 ^2 r^2-\frac{\alpha_2^2}{r^2}= 2 c_1=-v'(Z)^2 -4 \alpha_1 ^2 Z^2 - 8 W_2 Z - \kappa_3 p_\phi + 2 E,
\end{equation}
where we define the constants as follows
\begin{equation}
\begin{split}
\alpha_1  = \sqrt{\frac{\kappa_1 p_\phi}{4} + \frac{\kappa_3^2}{4} + W_1},\quad \alpha_2 = \sqrt{W_3-p_\phi^2},\quad \alpha_3=2E -2 c_1-\kappa_3 p_\phi.
\end{split}
\end{equation}
The results are
\begin{align}
	\begin{split}
	u(r) ={}& \frac{c_1}{2\alpha_1 } \arctan\left(\frac{\alpha_1 ^2 r^2 - c_1}{\alpha_1  \sqrt{-\alpha_1 ^2 r^4 + 2 c_1 r^2 + \alpha_2^2}}\right)+ \frac{\sqrt{-\alpha_1 ^2 r^4 + 2 c_1 r^2 + \alpha_2^2}}{2}-\\
	& - \frac{\alpha_2}{2} \ln\left(\frac{\alpha_2 \sqrt{-\alpha_1 ^2 r^4 + 2 c_1 r^2 + \alpha_2^2} + c_1 r^2 + \alpha_2^2}{r^2}\right) +c_2,\raisetag{1.25\baselineskip}
	\end{split}\\
	\begin{split}
	v(Z)={}&\left(\frac{\alpha_3\alpha_1 }{4}+\frac{W_2^2}{\alpha_1 ^3}\right) \arctan\left(\frac{2 \alpha_1 ^2 Z + 2 W_2}{\alpha_1  \sqrt{-4 \alpha_1 ^2 Z^2 -8 W_2 Z+\alpha_3}}\right) +\\
	&+ \left(\frac{Z}{2}+\frac{W_2^2}{\alpha_1 ^2}\right) \sqrt{-4 \alpha_1 ^2 Z^2 - 8 W_0 Z +\alpha_3}+c_3.
	\end{split}
\end{align}
As in the previous cases, the constants $E$ and $c_1$ must be such that the square roots are well defined for the ranges of $r$ and $Z$ we choose (if possible).
\end{enumerate}

To sum up: In this subsection we analysed 4 cylindrical superintegrable systems with additional integral of type $L_x p_y-L_y p_x+\ldots$, all of them only minimally superintegrable with integrals of order at most 2. For system \ref{RL II}, see \eqref{RLII}, the Hamilton-Jacobi equation separates, but we were not able to solve the resulting quadratures with an exception of a very special case. Numerical integration showed that bounded trajectories are not closed in general (for all initial values), thus this system is not maximally superintegrable. System \ref{RL IV} and its special case system \ref{RL III} are in general only minimally superintegrable.  We were able to calculate the trajectories analytically, but the results for $\phi$ contain singular functions $\arctan(k_1 \tan(\omega t)+k_2)$. In system \ref{RL IV} the angular frequency of the trajectories $\omega$ depends on the initial values, so this system is at most particularly maximally superintegrable \cite{Turbiner}. Even though $\omega$ does not depend on initial values for system \ref{RL III}, the singular nature of $\phi(t)$ in eq.~\eqref{phi} together with its definition mod $2\pi$ causes some problems with interpretation of the results. Numerical integration of the Hamilton's equations showed us that the trajectories are not closed if $\frac{p_\phi}{b}\neq\frac{n}{m}$, $n,m\in \N$. (The fraction $\frac{p_\phi}{b}$ is the coefficient in front of the arctan in eq.~\eqref{phi}.) Assuming we can trust the numerical results, this shows that the system is not maximally superintegrable because the trajectories are closed for some initial values only, but can be particularly maximally superintegrable~\cite{Turbiner}.


	%

\chapter*{Conclusions} 
\addcontentsline{toc}{chapter}{Conclusions} 
This thesis is a contribution to the field of integrability and superintegrability focusing on systems with magnetic field of the so called-cylindrical type, i.e. with integrals of the type $(p_\phi^A)^2+\ldots$ and $(p_Z^A)^2+\ldots$, where $p_i^A=p_i+A_i$ are so-called gauge covariant momenta, see \eqref{cov mom CM}.

In Chapter \ref{kap corr} we considered integrable systems. We extended the previously known classical determining equation \eqref{ord3}--\eqref{ord0} to quantum mechanics (Section~\ref{sec: det eq QM}) and found that the quantum correction does not modify equations for first order integrals. For second order integrals it modifies the zeroth order equations only. We also obtained the general form of the correction in cylindrical coordinates, see eq.~\eqref{corr cyl}, but it is not very suitable for non-cylindrical integrals.
	Considering the cylindrical integrals \eqref{cyl integrals}, only the zeroth order equation for $X_1$ is modified, the correction being \eqref{corr pphi}. The apparent corrections from the involutivity $[\hat{X}_1,\hat{X}_2]=0$ in the lower order equations vanish as a consequence of the equations of higher order. (We do not need to use the determining equations coming from the commutator with the Hamiltonian).

The main results of Chapter \ref{kap corr} are in Section~\ref{sec:quantum integr}, where we obtained all cylindrical quantum integrable systems by solving the corresponding determining equations. We followed the analysis from the classical case \cite{Fournier2019} because the only determining equation differing from the classical case is the zeroth order equation~\eqref{cyl0}. The determining equations were reduced to equations \eqref{reducedAa}--\eqref{matrixform2} containing 5 auxiliary functions of one variable $\rho(r)$, $\sigma(r)$, $\psi(\phi)$, $\tau(\phi)$ and $\mu(Z)$ which determine the first order terms of the cylindrical integrals \eqref{cyl integrals} and the magnetic field $\vec{B}$, see	eq.~\eqref{scond} and \eqref{Bcond}. Equation \eqref{matrixform2} contains a matrix $M$ which depends on the auxiliary functions only and allows us to split the considerations according to its rank. In \cite{Fournier2019} it was shown that rank 0 and 3 are either impossible (assuming non-vanishing magnetic field) or inconsistent with the other equations and the arguments remain valid in quantum mechanics. Ranks 1 and 2 split further into subcases. 
In subcase a) the quantum correction vanishes and the obtained systems are therefore the same as in classical mechanics in \cite{Fournier2019}. (The key results are cited in the corresponding subsections of Section~\ref{sec:first ord}.) In subcase b) the quantum correction is \emph{a priori} non-trivial, but vanishes in some cases due to the consistency conditions on the scalar potential $W$.
 We have obtained 3 systems with non-vanishing correction:

	\begin{enumerate}
		\item 
		The first system has the magnetic field $\vec{B}$ in the $z$-direction given in \eqref{11b} and scalar potential $W$ not depending on $Z$ in \eqref{1.1W}. The integral $X_1$ is determined from \eqref{1.1 sm}, the cylindrical integral $X_2$ reduces to the first order integral which in suitable gauge reads $X_2=p_z$. We therefore have an integrable motion in 2D (studied in \cite{Berube}) complemented by free motion in the $Z$-direction.
		\item 
		This is the previous system with the 
		scalar potential modified to $W=W_{12}(r,\phi)+W_3(Z)$ with $W_{12}(r,\phi)$ from eq.~\eqref{1.1W} and $W_3(Z)$ unconstrained. The 2D integrable motion is complemented by the motion in the $Z$-direction ruled by $W_3(Z)$, which has the second order cylindrical integral $X_2=(p_Z^A)^2+2 W_3(Z).$
		\item 
		The system with the magnetic field $\vec{B}$ from eq.~\eqref{12B}  and the scalar potential \eqref{12W} not depending on $Z$. Both are expressed in terms of one function $\beta(\phi)$ which satisfies the nonlinear ODE \eqref{betasimple}. We again have $X_2=p_z$ in suitably chosen gauge.
	\end{enumerate}

All quantum cylindrical quadratic integrable systems have the same magnetic field as their classical counterparts. The scalar potential $W$ is in general modified by an $\hbar^2$-dependent correction, although it sometimes vanishes due to other constraints.

In Chapter \ref{kap:sup} we looked for superintegrable systems among the known cylindrical integrable cases.
	Because the analysis of the second order determining equations \eqref{ord3}--\eqref{ord0} was too computationally demanding even with restriction to the integrable cases, we had to restrict to a less ambitious goal.
	
In Section~\ref{sec:first ord} we found all first order superintegrable cylindrical systems, all of which were already known to the literature, and excluded any further cases. The 3 found systems are as follows: In Subsection~\ref{sec:case 2a lin} we have the maximally superintegrable system with constant magnetic field and potential studied already in \cite{Landau}, see \cite{Marchesiello2015} as well. In Subsection~\ref{sec: 3a 1a} there is a class of systems with constant magnetic field and $W(z)$ from \cite{Marchesiello2019}, which reduces to a 2D systems without magnetic field and in Subsection~\ref{sec:2b 1.2} subcase \ref{system 2b 1.2} we have another system separable in the Cartesian as well as cylindrical coordinates from \cite{Marchesiello2019}.

As was shown in Subsection~\ref{sec:general}, the additional integrals must be $p_x^A+m(\vec{x})$ or $p_y^A+m(\vec{x})$ (in a suitably chosen reference frame), where $m$ is a function. In all three cases we can choose the gauge so that they are $p_x$ or $p_y$. The same is true for the first order reduced cylindrical integral $\tilde{X}_1$, which in suitable gauge reads $\tilde{X}_1=p_\phi$. The corresponding Hamilton-Jacobi and Schr\"odinger equations separate in the cylindrical as well as Cartesian coordinates and the equations are solved (sometimes in terms of special functions) in the corresponding subsections. The spectrum of the Hamiltonian is continuous in all three cases. We also solved the Hamilton's equations of motion and analysed the trajectories.

In Section~\ref{sec:second ord} we searched for second order superintegrable systems. In Subsection~\ref{sec:biquadr} we present the system which we found during our attempt to solve the second order determining equations in general. It is a special case of Case I d) in \cite{Marchesiello2019}, is second order minimally superintegrable and separates both in 
the Cartesian and cylindrical coordinates. The trajectories are unbounded with an exception of some singular values of the constants, so we do not obtain any information on higher order integrals.

In Subsections~\ref{sec:L^2} and \ref{sec:Runge-Lenz} we used the physically motivated \emph{ans\"atze} $L^2+\ldots$ and $L_x p_y-L_y p_x+\ldots$, respectively. We obtained all classical cylindrical systems with additional integrals of this type, see the list in the corresponding subsections. (Quantum corrections for the considered type of integrals are non-trivial, so we postpone the quantum cases to a later work.) Almost all the systems were new to the literature. All found system were only minimally quadratically superintegrable but had at least one first order integral, namely $\tilde{X}_1=p_\phi$ in suitably chosen gauge. The Hamilton-Jacobi equations were separable in the cylindrical coordinates. (They have only one Cartesian integral $X_2=(p_z^A)^2+\ldots$, which is not sufficient for separability in the Cartesian coordinates.) Despite this fact we were not able to obtain the trajectories analytically in all cases because they sometimes lead to quadratures not solvable in terms of known functions (as far as we know). We analysed the systems numerically and there is only one system, namely system \ref{L^2 II} in Subsection~\ref{sec:L^2}, whose bounded trajectories are closed and is therefore a promising candidate for search for higher order integrals, which we postponed to a later work. Systems \ref{RL III} and \ref{RL IV} in Subsection~\ref{sec:Runge-Lenz} had satisfied the condition for some initial values only, and are not maximally superintegrable, but they could be so called particularly maximally superintegrable \cite{Turbiner}, i.e. could have 5 independent integrals on some common invariant subspace of the phase space.

There are a lot of aspects to investigate further. Those we plan to study in our future work include the following: We will analyse the quantum version of the found systems in order to study the classical-quantum correspondence and look for purely quantum systems, find all second order superintegrable systems of the cylindrical type and search for higher order integrals, and study solving of Hamilton-Jacobi and Schr\"odinger equations for non-separable systems (quasi-separation or separation though canonical transformations mixing coordinates and momenta).

	\clearpage 
	\addcontentsline{toc}{chapter}{References} 

\bibliographystyle{abbrv}
	\bibliography{diplomka}

\begin{thebibliography}{10}

\bibitem{AbramowitzStegun}
M.~Abramowitz and I.~A. Stegun.
\newblock {\em Handbook of mathematical functions with formulas, graphs, and
  mathematical tables}, volume~55 of {\em National Bureau of Standards Applied
  Mathematics Series}.
\newblock U.S. Government Printing Office, Washington, D.C., 1972.

\bibitem{Benenti}
S.~Benenti, C.~Chanu, and G.~Rastelli.
\newblock Variable separation for natural {H}amiltonians with scalar and vector
  potentials on {R}iemannian manifolds.
\newblock {\em J. Math. Phys.}, 42(5):2065--2091, 2001.

\bibitem{Bertrand}
J.~Bertrand.
\newblock Th\'eor\`eme relatif au mouvement d'un point attir\'e vers un centre
  fixe.
\newblock {\em C. R. Acad. Sci}, 77:849--853, 1873.

\bibitem{BertrandSnobl}
S.~Bertrand and L.~\v{S}nobl.
\newblock On rotationally invariant integrable and superintegrable classical
  systems in magnetic fields with non-subgroup type integrals.
\newblock {\em J. Phys. A}, 52(19):195201, 25, 2019.

\bibitem{Berube}
J.~B\'{e}rub\'{e} and P.~Winternitz.
\newblock Integrable and superintegrable quantum systems in a magnetic field.
\newblock {\em J. Math. Phys.}, 45(5):1959--1973, 2004.

\bibitem{Charest}
F.~Charest, C.~Hudon, and P.~Winternitz.
\newblock Quasiseparation of variables in the {S}chr\"{o}dinger equation with a
  magnetic field.
\newblock {\em J. Math. Phys.}, 48(1):012105, 16, 2007.

\bibitem{DLMF}
{\it NIST Digital Library of Mathematical Functions}.
\newblock http://dlmf.nist.gov/, Release 1.0.25 of 2019-12-15.
\newblock F.~W.~J. Olver, A.~B. {Olde Daalhuis}, D.~W. Lozier, B.~I. Schneider,
  R.~F. Boisvert, C.~W. Clark, B.~R. Miller, B.~V. Saunders, H.~S. Cohl, and
  M.~A. McClain, eds.

\bibitem{Dorizzi}
B.~Dorizzi, B.~Grammaticos, A.~Ramani, and P.~Winternitz.
\newblock Integrable {H}amiltonian systems with velocity-dependent potentials.
\newblock {\em J. Math. Phys.}, 26(12):3070--3079, 1985.

\bibitem{Eisenhart}
L.~P. Eisenhart.
\newblock Enumeration of potentials for which one-particle {S}chroedinger
  equations are separable.
\newblock {\em Phys. Rev. (2)}, 74:87--89, 1948.

\bibitem{Evans}
N.~W. Evans.
\newblock Superintegrability in classical mechanics.
\newblock {\em Phys. Rev. A (3)}, 41(10):5666--5676, 1990.

\bibitem{Fournier2019}
F.~Fournier, L.~\v{S}nobl, and P.~Winternitz.
\newblock Cylindrical type integrable classical systems in a magnetic field.
\newblock {\em Journal of Physics A: Mathematical and Theoretical},
  53(8):085203, 2020.

\bibitem{FrisMandrosov}
J.~Fri\v{s}, V.~Mandrosov, Y.~A. Smorodinsky, M.~Uhl\'{\i}\v{r}, and
  P.~Winternitz.
\newblock On higher symmetries in quantum mechanics.
\newblock {\em Phys. Lett.}, 16:354--356, 1965.

\bibitem{Goldstein2001}
H.~Goldstein, C.~P. Poole, and J.~L. Safko.
\newblock {\em Classical Mechanics, 3rd Edition}.
\newblock New York: {P}earson, 3rd edition, 2001.

\bibitem{Jauch}
J.~M. Jauch and E.~L. Hill.
\newblock On the problem of degeneracy in quantum mechanics.
\newblock {\em Phys. Rev.}, 57:641--645, 1940.

\bibitem{Landau}
L.~D. Landau and E.~M. Lifshitz.
\newblock {\em Quantum mechanics: non-relativistic theory. {C}ourse of
  {T}heoretical {P}hysics, {V}ol. 3}.
\newblock Addison-Wesley Series in Advanced Physics. Pergamon Press Ltd.,
  London-Paris, 1958.
\newblock Translated from the Russian by J. B. Sykes and J. S. Bell.

\bibitem{Maple}
Maplesoft.
\newblock Maple 2019.2, 2019.
\newblock [software], Maplesoft, a division of Waterloo Maple Inc., Waterloo,
  Ontario.

\bibitem{Marchesiello2017}
A.~Marchesiello and L.~\v{S}nobl.
\newblock Superintegrable 3{D} systems in a magnetic field corresponding to
  {C}artesian separation of variables.
\newblock {\em J. Phys. A}, 50(24):245202, 24, 2017.

\bibitem{Marchesiello2018}
A.~Marchesiello and L.~\v{S}nobl.
\newblock An infinite family of maximally superintegrable systems in a magnetic
  field with higher order integrals.
\newblock {\em SIGMA Symmetry Integrability Geom. Methods Appl.}, 14:092, 11,
  2018.

\bibitem{Marchesiello2019}
A.~Marchesiello and L.~\v{S}nobl.
\newblock Classical superintegrable systems in a magnetic field that separate
  in {C}artesian coordinates.
\newblock {\em SIGMA Symmetry Integrability Geom. Methods Appl.}, 16:015, 35
  pages, 2020.

\bibitem{Marchesiello2015}
A.~Marchesiello, L.~\v{S}nobl, and P.~Winternitz.
\newblock Three-dimensional superintegrable systems in a static electromagnetic
  field.
\newblock {\em J. Phys. A}, 48(39):395206, 24, 2015.

\bibitem{Marchesiello2018Sph}
A.~Marchesiello, L.~\v{S}nobl, and P.~Winternitz.
\newblock Spherical type integrable classical systems in a magnetic field.
\newblock {\em J. Phys. A}, 51(13):135205, 24, 2018.

\bibitem{McSween2000}
E.~McSween and P.~Winternitz.
\newblock Integrable and superintegrable {H}amiltonian systems in magnetic
  fields.
\newblock {\em J. Math. Phys.}, 41(5):2957--2967, 2000.

\bibitem{Miller2013}
W.~Miller, Jr., S.~Post, and P.~Winternitz.
\newblock Classical and quantum superintegrability with applications.
\newblock {\em J. Phys. A}, 46(42):423001, 97, 2013.

\bibitem{Newton}
I.~Newton.
\newblock {\em A treatise of the system of the world}.
\newblock Translated from the Latin. With an introduction by I. Bernard Cohen.
  Dawsons of Pall Mall, London, 1969.

\bibitem{Shankar}
R.~Shankar.
\newblock {\em Principles of quantum mechanics}.
\newblock Plenum Press, New York, second edition, 1994.

\bibitem{TTW}
P.~Tempesta, A.~V. Turbiner, and P.~Winternitz.
\newblock Exact solvability of superintegrable systems.
\newblock {\em J. Math. Phys.}, 42(9):4248--4257, 2001.

\bibitem{Turbiner}
A.~V. Turbiner.
\newblock Particular integrability and (quasi)-exact-solvability.
\newblock {\em J. Phys. A}, 46(2):025203, 9, 2013.

\bibitem{WinternitzSmorodinski}
P.~Winternitz, Y.~A. Smorodinsky, M.~Uhl\'{\i}\v{r}, and I.~Fri\v{s}.
\newblock Symmetry groups in classical and quantum mechanics.
\newblock {\em Soviet J. Nuclear Phys.}, 4:444--450, 1967.

\bibitem{Mathematica}
{Wolfram Research{,} Inc.}
\newblock Mathematica, {V}ersion 10.4, 2016.
\newblock [software], Champaign, IL, 2016.

\bibitem{Zhalij_2015}
A.~Zhalij.
\newblock Quantum integrable systems in three-dimensional magnetic fields: the
  cartesian case.
\newblock {\em Journal of Physics: Conference Series}, 621:012019, 2015.

\end{thebibliography}


	
	%
	%
	
\end{document}